\documentclass[journal]{IEEEtran}

\usepackage[cmex10]{amsmath}
\usepackage{amsthm}
\usepackage{bbm}
\usepackage{color}
\usepackage{cite}

\usepackage{amssymb}
\usepackage{stmaryrd}
\usepackage{epic,eepic}

\usepackage{algorithm}
\usepackage{algorithmic}
\usepackage{mathtools}

\usepackage{epsf}           
\usepackage{graphicx}       
\usepackage{epsfig}         
\usepackage{pstricks}
\usepackage{pstricks-add}
\usepackage{pst-plot}
\usepackage{dsfont}
\usepackage{hypernat}
\usepackage{hyperref}

\usepackage{shortvrb,psfrag}
\usepackage{dsfont}
\usepackage{tikz}
\usepackage[straightvoltages]{circuitikz}
\usepackage{pgfplots}
\usepackage{tikz-cd}
\usetikzlibrary{arrows, positioning, arrows.meta, calc}
\usetikzlibrary{patterns, shapes.geometric}

\tikzset{
  shift left/.style ={commutative diagrams/shift left={#1}},
  shift right/.style={commutative diagrams/shift right={#1}}
}

\tikzstyle myBG=[line width=3pt,opacity=1]

\newtheorem{theorem}{Theorem}
\newtheorem{remark}{Remark}
\newtheorem{definition}{Definition}
\newtheorem{proposition}{Proposition}
\newtheorem{lemma}{Lemma}
\newtheorem{corollary}{Corollary}

\newcommand{\mc}{\mathcal}

\font\dsrom=dsrom10 scaled 1200 \def \indic{\textrm{\dsrom{1}}}
\newcommand{\UN}{\indic}

\newcommand{\N}{\mathbb{N}}

\DeclareMathOperator*{\argmin}{\arg\!\min}
\DeclareMathOperator*{\supp}{supp}
\DeclareMathOperator*{\Aut}{Aut}
\DeclareMathOperator*{\Unif}{Unif}
\DeclareMathOperator*{\Ima}{Im}

\DeclareMathOperator*{\argmax}{\arg\!\max}
\DeclareMathOperator*{\interior}{int}

\newcommand{\mlt}[1]{\textcolor{black}{#1}}
\newcommand{\ar}[1]{\textcolor{black}{#1}}

\begin{document}

\title{On the \mlt{Additivity} of Optimal Rates for Independent Zero-Error Source and Channel Problems}

\author{
\IEEEauthorblockN{Nicolas~Charpenay,~
\IEEEmembership{Student member,~IEEE},
}
\IEEEauthorblockN{Ma\"{e}l Le~Treust,~
\IEEEmembership{Member,~IEEE},
}
and
\IEEEauthorblockN{Aline~Roumy,~
\IEEEmembership{Member,~IEEE}
}
\thanks{The work of Nicolas~Charpenay was conducted during his PhD at IRISA UMR 6074 and Centre Inria de l'Universit\'{e} de Rennes, funded by CDSN ENS Paris-Saclay. The work of Ma\"{e}l Le Treust is funded by the French government under the France 2030 ANR program “PEPR Networks of the Future” (ref. ANR-22-PEFT-0010).}
\thanks{This work was presented in part at the IEEE International Symposium on Information Theory (ISIT) 2023 \cite{CharpenayISIT23}, and in part at the IEEE Information Theory Workshop (ITW) 2023 \cite{CharpenayITW23}.}
\thanks{Nicolas~Charpenay is with Univ Rennes, CNRS, IRMAR UMR 6625, F-35000 Rennes, France (e-mail: nicolas.charpenay@univ-rennes.fr).}
\thanks{Ma\"{e}l Le Treust is with Univ. Rennes, CNRS, Inria, IRISA UMR 6074, F-35000 Rennes, France (e-mail: mael.le-treust@cnrs.fr).}
\thanks{Aline Roumy is with Inria Center at Rennes University, France (e-mail: aline.roumy@inria.fr).}
}

\maketitle

\begin{abstract}

Zero-error coding encompasses a variety of source and channel problems where the probability of error must be exactly zero. This condition is stricter than that of the vanishing error regime, where the error probability goes to zero as the code blocklength goes to infinity. 
In general, zero-error coding is an open combinatorial question. 
%
We investigate two unsolved zero-error problems: the source coding problem with side information and the channel coding problem. We focus our attention on families of independent problems for which the \mlt{probability} distribution decomposes into a product of \mlt{probability} distributions. A crucial step is the \mlt{additivity} property of the optimal rate, which does not always hold in the zero-error regime, unlike in the vanishing error regime. \mlt{When the additivity holds, the concatenation of optimal codes is optimal.} We derive a condition under which the \mlt{additivity} \ar{of} the complementary graph entropy $\overline{H}$ for the AND product of \mlt{graphs} and for the disjoint union of graphs are equivalent. Then we establish the connection with a recent result obtained by Wigderson and Zuiddam and by Schrijver, for the zero-error capacity $C_0$. As a consequence, we provide new single-letter characterizations of $\overline{H}$ and $C_0$, for example when the graph is a product of perfect graphs, which is not perfect in general, and for the class of  graphs obtained by the product of a perfect graph $G$ with the pentagon graph $C_5$. By building on Haemers result for $C_0$, we also show that the \mlt{additivity} of $\overline{H}$ does not hold for the product of the Schl\"{a}fli graph with its complementary graph. \end{abstract}

\section{Introduction}


Transmitting information without any errors has been a concern for Shannon since the beginning of his work. In his seminal paper \cite{shannon1948mathematical}, Shannon proposed a construction for zero-error source coding, a problem soon solved by Huffman in \cite{huffman1952method}. Shortly after establishing the channel capacity in \cite{shannon1948mathematical}, Shannon turned his attention to channel coding with zero-error in \cite{shannon1956zero}, instead of vanishing error. This subtle difference radically changes the nature of the problem, essentially combinatorial rather than probabilistic. The single-letter characterization of the zero-error capacity is a notoriously difficult open problem. 
For example, the zero-error capacity of the noisy-typewriter channel with $7$ letters
is unknown \cite{de2024asymptotic}, some lower and upper bounds are stated in \cite{vesel2002improved}, \cite{codenotti03}, \cite{polak19}, \mlt{\cite{BuysPolakZuiddam_EuroComb25}}, see also \cite[pp.~122]{jurkiewicz2014survey}. In the zero-error regime, the exact values of the non-zero channel transition probabilities are irrelevant, only the support of the channel distribution matters. The zero-error property is translated into the \emph{characteristic graph} that encompasses the problem data in its structure: the vertices are the channel inputs $\mc{X}$, and two symbols $x$ and $x'$ are adjacent if they are ``confusable'', i.e. if they can produce the same channel output $y$ with positive probability. For sequences of symbols $x^n$, the characteristic graph is obtained by taking iteratively the AND product ($\wedge$), \mlt{where two sequences $x^n$ and ${x'}^n$ are adjacent if for all $t\in\{1,\ldots,n\}$ either $x_t={x'}_{\!\!t}$ or the symbols $x_t$ and ${x'}_{\!\!t}$ are adjacent}. In order to prevent any decoding error, a zero-error codebook must be composed of non-adjacent codewords. Thus, the size of the optimal codebook is given by the size of the maximal independent set, called the \emph{independence number}. In other words, the zero-error capacity $C_0$ is the asymptotic limit of the independence number of iterated AND product of the characteristic graph. 
Determining this independence number is an open question that has attracted a lot of attention in Information Theory \cite[Chap.~11]{csiszar2011information} and in Combinatorics and Graph Theory, see \cite[Chap. 27]{klavzar2011handbook}.

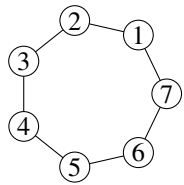
\begin{figure}[h!]
\centering
\begin{tikzpicture}
\foreach \i in {1,2,3,4,5,6,7}{
        \draw[] ($(360/7*\i-360/7:1)$) -- ($(360/7*\i:1)$);
        }
\foreach \i in {1,2,3,4,5,6,7}{
    \node (\i) at (\i*360/7:1) [shape=circle, fill = white, draw=black, inner sep=1pt] {\i};
    }
\end{tikzpicture}
    \caption{The characteristic graph $C_7$ of the noisy-typewriter channel with $7$ letters.}\label{fig:graphC7}
\end{figure}

This problem inspired Berge's notion of perfect graphs \cite[pp.~382]{berge73}, for which the zero-error capacity is given by the one-shot independence number \cite[Theorem 4.18]{grotschel1984polynomial}. Graphs with odd cycles are also related to Berge's conjecture \cite{Berge1961}, later proved  in \cite{Chudnovsky03} by Chudnovsky et al., namely ``a graph $G$ is perfect if and only if neither $G$ nor its complementary graph $\overline{G}$, have odd cycles of length $5$ or more.'' Since the zero-error capacity of the pentagon graph $C_5$ has been characterized by 
Lov{\'a}sz \cite{Lovasz79}, 
as well as the zero-error capacity for perfect graphs, see \cite[Theorem 4.18]{grotschel1984polynomial}, the graph $C_7$ depicted in Fig.~\ref{fig:graphC7}, is the minimal connected graph for which the zero-error capacity is an open problem.

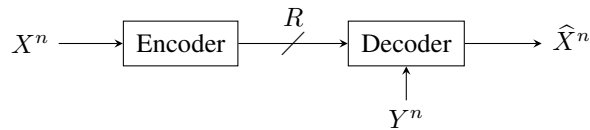
\begin{figure}[h!]
\centering
\begin{tikzpicture}[xscale=0.55,yscale=0.6]
\draw [>=stealth,->] (0,0.5) -- (2,0.5) ;
\draw  (1,1) node {$X^n$} ;
\draw (2,0) rectangle (5,1);
\draw (3.5,0.5) node {Encoder} ;
\draw [,>=stealth,->] (5,0.5) -- (8,0.5) ;
\draw (6.5,0.5) node {$\diagup$} ;
\draw (6.5,1.2) node {$R$} ;
\draw (8,0) rectangle (11,1);
\draw (9.5,0.5) node {Decoder} ;
\draw [>=stealth,->] (11,0.5) -- (13,0.5) ;
\draw  (12,1) node {$\widehat{X}^n$} ;
\draw [>=stealth,->] (9.5,-1) -- (9.5,0) ;
\draw  (9.5,-1.5) node {$Y^n$} ;
\end{tikzpicture}
\caption{The source coding problem with decoder side information, called \emph{side-information problem}.}
\label{fig:Zero-errorSW}
\end{figure}

In the source coding framework, Witsenhausen in \cite{witsenhausen1976zero} posed the question of the optimal compression rate when the decoder has side information. In this problem depicted in Fig.~\ref{fig:Zero-errorSW}, the encoder shares information about a source $X$, exploiting the side-information $Y$ observed by the decoder but not by itself. In the vanishing error regime, Slepian and Wolf in \cite{slepian1973noiseless} showed that the optimal rate is $H(X|Y)$. In the zero-error regime, no single-letter characterization is available. Similarly to the channel coding problem, the zero-error property is embedded into the \mlt{same} characteristic graph $G$, constructed with respect to the conditional distribution $P_{Y|X}$ of the ``side-information channel''. \mlt{The source distribution $P_X$ assigns a probability value to each vertex of the characteristic graph $G$, this defines the probabilistic graph $(G,P_X)$.} The main difference between \mlt{zero-error} source and \mlt{zero-error} channel coding problems is that the source distribution $P_X$ is given, \ar{whereas in channel coding} it is a degree of freedom allowing to maximize the \mlt{communication rate. This is a property common to both zero-error and vanishing-error regimes.} Witsenhausen showed in \cite{witsenhausen1976zero} that fixed-length \mlt{source} coding amounts to coloring the characteristic graph. In \cite{alon1996source}, Alon and Orlitsky studied variable-length coding and determined an asymptotic expression for the optimal rate based on the chromatic entropy. In \cite{koulgi2003zero}, Koulgi et al. proved that the optimal rate coincides with the \emph{complementary graph entropy} of the probabilistic graph $\overline{H}(G,P_X)$, 
introduced by K\"{o}rner and Longo in \cite{korner1973two} for the second step of the two-step source coding. These two expressions are asymptotic, as optimal source rates are determined by coloring an infinite product of graphs. \ar{Similarly to the zero-error capacity, single-letter characterizations are known only in a few cases, including} the pentagon graph $C_5$ \cite{koulgi2003zero} and perfect graphs \cite{csiszar1990entropy}. 
Another concept, called the \emph{K\"{o}rner graph entropy} \cite{Korner73}, provides a single-letter expression for the unrestricted input setting as considered by Alon and Orlitsky in \cite{alon1996source}. In this context, the zero-error constraint is met even outside the source's support, providing an upper bound on the optimal rate.

%
%
%
%

The difficulty in zero-error source and channel coding problems with side information lies in the asymmetry between the information known to the encoder and \mlt{to} the decoder. For example in the source problem, if the side information is known to both encoder and  decoder, the problem can be solved by using conditional Huffman coding \cite{huffman1952method}, with the same rate $H(X|Y)$ as in the vanishing error regime. Similarly, when the information known to the decoder is communicated to the encoder via a feedback channel, the zero-error capacity has a single-letter expression \mlt{established} by Shannon in \cite{shannon1956zero}.

\subsection{Related Literature}

Various questions have 
been addressed in point-to-point zero-error information theory. 
For instance, in \cite{Sason2024}, Sason characterized  the zero-error capacity of two subclasses of strongly regular graphs by leveraging the Lov\'{a}sz $\theta$-function \cite{Lovasz79}.
\mlt{In \cite{LaviSason_arXiv2025}, Lavi and Sason established
sufficient conditions under which the zero-error capacity of a polynomial in
graphs equals the corresponding polynomial of the individual capacities.}
The computability of zero-error capacity has been studied by Boche and Deppe in \cite{boche2021computability}. 
%
In \cite{guo90}, Guo and Watanabe examined a family of graphs where no finite-length code can achieve the zero-error capacity.
In \cite{Dalai2020}, Dalai et al. improved the upper bound on the zero-error list-decoding capacity 
for the 4/3 channel.
In \cite{Effros_TIT2021}, Noorzad et al. provided necessary and sufficient conditions for the randomly generated codebook to satisfy the zero-error property under list-decoding.
\mlt{In \cite{WangIrfanLangbergJaggi_ISIT25}, Wang et al. analyse the zero-error performance of superposition codes with expurgation processes.}
\mlt{In \cite{CaoChenBai_TIT25}, Cao et al. studied the zero-error capacity of channels with memory which are represented by graphs with one edge.}

Network zero-error information theory, have been studied, with significant recent advances.
In \cite{Devroye_IT17}, Devroye et al. investigated the zero-error capacity of the primitive relay channel, by proposing a one-shot relaying scheme, termed color-and-forward. The zero-error capacity with noisy channel feedback was studied in \cite{Devroye_Allerton17} and  \cite{ZhaoPermuter_IT10}, where dynamic programming provides lower and upper bounds.
Bracher and Lapidoth established, in \cite{BracherLapidoth2018}, the zero-error feedback capacity of the state-dependent channel. In \cite{LapidothYan2024}, Lapidoth and Yan characterized the zero-error helper capacity of modulo-additive noise channel in the presence of feedback, and derived lower and upper bounds when there is no feedback.
\mlt{In \cite{Zuiddam_TIT21}, Li and Zuiddam studied the asymptotic spectrum
of graphs and introduced the quantum version of the zero-error capacity.}
\mlt{In \cite{CaoYeung_TIT22}, Cao and Yeung determined the zero-error capacity regions of several noisy networks. 
In \cite{SaberiFarokhiNair_TIT22}, Saberi et al. introduced the zero-error feedback capacity of causal discrete channels with memory, and they provided a tight condition for bounded stabilization of unstable noisy linear systems.}

In \cite{wang2017graph}, Wang and Shayevitz investigated the combination of  zero-error source and channel coding schemes, by introducing the notion of ``graph information ratio'', which refers to the relative Shannon capacity of two graphs of K\"{o}rner and Marton in \cite{KornerMarton2001}. 
In \cite{Gohari_ZeroError_2023}, Alipour et al. defined a new quantity called the relative fractional independence number, which provides new upper and lower bounds on the zero-error capacity.
In \cite{gu2021non}, Gu and Shayevitz provided outer and inner bounds on the zero-error capacity region for the two-way channel.
In \cite{Shayevitz_IT17}, Shayevitz investigated the zero-error broadcasting problem by introducing the $\rho$-capacity function, for which upper and lower bounds are derived. 
In \cite{OrdentlichShayevitz_ISIT15}, Ordentlich and Shayevitz investigated the zero-error capacity region of the binary adder multiple-access channel. They provided a new outer bound that strictly improves upon the bound obtained by Urbanke and Li, in \cite{UrbankeLi_ITW98}.
In \cite{Wiese_TAC19}, Wiese et al. defined zero-error wiretap codes by requiring that every output at the eavesdropper can be generated by at least two inputs. They defined the zero-error secrecy capacity as the supremum of rates for which there exists a zero-error wiretap code. They showed \ar{that it is either zero} or the zero-error capacity of the channel between the encoder and the legitimate receiver. 
In \cite{Gohari_TIT2020_ZeroErrorMolecular}, Khooshemehr et al. characterized the zero-error capacity of the molecular delay channel by assuming that the maximum number of the released molecules is fixed. 

In \cite{OrlitskyRoche2001}, Orlitsky and Roche formulated the graph-based problem of coding for computing, which has also seen recent advances.
For instance, in \cite{MedardEffros2010}, Doshi et al. investigated the distributed functional compression problem and proposed a layered architecture that uses graph coloring and distributed source coding.
In \cite{MalakITW22} and \cite{MalakISIT2023}, Malak demonstrated that coding gains can be obtained by using fractional coloring instead of the traditional graph coloring.  \mlt{In \cite{DeylamMalak_ISIT25}, Deylam Salehi et al. determined achievability and converse bounds for the broadcast problem when receivers compute distinct functions.}

\subsection{Summary of the Contributions}

In this paper, we \mlt{investigate} the \mlt{additivity property} of optimal rates for zero-error \mlt{source and channel} problems. 
\mlt{First}, we consider a source coding problem with side-information that decomposes into a family of \emph{independent} subproblems $(X_a,Y_a)_{a\in\mc{A}}$ \mlt{with probabilistic graphs $(G_a,P_{X_a})_{a\in\mc{A}}$. This \mlt{problem has} a characteristic graph with a specific structure given by the  AND product $G = \wedge_{a\in\mc{A}} G_a$, and a product distribution $\bigotimes_{a\in \mathcal{A}} P_{X_a}$.} 
In the vanishing error regime, the optimal rate \emph{\mlt{is additive}}, meaning that it is equal to the sum of the optimal rates \mlt{of} each subproblem. \mlt{Consequently, the concatenation of optimal codes \ar{is optimal}.} On the contrary, by building on Haemers results for the zero-error capacity of the Schl\"{a}fli graph in \cite{haemers1979some}, we show that independence alone doesn't ensure the \mlt{additivity property} in the zero-error \mlt{source problem}. This inspired us to investigate the conditions under which \mlt{the \emph{complementary graph entropy} is additive, i.e. $\overline{H}\big(G_1 \wedge G_2, P_{X_1} \otimes P_{X_2}\big) = \overline{H}(G_1,P_{X_1}) + \overline{H}(G_2,P_{X_2})$.}



Our first contribution \mlt{Theorem~\ref{th:caraclin} in Sec.~\ref{sec:SWind} shows} that \mlt{$\overline{H}$ is additive} for the AND product ($\wedge$) of graphs if and only if it \mlt{is additive} for the disjoint union of graphs ($\sqcup$), also called ``sum of graphs" in \cite{tuncel2009complementary}. \mlt{The disjoint union is obtained by taking the union of the vertices, and two vertices are adjacent if they are adjacent in the original graph. \ar{Additivity for the disjoint union holds} if $\overline{H}\big(G_1 \sqcup G_2,\alpha P_{X_1} + (1-\alpha)P_{X_2}\big) = \alpha \overline{H}(G_1,P_{X_1}) + (1-\alpha)\overline{H}(G_2,P_{X_2})$. \ar{We demonstrate an equality that ensures the equivalence between the additivity of the AND product and that of the disjoint union. More precisely,} we show that all pairs of probabilistic graphs $(G_1,P_{X_1})$ and \mlt{$(G_2,P_{X_2})$} satisfy}
\begin{align}
&\overline{H}\bigg(G_1 \sqcup G_2,\frac12(P_{X_1} + P_{X_2})\bigg)\nonumber\\
=& \frac{1}{2}\cdot \overline{H}\left(G_1 \wedge G_2, P_{X_1} \otimes P_{X_2} \right), \label{eq:HbarEqual}
\end{align}
where $P_{X_1} \otimes P_{X_2}$ is the product of the probability distributions, and where the probability distribution over 
\mlt{the vertices of} $G_1 \sqcup G_2$ is \mlt{obtained} by taking the uniform distribution $P_A = (\frac12,\frac12)$ over the two probabilistic graphs $(G_1,P_{X_1})$ and $(G_1,P_{X_2})$. \ar{It follows directly from \eqref{eq:HbarEqual}, that the additivity of 
$\overline{H}(\wedge\:\cdot)$ is equivalent to that of $\overline{H}(\sqcup\:\cdot)$.}
Then, we extend \eqref{eq:HbarEqual} by showing that if the distribution \mlt{$P_A$ over the family of probabilistic graphs $(G_a,P_{X_a})_{a\in\mc{A}}$} is a type, \mlt{see \cite[Definition~2.1]{csiszar2011information}}, then $\overline{H}(\sqcup\:\cdot)$ is a function of the complementary graph entropy $\overline{H}(\wedge\:\cdot)$ of a specific product of graphs. Since the set of types is dense in the set of distributions, we show that \mlt{the additivity of $\overline{H}(\wedge\:\cdot)$ and of $\overline{H}(\sqcup\:\cdot)$} are equivalent for all distributions $P_A$.

\mlt{Second, we consider a channel coding problem that decomposes into a family of \emph{independent} zero-error subproblems.} Recently, Wigderson and Zuiddam in \cite{WigdersonZuiddam2023} and Schrijver in \cite{schrijver2023shannon} \mlt{showed} the equivalence of the \mlt{additivity properties} of $C_0(\wedge\:\cdot)$ and $C_0(\sqcup\:\cdot)$, where $C_0$ denotes the zero-error channel capacity. In \cite{schrijver2023shannon}, the key ingredients in the proof are the superadditivity of $C_0$, and the \mlt{additivity property} of the independence number $\alpha(G\sqcup G') = \alpha(G) + \alpha(G')$ with respect to the disjoint union of the graphs. Since the characteristic graph $G$ is defined similarly, we explore the connections and differences between the \mlt{additivity} results for $C_0(G)$ and for $\overline{H}(G,P_X)$. For instance, the probability distribution $P_X$ is given in the source problem, whereas it is \emph{a priori} not specified in the channel problem.

In order to relate $C_0(G)$ and $\overline{H}(G,P_X)$, we use the zero-error capacity $C(G,P_X)$ of a graph $G$ relative to a distribution $P_X$, introduced by Csisz\'{a}r and K\"{o}rner in \cite{csiszar1981capacity} to bound the capacity of the arbitrarily varying channel with maximal error probability. The zero-error capacity $C_0(G)$ and $C(G,P_X)$ are related via the result of \cite[\mlt{equation (11.2)}]{csiszar2011information} where
\begin{align}
C_0(G) = \max_{P_X} C(G,P_X).\label{eq:Max}
\end{align}
Moreover, Marton showed in \cite{marton1993shannon} that the complementary graph entropy satisfies
\begin{align}
C(G,P_X) + \overline{H}(G,P_X) = H(X).\label{eq:HX}
\end{align}
Equations \eqref{eq:Max} and \eqref{eq:HX} are the analogues of the channel capacity $C = \max_{P_X} I(X;Y)$ and the entropy property $I(X;Y)+ H(X|Y) = H(X)$ in the vanishing error regime. \mlt{Proposition~\ref{prop:marton} in Sec.~\ref{sec:ZECapacityDistribution} uses \eqref{eq:HX} to prove the equivalence between the additivity of $C(G,P_X)$ and that of $\overline{H}(G,P_X)$. In order to complete this study, it remains to relate the additivity of $C(G,P_X)$ to the one of $C_0(G)$. This requires careful analysis of the distribution that maximizes $C(G,P_X)$ and how it decomposes over the AND product and over the disjoint union of graphs. Suppose that for all subproblems the source distribution $P_X$ maximizes $C(G,P_X)$, Theorem~\ref{th:mainC0} in Sec.~\ref{sec:FinalEquivalenceAND} and Theorem~\ref{th:linC_0sqcup} in Sec.~\ref{sec:SumChannel} conclude} that the \mlt{additivity properties} of $C_0(G)$, $C(G,P_X)$ and $\overline{H}(G,P_X)$ for the AND product ($\wedge$) and for the disjoint union ($\sqcup$) are all equivalent. 

A crucial notion is the set of \emph{capacity-achieving distributions} that contains all the distributions $P_X$ for which $C_0(G)=C(G,P_X)$. We show that the uniform distribution is \emph{capacity-achieving} when the graph is vertex-transitive, i.e. when all vertices play the same role within the graph. Since the Schl\"{a}lfi graph $S$ and its complementary graph $\overline{S}$ are vertex-transitive, \ar{so is} their product $S\wedge \overline{S}$, and the uniform distribution is capacity achieving for $S$, $\overline{S}$ and $S\wedge \overline{S}$. Together with Haemers result \cite{haemers1979some}, this shows a counterexample where the \mlt{additivity property} of $C_0(G)$, $C(G,P_X)$ and $\overline{H}(G,P_X)$ for the AND product and for the disjoint union of $S$ and $\overline{S}$ does not hold. 

The \mlt{additivity property} enlarges the class of \ar{graphs} for which $C_0$ and $\overline{H}$ have a single-letter characterization. For perfect graphs, we show that the \mlt{additivity} of $C_0$ and $\overline{H}$ always hold for the AND product and for the disjoint union. This allows us to characterize new single-letter solutions for all products of perfect graphs, that are not perfect in general, and for the product of a perfect graph $G$ with the pentagon graph $C_5$, by building on the characterization of $\overline{H}(G\sqcup C_5)$ by Koulgi et al. in \cite{tuncel2009complementary}. 

\begin{figure*}[h!]
    \begin{tikzpicture}[scale=.85]
        \footnotesize
        \node[draw=black, fill= none] (C0P) at (0,0) {\mlt{additivity} of $C_0(\wedge \: \cdot)$};
        \node[draw=black, fill= none] (C0S) at (0,3) {\mlt{additivity} of $C_0(\sqcup \: \cdot)$};
        \node[draw=black, fill= none] (CGP) at (6,0) {\mlt{additivity} of $C(\wedge \:\cdot, P_X)$};
        \node[draw=black, fill= none] (CGS) at (6,3) {\mlt{additivity} of $C(\sqcup \:\cdot, P_X)$};
        \node[draw=black, fill= none] (HP) at (12,0) {\mlt{additivity} of $\overline{H}(\wedge \: \cdot, P_X)$};
        \node[draw=black, fill= none] (HS) at (12,3) {\mlt{additivity} of $\overline{H}(\sqcup \: \cdot, P_X)$};
        \node[draw=none, fill= none, rotate=90] (S) at (0,1.5) {$\Longleftrightarrow$};
        \node[draw=none, fill= none] (M1) at (9,0) {$\Longleftrightarrow$};
        \node[draw=none, fill= none] (M2) at (9,3) {$\Longleftrightarrow$};
        \node[draw=none, fill= none] (C1) at (3,0) {$\Longleftrightarrow$};
        \node[draw=none, fill= none] (C2) at (3,3) {$\Longleftrightarrow$};
        \node[draw=none, fill= none, rotate=90] (I1) at (6,1.5) {$\Longleftrightarrow$};
        \node[draw=none, fill= none, rotate=90] (I2) at (12,1.5) {$\Longleftrightarrow$};
        \node[draw=none, fill= none] (S_) at ($(S)+(-2.5,0)$) { 
            \begin{tabular}{r}
                {Wigderson and Zuiddam 2023}\\
                {\cite[Theorem 4.1]{WigdersonZuiddam2023}}\\
                {Schrijver 2023}\\
                 {\cite[Theorem 2]{schrijver2023shannon}}
            \end{tabular}};
        \node[draw=none, fill= none] (M1_) at ($(M1)+(0,-.75)$) { \begin{tabular}{c}
                Marton 1993\\
                {{\cite[Lemma 1]{marton1993shannon}}}
            \end{tabular}};
        \node[draw=none, fill= none] (M2_) at ($(M2)+(0,.75)$) { \begin{tabular}{c}
                Marton 1993\\
                {{\cite[Lemma 1]{marton1993shannon}}}
            \end{tabular}};
        \draw[dashed] (2.0,-.5) rectangle (4.0,3.5);
        \draw[dashed] (5.5,.5) rectangle (12.5,2.5);
        \node[draw=none, fill= none] (T1) at (10.3,1.5) { \begin{tabular}{r}
                Charpenay et al. 2023 \\
                \cite[Theorem IV.8]{CharpenayISIT23}\\
                Theorem~\ref{th:caraclin}
            \end{tabular}};
        \node[draw=none, fill= none] (T1_) at (7.5,1.5) { \begin{tabular}{c}
                Proposition~\ref{prop:marton}
            \end{tabular}};
        \node[draw=none, fill= none] (T2) at (3,.75) { \begin{tabular}{c}
                Theorem~\ref{th:mainC0}
            \end{tabular}};
        \node[draw=none, fill= none] (T2_) at (3,2.25) { \begin{tabular}{c}
                Theorem~\ref{th:linC_0sqcup}
            \end{tabular}};
    \end{tikzpicture}
    \caption{Equivalences of \mlt{additivity} properties for the zero-error capacity $C_0$, for the zero-error capacity relative to a distribution $C$, and for the complementary graph entropy $\overline{H}$. The equivalences of our  \mlt{Theorem~\ref{th:caraclin}, Proposition~\ref{prop:marton}, Theorem~\ref{th:mainC0} and Theorem~\ref{th:linC_0sqcup} are valid under specific conditions on the probability distributions.}}
    \label{fig:additivity}
\end{figure*}

\ar{The path taken to demonstrate all these equivalences is shown in Fig.~\ref{fig:additivity}.} In Sec.~\ref{sec:source}, we study the \mlt{additivity} of the complementary graph entropy $\overline{H}$ for the source problem with side information. The connection with the \mlt{additivity} of the zero-error capacity $C_0$ is investigated in Sec.~\ref{sec:channel}. New single-letter solutions for $C_0$ and $\overline{H}$ are provided in  Sec.~\ref{sec:Examples}, as well as the counter-example \mlt{for the additivity} based on Haemers result \cite{haemers1979some} for the Schl\"{a}fli graph.

\section{Zero-Error Source Coding With Decoder Side Information}\label{sec:source}

\subsection{\mlt{Results from the Literature and Open Problems}}\label{sec:PreliminarySource}

We investigate the source coding problem depicted in Fig.~\ref{fig:Zero-errorSW}, which we call the \emph{side-information problem}. This situation arises in data compression where the decoder has side-information $Y$ about the source $X$ that must be retrieved. This problem has been solved in the vanishing error probability regime by Slepian and Wolf \cite{slepian1973noiseless}, but remains open in the zero-error regime. Asymptotic expressions have been derived by Witsenhausen  \cite{witsenhausen1976zero} for fixed-length codes, and by Alon and Orlitsky \cite{alon1996source} for variable-length codes.

More formally, we assume that a sequence of length $n \in \mathbb{N}^\star$ of i.i.d. random variables $(X^n, Y^n)$ is drawn according to the probability distribution $P_{X,Y} \in \Delta(\mathcal{X} \times \mathcal{Y})$ where $\mathcal{X}$ and $\mathcal{Y}$ are finite sets and $\Delta(\mathcal{X} \times \mathcal{Y})$ denotes the set of probability distributions over $\mathcal{X} \times \mathcal{Y}$. We assume that the marginal distributions $P_{X} \in \Delta(\mathcal{X})$ and $P_{Y} \in \Delta(\mathcal{Y})$ have full support. \mlt{We denote by $\lbrace 0,1 \rbrace^*$ the set of binary words.} We consider variable-length source coding, which encompasses the special case of fixed-length source coding. An  $(n, \phi_e, \phi_d)$ \emph{variable-length side-information source code} for $X$ and $Y$ consists of
\begin{itemize}
\item[-] an encoder $\phi_e : \mathcal{X}^n \rightarrow \lbrace 0,1 \rbrace^*$ that assigns to each $x^n$ a binary \mlt{word} such that \mlt{its image} $\Ima \phi_e$ is prefix-free,
\item[-] a decoder $\phi_d : \mathcal{Y}^n \times \lbrace 0,1 \rbrace^* \rightarrow \mathcal{X}^n$ that assigns an estimate $\widehat{x}^n$ to each pair $(y^n,\phi_e(x^n))$.
\end{itemize}
The rate of the $(n, \phi_e, \phi_d)$-code is the average length of the codeword per source symbol, i.e. $R \doteq \frac{1}{n}\mathbb{E}[\ell \circ \phi_e(X^n)]$ \mlt{where $\ell$ denote the codeword length. We denote the probability of error by $P_e^{(n)} \doteq \mathbb{P}\big(\widehat{X}^n \neq X^n \big)$.}

\begin{definition}
        The \emph{optimal rate in the zero-error regime} is the minimal rate among all coding schemes that satisfy the zero-error constraint: 
    \begin{align}
         R^{\star}_{0} \doteq 
        \inf\limits_{\displaystyle (n, \phi_e, \phi_d): P_e^{(n)}=0 } \quad
         \frac{1}{n}\mathbb{E}\big[\ell \circ \phi_e(X^n)\big].
         \end{align}
\end{definition}

When the side-information $Y$ is available at both encoder and decoder, the optimal rate in the zero-error regime \mlt{coincides} with the optimal rate $H(X|Y)$ in the vanishing error regime. The zero-error coding construction relies on a conditional Huffman coding \cite{huffman1952method}. In the side-information problem, the encoder does not observes the side-information $Y$. \mlt{Due to this information asymmetry, the optimal rates in the zero-error and vanishing-error regimes are distinct.} 
The characterisation of the optimal rate $R^{\star}_{0}$ is a notoriously difficult open problem of combinatorial nature. The key features are captured by the ``characteristic graph''  of Witsenhausen \cite{witsenhausen1976zero}, which is constructed based on the support of the side information channel $P_{Y|X}$.

\begin{definition}[Characteristic graph]\label{def:chargraph}
Let $\mathcal{X}, \mathcal{Y}$ be two finite sets and $P_{Y|X}$ be a conditional distribution. The \emph{characteristic graph} $G = (\mc{X},\mc{E})$ associated to $P_{Y|X}$ is defined by:
\begin{itemize}
\item[-] $\mathcal{X}$ as set of vertices,
\item[-] \mlt{$x\neq x' \in \mathcal{X}$ are adjacent $xx'\in\mc{E}$, if for some $y \in \mathcal{Y}$, $P_{Y|X}(y|x)\cdot P_{Y|X}(y|x') > 0$.} 
\end{itemize}

We denote by $(G,P_X)$ the \emph{probabilistic graph} induced by the characteristic graph $G$ and the distribution $P_X$ on its vertices.
\end{definition}

When the side information $y$ does not allow to distinguish exactly between the source realizations $x$ and $x'$, then $x$ and $x'$ are adjacent, and must be mapped to different codewords. A zero-error encoding consists of a graph coloring where adjacent vertices are mapped to different colors. 

\begin{definition}[Coloring, chromatic number $\chi$]\label{def:chi}
Let $G = (\mc{X}, \mathcal{E})$ be a graph. A mapping $c : \mc{X} \rightarrow \mathcal{C}$ is a \emph{coloring} if for all adjacent vertices $x$, $x'$ with $xx'\in\mc{E}$, we have $c(x)\neq c(x')$. The \emph{chromatic number} $\chi(G)$ is the smallest $|\mathcal{C}|$ such that there exists a coloring $c : \mc{X} \rightarrow \mathcal{C}$ of $G$.
\end{definition}


For all graphs $G_1$ and $G_2$, we have $\chi(G_1\wedge G_2) \leq \chi(G_1) \cdot \chi(G_2)$. For sequences of symbols with underlying i.i.d. conditional distribution $P^{\otimes n}_{Y|X}(y^n|x^n) = \prod_{t=1}^nP_{Y|X}(y_t|x_t)$, two sequences of source \mlt{symbols} $x^n, x'^n$ are adjacent in the graph if $P^{\otimes n}_{Y|X}(y^n|x^n)P^{\otimes n}_{Y|X}(y^n|x'^n) > 0$ for some sequence of channel outputs $y^n$, i.e. if and only if either $x_t=x'_t$ or $x_tx_t'\in\mc{E}$, \mlt{for all} $\ 1\le t \le n$. This implies that for sequences of symbols, the characteristic graph is built by using the AND product of graphs, denoted by $\wedge$, and also called ``strong product'' or ``normal product'' in \cite{Lovasz79, marton1993shannon}, and defined below.

\begin{definition}[AND product $\wedge$]\label{def:ANDprod}
Let  $G_1 = (\mathcal{X}_1, \mathcal{E}_1)$, $G_2 = (\mathcal{X}_2, \mathcal{E}_2)$ be two graphs, their \emph{AND product} $G_1 \wedge G_2$ is defined by
\begin{itemize}
\item[-] $\mathcal{X}_1 \times \mathcal{X}_2$ as set of vertices,
\item[-] $(x_1 x_2),(x'_1 x'_2)$ are adjacent if 
$x_1x'_1 \in \mathcal{E}_1$ and $x_2x'_2 \in \mathcal{E}_2$, 
with the convention of self-adjacency for all vertices.
\end{itemize}
We denote by $G_1^{\wedge n}$ the $n$-th \emph{AND power}
$G_1^{\wedge n} = G_1 \wedge ... \wedge G_1$ $n$ times.

When $(G_1,P_{X_1})$ and $(G_2,P_{X_2})$ are two probabilistic graphs, the distribution induced on the product of vertices $\mathcal{X}_1 \times \mathcal{X}_2$ is the product of the two distributions $P_{X_1} \otimes P_{X_2}$.
\end{definition}

%


Alon and Orlitsky introduced an asymptotic expression in \cite{alon1996source}, for the optimal rate $R_0^{\star}$ in the ``restricted inputs'' setting. It relies on the notion of chromatic entropy $H_{\chi}(G^{\wedge n},P^{\otimes n}_X)$, which is the minimal entropy of a coloring of $G^{\wedge n}$ induced by the i.i.d. distribution $P^{\otimes n}_X$. 


\begin{theorem}[{{from \cite[Lemma 6]{alon1996source}}}]\label{prop:alon} Given a probabilistic graph $(G,P_X)$,
    \begin{align}
        R^{\star}_{0} = \lim_{n \rightarrow \infty} \frac{1}{n} H_{\chi}(G^{\wedge n},P^{\otimes n}_X), \label{eq:defhbar2}
    \end{align}
where the \emph{chromatic entropy} is defined by
\begin{align}
    H_{\chi}(G,P_X) \doteq \min \Big\{ H\big(c(X)\big) \Big| c \text{ is a coloring of } G \Big\}.
\end{align}
\end{theorem}

The sequence $\big( H_{\chi}(G^{\wedge n},P^{\otimes n}_X)\big)_{n\in\N^{\star}}$ is subadditive, Fekete's Lemma \cite[pp. 103]{van2001course} ensures the existence of the limit in \eqref{eq:defhbar2}, which coincides with the infimum. 

There is no single-letter expression for $R^{\star}_{0}$. In \cite{alon1996source}, Alon and Orlitsky provided a single-letter upper bound of $R^{\star}_{0}$ which corresponds to the ``unrestricted inputs'' scenario where the source $X$ has to be recovered with zero-error for all side information $y$, even when $(X,y)$ is outside the support of $P_{X,Y}$.


\begin{figure}[h!]
    \centering
    \begin{tikzpicture}
        \foreach \i in {0,1,2,3,4}{
        \draw[] ($(72*\i-36:0.75)+(-4,0)$) -- ($(72*\i + 36:0.75)+(-4,0)$);
        }
        \foreach \i in {1,2,3,4,5}{
        \node[shape = circle, draw = black, inner sep = 1pt, fill = white] (B) at ($(72*\i+36:0.75) + (-4,0)$) {\footnotesize$\i$};
        }
    \end{tikzpicture}
    \caption{The pentagon graphs $C_5$ with uniform distribution $P_X=\Unif \big(\lbrace 1, ...,5\rbrace \big)$ over the vertices.}
    \label{fig:Disjoint}
\end{figure}

With high probability, the source sequence $X^n$ is typical with respect to $P_X$. Let $G^{\wedge n}[\mathcal{T}^n_{\varepsilon}(P_X)]$ be the subgraph of $G^{\wedge n}$ induced by the set of \emph{typical  sequences} $\mathcal{T}^n_{\varepsilon}(P_X)$ with tolerance $\varepsilon>0$, see \cite[Definition 2.8]{csiszar2011information}. We can construct a zero-error code by taking the minimal coloring of this induced subgraph $\chi \big(G^{\wedge n}[\mathcal{T}^n_{\varepsilon}(P_X)]\big)$. The encoder sends the color index to the decoder if $X^n$ is typical, otherwise it sends the index of the sequence $X^n$ in $\mathcal{X}^n$. This coding strategy has a rate upper-bounded by
\begin{align}
    \frac{1}{n} + \mathbb{P}\big(X^n \notin \mathcal{T}^n_{\varepsilon}(P_X)\big) \log|\mathcal{X}| + \frac{1}{n} \log \chi \big(G^{\wedge n}[\mathcal{T}^n_{\varepsilon}(P_X)]\big).
    \label{eq:typical}
\end{align}
The zero-error property is satisfied since the decoder is able to retrieve $X^n$ thanks to $Y^n$ and the color symbol. Koulgi et al.  have shown in \cite[Theorem 1]{koulgi2003zero} that taking the limit when $n$ goes to infinity and $\varepsilon$ goes to $0$ yields the best achievable rate in the zero-error side-information problem. This quantity, introduced by K\"{o}rner and Longo in \cite{korner1973two} for the two-step source coding, is called the complementary graph entropy. Here, we adopt the formalism of \cite[Definition~7]{simonyi2003witsenhausen} and \cite[Definition~7]{tuncel2009complementary}.

\begin{definition}\label{def:Hbar}
Given a probabilistic graph $(G,P_X)$, the complementary graph entropy $\overline{H}(G,P_X)$ is defined by:
    \begin{align}
        \overline{H}(G,P_X) \doteq \lim_{\varepsilon \rightarrow 0} \limsup_{n \rightarrow \infty} \frac{1}{n}\log \chi \big(G^{\wedge n}[\mathcal{T}^n_{\varepsilon}(P_X)]\big).  \label{eq:defhbar}
    \end{align}
\end{definition}

The question of the subadditivity of the sequence $\big(\log \chi\big(G^{\wedge n}[\mc{T}^n_{\varepsilon}(P_X)]\big)\big)_{n\in\N^{\star}}$ is  open.


\begin{theorem}[{{from \cite[Theorem 1]{koulgi2003zero}}}]\label{prop:koulgi}
    The optimal rate in the zero-error regime \mlt{is}
    \begin{align}
        R^{\star}_{0} = \overline{H}(G,P_X),
    \end{align}
    where $(G,P_X)$ is the probabilistic graph formed of the characteristic graph associated to the distribution $P_{Y|X}$, with the underlying distribution $P_X$ on its vertices.
\end{theorem}

Note that when restricting to \emph{zero-error fixed-length} code, the optimal rate $H_{0}(G)$ introduced by Witsenhausen in \cite{witsenhausen1976zero} does not depend on the source distribution $P_X$, since
\begin{align}
H_{0}(G) \doteq \lim_{n\to+\infty} \frac1n \log\chi (G^{\wedge n}). \label{eq:Witsenhausen}
\end{align}
The limit exists and coincides with the infimum thanks to Fekete's Lemma \cite[pp. 103]{van2001course} for the subadditive sequence $\big(\log \chi(G^{\wedge n})\big)_{n\in\N^{\star}}$. In fact, the \emph{Witsenhausen rate} $H_{0}(G)$ is equal to the maximum of $\overline{H}$ over the set of source distributions \mlt{$Q_X\in\Delta(\mc{X})$}.

\begin{lemma}[{from \cite[Lemma~3]{simonyi2003witsenhausen}}]\label{lemma:H0}
Given the graph $G = (\mc{X}, \mathcal{E})$, the \emph{Witsenhausen rate} satisfies
\begin{align}
H_{0}(G)  = \mlt{\max_{Q_X}\; \overline{H}(G,Q_X).}\label{eq:WitsenhausenRate}
\end{align}
\end{lemma}
The proof relies on the type counting Lemma, see \cite[Lemma~2.2]{csiszar2011information}. \ar{Interestingly \eqref{eq:WitsenhausenRate} shows that the use of zero-error fixed-length codes is a strong restriction compared to the use of zero-error variable-length codes. Indeed, the optimal rate for variable-length code satisfies $R^{\star}_{0} = \overline{H}(G,P_X) \le \max_{Q_X} \; \overline{H}(G,Q_X) = H_{0}(G)$. Moreover, note that the rate $H_{0}(G)$ corresponds to a universal source coding problem in which the characteristic graph is known, but the source distribution is unknown.}

\mlt{We now} focus on variable-length coding. The optimal rate $R^{\star}_{0}$ has a trivial single-letter upper bound given by $H(X)$, where the zero-error coding construction relies on Huffman coding and the decoder ignores the side information $Y$. In fact, this upper bound is tight for a dense subset of distributions in $\Delta(\mc{X}\times \mc{Y})$.

\begin{proposition}[Full support, from \cite{witsenhausen1976zero}]\label{prop:FullSupport}
If the distribution $P_{X,Y}$ has full support, then $R^{\star}_{0}=H(X)$.
\end{proposition}

Indeed, since the distribution $P_{X,Y}$ has full support, the characteristic graph $G$ is complete, i.e. every pair of symbols $x\in\mc{X}$, $x'\in\mc{X}$ are adjacent in $G$, thus $H_{\chi}(G,P_X) = H(X)$.

There are a few other cases where the optimal zero-error rate is known such as perfect graphs \cite[Corollary~12]{csiszar1990entropy}, or the pentagon $C_5$ with uniform distribution shown in Fig.~\ref{fig:Disjoint} where $R^{\star}_{0}=\frac{1}{2}\log_2(5)$, see \cite[Example~1]{koulgi2003zero}. In general, the single-letter characterization of $R^{\star}_{0}$ remains a difficult open question.


\subsection{\mlt{Problem Statement for Independent Sources}}\label{sec:SWind}

In order to understand \mlt{this difficulty}, we examine a specific scenario where the source and the side information decompose into \emph{independent} variables. In the vanishing error regime, independence is a key assumption that \mlt{ensures the additivity} of optimal rates, shedding light on practical coding techniques. In the zero-error regime, the independence hypothesis alone is not sufficient to ensure the \mlt{additivity} of optimal rates. We formulate a condition that implies \mlt{additivity}, enabling us to enlarge the set of problems for which the optimal rate has a \mlt{known} single-letter characterization.

%
%
%

\begin{figure}[h!]
\centering
\begin{tikzpicture}[xscale=0.55,yscale=0.6]
\draw [>=stealth,->] (0,0.5) -- (2,0.5) ;
\draw  (0.15,1) node {$X_1^n, ..., X_{|\mc{A}|}^n$} ;
\draw (2,0) rectangle (5,1);
\draw (3.5,0.5) node {Encoder} ;
\draw [,>=stealth,->] (5,0.5) -- (8,0.5) ;
\draw (6.5,0.5) node {$\diagup$} ;
\draw (6.5,1.2) node {$R$} ;
\draw (8,0) rectangle (11,1);
\draw (9.5,0.5) node {Decoder} ;
\draw [>=stealth,->] (11,0.5) -- (13,0.5) ;
\draw  (12.85,1) node {$\widehat{X}_1^n, ..., \widehat{X}_{|\mc{A}|}^n$} ;
\draw [>=stealth,->] (9.5,-1) -- (9.5,0) ;
\draw  (9.5,-1.5) node {$Y_1^n, ..., Y_{|\mc{A}|}^n$} ;
\end{tikzpicture}
    \caption{Independent side-information problems}
    \label{fig:Linearsourceprod}
\end{figure}
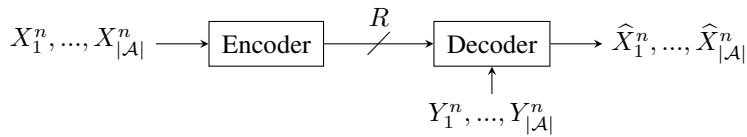

More formally, for a finite set $\mathcal{A}$, we assume a set of pairs $(X_a, Y_a)_{a\in \mathcal{A}}$, referred to as an \emph{independent family}, that consists of $|\mathcal{A}|$ pairs with \mlt{a} joint distribution that decomposes \mlt{into} a product of distributions. This independent family generates sequences of i.i.d. random variables, 
\begin{align}
\big(X_1^n,Y_1^n, \ldots, X_{|\mc{A}|}^n, Y_{|\mc{A}|}^n\big) \sim \Big(P_{X_1,Y_1}\otimes\ldots\otimes P_{X_{|\mc{A}|},Y_{|\mc{A}|}}\Big)^{\otimes n}.\nonumber
\end{align}
\emph{Independent side-information problems} correspond to a side-information problem in which the source and the side information \mlt{form} an independent family, as shown in  Fig.~\ref{fig:Linearsourceprod}. In the vanishing error regime, the optimal rate \mlt{is additive},
\begin{align}
    R^{\star} =& H\big(X_1,\ldots,X_{|\mc{A}|}\big|Y_1,\ldots,Y_{|\mc{A}|}\big)\\
    =& \sum_{a\in\mc{A}} H(X_a|Y_a) = \sum_{a\in\mc{A}} R_a^{\star} .\label{eq:additivityVanishing}
\end{align}
This property is fundamental to guarantee the optimality of the codebook constructed by concatenating the codewords of the optimal codebooks \mlt{of} each subproblem.


But does the \emph{\mlt{additivity}} property also hold in the zero-error regime for independent side-information problems? To answer this question, we first derive an asymptotic expression for the optimal zero-error rate. This derivation follows from the fact that the independent family can be characterized by a product of graphs.

\begin{proposition}\label{prop:charprod}
Let $(X_a, Y_a)_{a\in \mathcal{A}}$ be an independent family. The optimal rate for the independent zero-error side-information problem is
\begin{align}
R^{\star}_0 = \overline{H}\left(\bigwedge_{a\in \mathcal{A}} G_a , \bigotimes_{a\in \mathcal{A}} P_{X_a} \right),
\end{align}
where for all $a\in \mathcal{A}$, $G_a$ is the characteristic graph associated to the conditional distribution $P_{Y_a|X_a}$, with the underlying probability distribution $P_{X_a}$ on its vertices. 
\end{proposition}

It is known that the complementary graph entropy $\overline{H}$ is sublinear {with respect to the AND product}. Indeed, \cite[Theorem 2]{tuncel2009complementary} states that for all probabilistic graphs $(G,P_X)$ and $(G',P'_X)$
\begin{align}
\overline{H}(G \wedge G', P_X\otimes P'_X) \leq& \,\overline{H}(G,P_X) + \overline{H}(G',P'_X).\label{eq:linprodTuncel}
\end{align}

However, $\overline{H}$ \mlt{is not additive} in general. Inspired by Haemers result \cite{haemers1979some}, we show in Theorem~\ref{th:maincontex}, that the inequality \eqref{eq:linprodTuncel} is strict for \mlt{the product of} the Schl\"{a}fli graph $S$ \mlt{with} its complement $\overline{S}$.
In the following, we study a condition that \mlt{ensures the additivity} of $\overline{H}$, i.e. where \eqref{eq:linprodTuncel} holds with equality.
To do this, we introduce the disjoint union of graphs, also called ``sum of graphs" in \cite{tuncel2009complementary}.

\begin{definition}[Disjoint union of probabilistic graphs $\sqcup$]\label{def:sqcup}
Let $\mathcal{A}$ be a finite set, let $P_A \in \Delta(\mathcal{A})$, and for all $a \in \mathcal{A}$, let $(G_a,P_{X_a})$ be a probabilistic graph with $G_a = (\mathcal{X}_a, \mathcal{E}_a)$. The disjoint union with respect to $P_A$ is a probabilistic graph denoted by 
$\Big(\bigsqcup_{a \in \mathcal{A}} G_a, \sum_{a \in \mathcal{A}} P_A(a) P_{X_a}\Big)$,
where the vertices $\mathcal{X}$ and the edges $\mathcal{E}$ 
are defined by:
\begin{itemize}
    \item[-] $\mathcal{X} = \bigsqcup_{a \in \mathcal{A}} \mathcal{X}_a$ is the disjoint union of the sets $(\mathcal{X}_a)_{a \in \mathcal{A}}$,
    \item[-] For all $x,x' \in \mathcal{X}$, $xx' \in \mathcal{E}$ if and only if they both belong to the same $\mathcal{X}_a$ and $xx' \in \mathcal{E}_a$.
\end{itemize}
The distribution $P_X = \sum_{a \in \mathcal{A}} P_A(a) P_{X_a}$ is constructed with respect to $P_A$ and the family of distributions $(P_{X_a})_{a \in \mathcal{A}}$ that have disjoint support in $\mathcal{X}$. 
\end{definition}

\begin{figure}[h!]
    \centering
    \begin{tikzpicture}[scale=0.78]
        \node[shape=circle,draw=black, inner sep = .5pt,fill=white] (0) at ($(0,0)$) {\footnotesize$1/4$};
        \node[shape=circle,draw=black, inner sep = .5pt,fill=white] (1) at ($(0) + (1,0)$) {\footnotesize$1/2$};
        \node[shape=circle,draw=black, inner sep = .5pt,fill=white] (2) at ($(0) + (2,0)$) {\footnotesize$1/4$};
        \node[draw=none] (N) at ($(0)+(-1.25,0)$) {$G_1 = $};
        
        \node[shape=circle,draw=black, inner sep = .5pt,fill=white] (3) at ($(5,0)$) {\footnotesize$1/3$};
        \node[shape=circle,draw=black, inner sep = .5pt,fill=white] (4) at ($(3)+(1,0)$) {\footnotesize$2/3$};
        \node[draw=none] (K) at ($(3)+(-1.25,0)$) {$G_2 = $};
        \draw[draw=black,thick] (4) edge (3);
        
        \node[shape=circle,draw=black, inner sep = .5pt,fill=white] (04) at ($(-.5,-2.25)$) {\footnotesize$1/6$};
        \node[shape=circle,draw=black, inner sep = 0pt,fill=white] (03) at ($(04)+(0,1)$) {\footnotesize$1/12$};
        \node[shape=circle,draw=black, inner sep = .5pt,fill=white] (14) at ($(04)+(1,0)$) {\footnotesize$1/3$};
        \node[shape=circle,draw=black, inner sep = .5pt,fill=white] (13) at ($(04)+(1,1)$) {\footnotesize$1/6$};
        \node[shape=circle,draw=black, inner sep = .5pt,fill=white] (24) at ($(04)+(2,0)$) {\footnotesize$1/6$};
        \node[shape=circle,draw=black, inner sep = 0pt,fill=white] (23) at ($(04)+(2,1)$) {\footnotesize$1/12$};
        \node[draw=none] (W) at ($(04)+(-1.5,0.5)$) {$G_1 \wedge G_2 = $};
        \draw[draw=black,thick] (04) edge (03);
        \draw[draw=black,thick] (14) edge (13);
        \draw[draw=black,thick] (24) edge (23);
        
        \node[shape=circle,draw=black, inner sep = 0pt,fill=white] (0_) at ($(5.5,-1.25)$) {\footnotesize$1/16$};
        \node[shape=circle,draw=black, inner sep = .5pt,fill=white] (1_) at ($(0_) + (1,0)$) {\footnotesize$1/8$};
        \node[shape=circle,draw=black, inner sep = 0pt,fill=white] (2_) at ($(0_) + (2,0)$) {\footnotesize$1/16$};
        \node[shape=circle,draw=black, inner sep = .5pt,fill=white] (3_) at ($(0_) + (.5,-1)$) {\footnotesize$1/4$};
        \node[shape=circle,draw=black, inner sep = .5pt,fill=white] (4_) at ($(0_)+(1.5,-1)$) {\footnotesize$1/2$};
        \node[draw=none] (U) at ($(0_)+(-1.75,-.25)$) {$G_1\!  \sqcup  G_2 = $};
        \draw[draw=black,thick] (3_) edge (4_);

    \end{tikzpicture}
    \caption{An empty graph $(G_1,P_{X_1}) = (N_3, (\frac{1}{4},\frac{1}{2}, \frac{1}{4}) )$ and a complete graph $\big(G_2,P_{X_2}) = (K_2, (\frac{1}{3}, \frac{2}{3}) \big)$, along with their AND product \mlt{$\big(G_1 \wedge G_2,P_{X_1}\otimes P_{X_2}\big)$} and their disjoint union \mlt{$(G_1 \sqcup G_2,\frac{1}{4}P_{X_1}+ \frac{3}{4}P_{X_2})$.} The values on each vertex correspond to the underlying probability distributions $P_{X_1}\otimes P_{X_2}$ and $\frac14 P_{X_1}+ \frac34 P_{X_2}$.}
    \label{fig:exprod}
\end{figure}
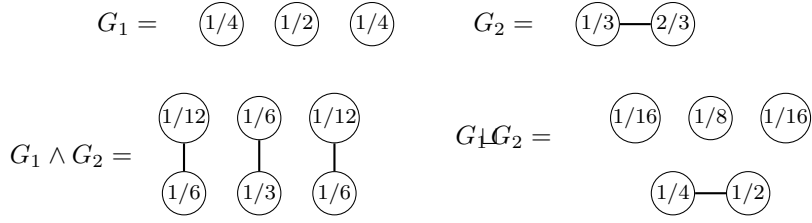

\mlt{In Fig.~\ref{fig:exprod}, we provide an example of an AND product and of a disjoint union of probabilistic graphs.} As \mlt{for} the AND product, the complementary graph entropy $\overline{H}$ is sublinear with respect to the disjoint union. Indeed, \cite[Theorem 2]{tuncel2009complementary} states that for all probabilistic graphs $(G,P_X)$ and $(G',P'_X)$ and the distribution $P_A = (s,1-s)$ with $s \in [0,1]$,
\begin{align}
&\overline{H}(G \sqcup G',sP_X + (1-s)P'_X) \nonumber\\
\leq&\; s \overline{H}(G,P_X) + (1-s) \overline{H}(G',P'_X)\label{eq:linUnionTuncel}.
\end{align}

\mlt{We present our first contribution, which establishes the equivalence between the additivity of the AND product and the additivity of the the disjoint union.} An important consequence  \mlt{when the additivity holds} is that the concatenation of \mlt{optimal codes of each subproblems is optimal.}

\begin{theorem}[Equivalence of the \mlt{additivity for $\wedge$ and for $\sqcup$}, from {\cite[Theorem IV.8]{CharpenayISIT23}}] 
\label{th:caraclin}
Let $\mathcal{A}$ be a finite set, $P_A$ a distribution with full-support, and let $(G_a,P_{X_a})_{a \in \mathcal{A}}$ a family of probabilistic graphs. The following equivalence holds:
\begin{align}
&\overline{H}\left(\bigwedge_{a \in \mathcal{A}} G_a, \bigotimes_{a \in \mathcal{A}} P_{X_a} \right) = \sum_{a \in \mathcal{A}} \overline{H}(G_a,P_{X_a}),\label{eq:LinHbar2}\\
\Longleftrightarrow \;\; & \overline{H}\left(\bigsqcup_{a \in \mathcal{A}} G_{a} ,\!\sum_{a \in \mathcal{A}} P_A(a) P_{X_a}\! \right)\! = \!\sum_{a \in \mathcal{A}} P_{A}(a) \overline{H}(G_{a},P_{X_a}). \label{eq:LinHbar1}
\end{align}
We say that the \mlt{additivity} property holds \mlt{for the AND product if \eqref{eq:LinHbar2} is satisfied. Similarly, the additivity holds for the disjoint union  if \eqref{eq:LinHbar1} is satisfied.}
\end{theorem}
This result allows to characterize new single-letter solutions, as discussed in Sec.~\ref{sec:Examples}. \mlt{By using Theorem~\ref{th:caraclin}}, we characterize the optimal rate of all products of perfect graphs, which are not perfect in general. \mlt{Since} the disjoint product of perfect graphs is perfect, the optimal rates \mlt{add up}. 

Without loss of generality, we consider that $\mathcal{A}$ is the support of $P_A$. We observe that \eqref{eq:LinHbar2} does not depend on the distribution $P_A$, therefore if \eqref{eq:LinHbar1} holds for \mlt{one} distribution $P_A$ with full support, then it holds for all distributions with full support. This remark 
allows us to state the following corollary.

\begin{corollary}\label{coro:Subset}
If the \mlt{additivity} properties \eqref{eq:LinHbar2} or \eqref{eq:LinHbar1} hold for a family of probabilistic graphs $(G_a,P_{X_a})_{a \in \mathcal{A}}$, then \eqref{eq:LinHbar2} and \eqref{eq:LinHbar1} also hold for any sub-family of probabilistic graphs $(G_{\tilde{a}},P_{X_{\tilde{a}}} )_{\tilde{a} \in \widetilde{\mathcal{A}}}$ with $\widetilde{\mathcal{A}}\subset \mathcal{A}$.
\end{corollary}

\subsection{Proof of Theorem~\ref{th:caraclin}}

\mlt{In this section, we present the three main ingredients for the proof of Theorem~\ref{th:caraclin}.} In  Lemma~\ref{lemma:partvaluesA}, we provide an asymptotic formulation that relies solely on the AND product of graphs. 

\begin{lemma}\label{lemma:partvaluesA}
Let $P_A \in \Delta(\mathcal{A})$ a probability distribution with full-support and let  $(\overline{a}_n)_{n \in \mathbb{N}^\star} \in \mathcal{A}^{\mathbb{N}^\star}$ be any sequence such that its type $T_{\overline{a}^n} \rightarrow P_A$ when $n \rightarrow \infty$. Then we have
\begin{align}
&\overline{H}\left(\bigsqcup_{a \in \mathcal{A}} G_a,\sum_{a \in \mathcal{A}} P_A(a) P_{X_a} \right) \nonumber\\
=& \lim_{n \rightarrow \infty} \frac{1}{n}H_\chi\left(\bigwedge_{a \in \mathcal{A}} G_a^{\wedge n T_{\overline{a}^n} (a)} , \bigotimes_{a\in \mc{A}} P_{X_a}^{\otimes n T_{\overline{a}^n} (a)}\right).\label{eq:lemmapartvaluesA}
\end{align}
As a consequence, if $P_A$ belongs to the set $\Delta_k(\mc{A})$ of types of sequences of length $k \in \mathbb{N}^\star$, we have
\begin{align}
&\overline{H}\left(\bigsqcup_{a \in \mathcal{A}} G_a,\sum_{a \in \mathcal{A}} P_A(a) P_{X_a}\right) \nonumber\\
=& \frac{1}{k}\overline{H}\left(\bigwedge_{a \in \mathcal{A}} G_a^{\wedge k P_A(a)}, \bigotimes_{a\in \mc{A}} P_{X_a}^{\otimes k P_A(a)}\right).\label{eq:corpartvalues}
\end{align}
Moreover, if $P_A$ is the uniform distribution, we have
\begin{align}
\overline{H}\left(\bigsqcup_{a \in \mathcal{A}} G_a,\sum_{a \in \mathcal{A}} \frac{1}{|\mc{A}|} P_{X_a}\right) = \frac{1}{|\mathcal{A}|}\overline{H}\left(\bigwedge_{a \in \mathcal{A}} G_a , \bigotimes_{a\in \mc{A}} P_{X_a} \right).\label{eq:corpartvaluesUnif}
\end{align}
\end{lemma}

The proof of Lemma \ref{lemma:partvaluesA} is  stated in App.~\ref{section:prooflemmapartvalues}. It  follows from the properties of the product of graphs 
induced by typical sequences $a^n\in \mc{T}_{\varepsilon}^n(P_A)$, such as the distributivity of $\wedge$ with respect to $\sqcup$, the subadditivity of the sequence of chromatic entropy $\big( H_{\chi}(G^{\wedge n},P^{\otimes n}_X)\big)_{n\in\N^{\star}}$, and the property $H_{\chi}\big(G_1 \sqcup G_2, P_A(a_1)P_{X_1}+ P_A(a_2)P_{X_2}\big) = H_{\chi}(G_1,P_{X_1})= H_{\chi}(G_2,P_{X_2})$, for all $P_A$ when the probabilistic graphs $(G_1,P_{X_1})$ and $(G_2,P_{X_2})$ are isomorphic. Lemma \ref{lemma:partvaluesA} shows that the \mlt{complementary graph entropy of the disjoint union of graphs $\overline{H}(\sqcup\:\cdot)$ is a function of the complementary graph entropy} of the AND product of graphs $\overline{H}(\wedge\:\cdot)$, on a dense subset of distributions. In that case, the equivalence of the \mlt{additivity with respect to} $\wedge$ and $\sqcup$ is direct.

A consequence of  \eqref{eq:corpartvaluesUnif} is that both \mlt{additivity properties} of \eqref{eq:LinHbar2} and \eqref{eq:LinHbar1} are equivalent when $P_A=\Unif(\mc{A})$. In order to prove Theorem~\ref{th:caraclin}, we extend this property to all distributions $P_A \in \Delta(\mc{A})$ by using the convexity \mlt{property} of the function 
\begin{align}
    \eta : P_A \mapsto \overline{H}\Bigg(\bigsqcup_{a \in \mathcal{A}} G_a,\sum_{a \in \mathcal{A}} P_A(a) P_{X_a}\Bigg).\label{eq:EtaFunction}
\end{align}

\begin{lemma}\label{th:etaconvex}
Given a family of graphs $(G_a)_{a\in\mc{A}}$, the function $\eta$ 
is convex in $P_A$ 
and $(\log \max_a |\mathcal{X}_a|)$-Lipschitz.
\end{lemma}
The proof of Lemma~\ref{th:etaconvex} is stated in App.~\ref{section:proofetaconvex}. Now, since $\eta$ is convex, if it meets the linear interpolation of $(\eta(\mathds{1}_a))_{a \in \mathcal{A}}$, where $(\mathds{1}_a)_{a \in \mathcal{A}}$ are the extreme points of the set of distributions $\Delta(\mathcal{A})$, then $\eta$ is linear. This statement is provided by Lemma~\ref{lemma:convexanalysis2} whose proof is given in App.~\ref{section:prooflemmaconvexanalysis}. 

\begin{lemma}\label{lemma:convexanalysis2}
Let $\mathcal{A}$ be a finite set, and $\gamma : \Delta(\mathcal{A}) \rightarrow \mathbb{R}$ be a convex function, and for all $a \in \mathcal{A}$, let $\mathds{1}_a$ be the distribution that assigns 1 to the symbol $a$ and 0 to the others. Then the following holds:
\begin{align}
& \exists P_A \in \interior(\Delta(\mathcal{A})), \; \gamma(P_A) = \sum_{a \in \mathcal{A}} P_A(a) \gamma(\mathds{1}_a) \label{eq:lemma1AA}\\
\Longleftrightarrow \;\; & \forall P_A \in \Delta(\mathcal{A}), \qquad \; \gamma(P_A) = \sum_{a \in \mathcal{A}} P_A(a) \gamma(\mathds{1}_a). \label{eq:lemma1BA}
\end{align}
where $\interior(\Delta(\mathcal{A}))$ is the interior of $\Delta(\mathcal{A})$, i.e. the set of full-support distributions on $\mathcal{A}$.
\end{lemma}

Now let us prove Theorem~\ref{th:caraclin}.

$(\Longrightarrow)$ \mlt{We} assume that $\overline{H}\left(\textstyle\bigwedge_{a \in \mathcal{A}} G_{a}, \textstyle\bigotimes_{a\in \mc{A}} P_{X_a}\right) = \textstyle\sum_{a \in \mathcal{A}} \overline{H}(G_{a}, P_{X_a})$. By taking $P_A =\Unif(\mathcal{A})$, the equation \eqref{eq:corpartvaluesUnif} in Lemma~\ref{lemma:partvaluesA} implies that
\begin{align}
&\overline{H}\left(\bigsqcup_{a \in \mathcal{A}} 
G_a,\sum_{a \in \mathcal{A}} \frac{1}{|\mathcal{A}|} P_{X_a} \right)\nonumber\\
=& \frac{1}{|\mathcal{A}|}\overline{H}\left(\bigwedge_{a \in \mathcal{A}} G_a , \bigotimes_{a \in \mathcal{A}} P_{X_a} \right)\nonumber\\
=& \sum_{a \in \mathcal{A}} \frac{1}{|\mathcal{A}|}\overline{H}(G_{a},P_{X_a}).
\end{align}
The function $\eta$ is convex by Lemma~\ref{th:etaconvex}, and  satisfies \eqref{eq:lemma1AA} with the interior point $P_A = \Unif(\mathcal{A})$, thus Lemma~\ref{lemma:convexanalysis2} implies \mlt{that for all $P_A \in \Delta(\mathcal{A})$,}
\begin{align} 
\overline{H}\left(\bigsqcup_{a \in \mathcal{A}}%
G_{a} ,\sum_{a \in \mathcal{A}} P_A(a) P_{X_a} \right) = \sum_{a \in \mathcal{A}} P_{A}(a) \overline{H}(G_{a},P_{X_a}).\label{eq:proofcormainBB}
\end{align}

$(\Longleftarrow)$ Conversely, \mlt{we} assume \eqref{eq:proofcormainBB} and \mlt{we} consider $P_A=\Unif(\mathcal{A})$. According to \eqref{eq:corpartvaluesUnif} in Lemma~\ref{lemma:partvaluesA}, we conclude that
\begin{align}
&\overline{H}\left(\bigwedge_{a \in \mathcal{A}} G_a, \bigotimes_{a \in \mathcal{A}} P_{X_a}  \right) \nonumber\\ 
=& |\mathcal{A}|\cdot \overline{H}\left(\bigsqcup_{a \in \mathcal{A}}
G_a, \sum_{a \in \mathcal{A}} \frac{1}{|\mathcal{A}|}
P_{X_a} \right) \nonumber\\
=& \sum_{a \in \mathcal{A}} \overline{H}(G_{a},P_{X_a}). 
\end{align}

This concludes the proof of Theorem~\ref{th:caraclin}.

\subsection{\mlt{Problem Statement with Partial Side Information at the Encoder}}\label{sec:partialSideInfoPb}

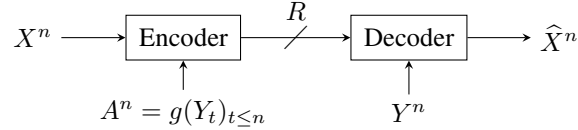
\begin{figure}[h!]
\centering
\begin{tikzpicture}[xscale=0.55,yscale=0.6]
\draw [>=stealth,->] (0,0.5) -- (2,0.5) ;
\draw  (1,1) node {$X^n$} ;
\draw (2,0) rectangle (5,1);
\draw (3.5,0.5) node {Encoder} ;
\draw [,>=stealth,->] (5,0.5) -- (8,0.5) ;
\draw (6.5,0.5) node {$\diagup$} ;
\draw (6.5,1.2) node {$R$} ;
\draw (8,0) rectangle (11,1);
\draw (9.5,0.5) node {Decoder} ;
\draw [>=stealth,->] (11,0.5) -- (13,0.5) ;
\draw  (12,1) node {$\widehat{X}^n$} ;
\draw [>=stealth,->] (9.5,-1) -- (9.5,0) ;
\draw  (9.5,-1.5) node {$Y^n$} ;
\draw [>=stealth,->] (3.5,-1) -- (3.5,0) ;
\draw  (3.5,-1.5) node {$A^n = g(Y_t)_{t \leq n}$} ;
\end{tikzpicture}
    \caption{The partial-side-information problem.}
    \label{fig:Linearsourcesum}
\end{figure}

\mlt{In this section, we highlight the operational significance of the disjoint union of graphs. Recall that Theorem~\ref{th:caraclin} uses the \mlt{additivity} for the disjoint union in order to establish the \mlt{additivity} for the AND product of graphs.} This is \mlt{generally simpler because} the disjoint union has fewer vertices and edges than the AND product. Yet, the usefulness extends further as the disjoint union corresponds to the problem in Fig.~\ref{fig:Linearsourcesum}, where the encoder has partial information about the decoder's side information, obtained through the deterministic function $g : \mathcal{Y} \rightarrow \mathcal{A}$. This setting in Fig.~\ref{fig:Linearsourcesum}, is a specific case of \mlt{the setting of Fig.~\ref{fig:Zero-errorSW} where the decoder must retrieve the source $(X,g(Y))$.} We refer to this as the \emph{partial-side-information problem}.

%
%
%
%
%
%

\begin{proposition}\label{prop:charsum}
When the encoder has partial side information, the optimal rate \mlt{is}
\begin{align}
R^{\star}_0  = \overline{H}\left(\bigsqcup_{a \in \mathcal{A}} G_a , \sum_{a \in \mathcal{A}} P_A(a) P_{X_a} 
\right),
\end{align}
where for all $a \in \mathcal{A}$, $G_a$ is the characteristic graph associated to the conditional distribution
{$P_{Y|X,A=a}$ with the underlying probability distribution $P_{X|A=a}$ on its vertices. These conditional distributions and $P_A$ are obtained from the joint distribution $P_{XY}\UN_{A=g(Y)}$ depending on the deterministic function $g : \mathcal{Y} \rightarrow \mathcal{A}$.}
\end{proposition}
Indeed, for each realization of the encoder side information $a=g(y)$, we construct a characteristic graph $G_a$ to model the sub-problem indexed by $a\in\mc{A}$. Since, both encoder and decoder have access to $a \in \mc{A}$, the characteristic graph consists in the  disjoint union \mlt{of} the graphs $(G_a)_{a\in\mc{A}}$. Moreover, each $G_a$ contains all realizations $x \in \mathcal{X}$, and there is an edge between two vertices $x,x'$ if and only if $P_{Y|X,a}(y|x,a)P_{Y|X,a}(y|x',a) > 0$ for some $y \in g^{-1}(a)$.

If \mlt{$\overline{H}$ is additive for the disjoint union} in the sense of \eqref{eq:LinHbar1}, then the following coding scheme \mlt{is optimal}: 
\begin{itemize}
    \item For each symbol $a\in\mc{A}$, \mlt{we} select the indices $t\in\{1,\ldots,n\}$ of the sequence $a^n$ such that $a_t =a$. \mlt{We} denote by $(X^{n_a},Y^{n_a})$ the corresponding subsequences  of length $n_a$, extracted from $(X^n,Y^n)$.
    \item For each $a\in\mc{A}$, \mlt{we} use the optimal codebook for the independent sources $(X^{n_a},Y^{n_a})_{a\in\mc{A}}$ with distinct length of sequences $(n_a)_{a\in\mc{A}}$ and \mlt{we} concatenate the codewords obtained.
\end{itemize} 
With high probability, the sequence $A^n$ belongs to the set of typical sequences $\mathcal{T}^n_{\varepsilon}(P_A)$, therefore the empirical distribution $(\frac{n_a}{n})_{a\in\mc{A}}$ converges to $P_A$ in probability. The coding rate of the above scheme converges to
\begin{align}
\sum_{a \in \mathcal{A}} P_{A}(a) \overline{H}(G_{a},P_{X_a}) = \overline{H}\left(\bigsqcup_{a \in \mathcal{A}} G_{a}
, \sum_{a \in \mathcal{A}} P_A(a) P_{X_a} \right),\nonumber
\end{align}
which is optimal for the partial-side-information problem when \mlt{$\overline{H}$ is additive.}

\section{Zero-Error Channel Coding Problem}\label{sec:channel}

Recently, Wigderson and Zuiddam in \cite{WigdersonZuiddam2023} and Schrijver in \cite{schrijver2023shannon} \mlt{established} the equivalence between the \mlt{additivity} of the zero-error channel capacity \mlt{for} the AND product $C_0(\wedge\:\cdot)$ and \mlt{for} the disjoint union of graphs $C_0(\sqcup\:\cdot)$. The main difference with the side information problem is that the channel input distribution $P_X$ is \emph{\mlt{a priori}} not specified in the zero-error channel coding problem.

In order to establish the equivalence between the \mlt{additivity} properties  of $\overline{H}$ and $C_0$, we introduce the \emph{zero-error capacity $C(G,P_X)$ of a graph $G$ relative to a distribution $P_X$} due to Csisz\'{a}r and K\"{o}rner \cite{csiszar1981capacity}. We show the equivalence of \mlt{all additivity properties, i.e. those of  \cite[Theorem 4.1]{WigdersonZuiddam2023}, of \cite[Theorem 2]{schrijver2023shannon}, and of Theorem~\ref{th:caraclin},} provided that the source distribution maximizes the zero-error capacity relative to a distribution of the AND product of graphs.

\subsection{\mlt{Open Problem of Zero-Error Channel Capacity}}

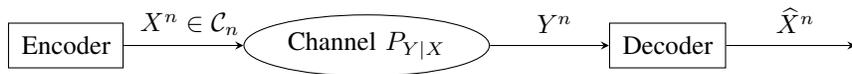
\begin{figure}[h!]
    \centering
    \begin{tikzpicture}[xscale=0.85,yscale=1]
    \node[shape=rectangle, draw=black, fill=white, inner sep=5pt] (E) at (0,0) {Encoder};
    \node[shape=ellipse, draw=black, fill=white, inner sep=3pt] (W) at (4,0) {Channel $P_{Y|X}$};
    \node[shape=rectangle, draw=black, fill=white, inner sep=5pt] (D) at (8,0) {Decoder};
    
    \node[draw=none] (X_1) at ($(D)+(1.35,13pt)$) {$\widehat{X}^n $};
    
    \draw[->, >=stealth] (D) edge ($(D)+(1.6,0)$);
    \draw[->, >=stealth] (W) edge (D);
    \draw[->, >=stealth] (E) edge (W);
    
    \node[draw=none] (X0) at ($(1.5,0)$) {};
    \node[draw=none] (X) at ($(X0)+(0.2,13pt)$) {$X^n \in \mc{C}_n$};
    \node[draw=none] (Y0) at ($(6.0,0)$) {};
    \node[draw=none] (Y) at ($(Y0)+(0.5,13pt)$) {$Y^n$};
    \end{tikzpicture}
    \caption{The channel coding problem.}
    \label{fig:Zero-errorC}
\end{figure}

The channel coding problem in Fig. \ref{fig:Zero-errorC} is introduced in \cite{shannon1948mathematical} in the vanishing error regime, and in \cite{shannon1956zero} in the zero-error regime. 
We consider a \emph{discrete memoryless channel} that consists of an input alphabet $\mathcal{X}$, a {finite} output alphabet $\mathcal{Y}$ and a conditional distribution $P_{Y|X}$. A $(n,\mathcal{C}_n,\phi_d)$-\emph{code} consists of 
\begin{itemize}
\item[-] an encoder that selects uniformly a codeword $x^n$ from the codebook $\mathcal{C}_n \subseteq \mathcal{X}^n$, and sends it over the channel, 
\item[-] a decoder $\phi_d$ that assigns an estimate $\widehat{x}^n$ to each received \mlt{sequence} $y^n$.
\end{itemize}
The rate of the $(n,\mathcal{C}_n,\phi_d)$-code is the average number of messages transmitted per channel use, i.e. $\frac{1}{n}\log|\mathcal{C}_n|$, and the probability of error is $P_e^{(n)} \doteq \mathbb{P}\big(\widehat{X}^n \neq X^n \big)$.

\begin{definition}
The \emph{zero-error channel capacity} is the maximal rate among all coding schemes that satisfy the zero-error property:
\begin{align}
C_{0}^{\star} \doteq \sup\limits_{ \displaystyle (n, \mathcal{C}_n, \phi_d): P_e^{(n)}=0} \quad \frac{1}{n}\log|\mathcal{C}_n|.
\end{align}
\end{definition}


In the zero-error regime, the capacity depends on the independence number of the characteristic graph, defined below.

\begin{definition}[Independent subset, independence number $\alpha$]
\label{def:independent}
Let $G = (\mathcal{X}, \mathcal{E})$ be a graph. A subset $\mathcal{S} \subseteq \mathcal{X}$ is \emph{independent} in $G$ if $xx' \notin \mathcal{E}$ for all $x\neq x' \in \mathcal{S}$. The \emph{independence number} is the maximal size of an independent set in $G$, and is denoted by $\alpha(G)$. 
\end{definition}

The zero-error channel capacity relies on the same \emph{characteristic graph}, \mlt{see Definition~\ref{def:chargraph}.}

\begin{theorem}[from \cite{shannon1956zero}]\label{prop:shannonC_0}
Let $G$ be the characteristic graph corresponding to the discrete memoryless channel $(\mathcal{X},\mathcal{Y},P_{Y|X})$. 
The zero-error channel capacity satisfies
\begin{align}
C_0^{\star} = C_0(G) \doteq \lim_{n \rightarrow \infty} \frac{1}{n} \log \alpha(G^{\wedge n}).\label{eq:ZeroErrorCapacity}
\end{align}
\end{theorem}
Since $\alpha(G_1\wedge G_2) \geq \alpha(G_1) \cdot \alpha(G_2)$ for all graphs $G_1$ and $G_2$, the sequence $\big( \log\alpha(G^{\wedge n})\big)_{n\in\N^{\star}}$ is superadditive, Fekete's Lemma \cite[pp. 103]{van2001course} ensures the existence of the limit in \eqref{eq:ZeroErrorCapacity}, which \mlt{coincides} with the supremum. 


\begin{remark}
Note that, by convention, we define the zero-error capacity with the logarithm. Another existing convention (for example in \cite{Lovasz79}) for the zero-error capacity is $\Theta(G) \doteq \lim_{n \rightarrow \infty} \sqrt[n]{\alpha(G^{\wedge n})}$, which is equivalent in the sense that $C_0 = \log \Theta$. 
\end{remark}

We present in Sec.~\ref{sec:PerfectGraphs} some examples from the literature where $C_0(G)$ is known, in particular when $G$ is a perfect graph. The Lov{\'a}sz $\theta$-function, 
introduced in \cite{Lovasz79}, is an upper bound on the zero-error capacity. This function is used to show that 
$C_0(C_5) = \frac{1}{2}\log 5$, which makes $C_5$ the smallest non-perfect graph for which $C_0$ is known. Further observations on the $\theta$-function are derived by Sason in \cite{sason2023observations}, and new zero-error capacity results are characterized for two subclasses of strongly regular graphs in \cite{Sason2024}. The zero-error capacity of $C_7$ is still unknown. Several existing lower bounds on $C_0(C_7)$ were found via computer programming, in particular in \cite{vesel2002improved}, \cite{polak19} and \cite{mathew2017new}.

\subsection{\mlt{Problem Statement for Independent Channels}}

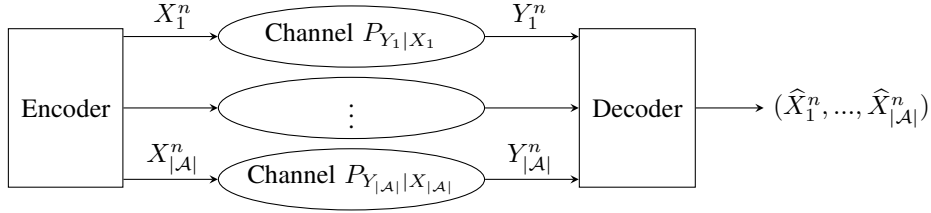
\begin{figure*}[h!]
\centering
\begin{tikzpicture}[xscale=0.9,yscale=0.98]
    \node[shape=ellipse, draw=black, fill=white, inner sep=3pt] (W1) at (4,0) {Channel $P_{Y_1|X_1}$};
    \node[shape=ellipse, draw=black, fill=white, inner sep=3pt] (W2) at (4,-1) {\textcolor{white}{Channel $P_{Y_1|X_1}$}};
    \node[shape=ellipse, draw=none, fill=none, inner sep=3pt] (W2_) at (4,-1) {$\vdots$};
    \node[shape=ellipse, draw=black, fill=white, inner sep=3pt] (W3) at (4,-2) {\textcolor{white}{Channel $P_{Y_1|X_1}$}};
    \node[shape=ellipse, draw=none, fill=none, inner sep=3pt] (W3_) at (4,-2) {Channel $P_{Y_{|\mathcal{A}|}|X_{|\mathcal{A}|}}$};
    \node[shape=rectangle, draw=none, fill=none, inner sep=5pt, minimum height = 6em] (D-) at (8,-2) {\textcolor{white}{Decoder}};
    \node[shape=rectangle, draw=none, fill=none, inner sep=5pt, minimum height = 6em] (D+) at (8,0) {\textcolor{white}{Decoder}};
    \node[shape=rectangle, draw=black, fill=white, inner sep=5pt, minimum height = 6em] (D) at (8,-1) {Decoder};
    \node[shape=rectangle, draw=none, fill=none, inner sep=5pt, minimum height = 6em] (E-) at (0,-2) {\textcolor{white}{Encoder}};
    \node[shape=rectangle, draw=none, fill=none, inner sep=5pt, minimum height = 6em] (E+) at (0,0) {\textcolor{white}{Encoder}};
    \node[shape=rectangle, draw=black, fill=white, inner sep=5pt, minimum height = 6em] (E) at (0,-1) {Encoder};
    
    \node[draw=none] (X_1) at ($(D)+(3,0)$) 
    {$(\widehat{X}_1^n, ..., \widehat{X}_{|\mathcal{A}|}^n)$};
    
    \draw[->, >=stealth] (D) edge (X_1);
    \draw[->, >=stealth] (W1) -- (D+);
    \draw[->, >=stealth] (W2) -- (D);
    \draw[->, >=stealth] (W3) -- (D-);
    \draw[->, >=stealth] (E+) -- (W1);
    \draw[->, >=stealth] (E) -- (W2);
    \draw[->, >=stealth] (E-) -- (W3);
    
    \node[draw=none] (X0) at ($(1.5,0)$) {};
    \node[draw=none] (X) at ($(X0)+(0,8pt)$) {$X_1^n$};
    \node[draw=none] (Y0) at ($(6.5,0)$) {};
    \node[draw=none] (Y) at ($(Y0)+(0,8pt)$) {$Y_1^n$};
    \node[draw=none] (X0) at ($(1.5,-2)$) {};
    \node[draw=none] (X) at ($(X0)+(0,8pt)$) {$X_{|\mathcal{A}|}^n$};
    \node[draw=none] (Y0) at ($(6.5,-2)$) {};
    \node[draw=none] (Y) at ($(Y0)+(0,8pt)$) {$Y_{|\mathcal{A}|}^n$};
    \end{tikzpicture}
    \vspace{-2em}
    \caption{The independent channel coding problem: the information is transmitted via $|\mc{A}|$ parallel channels $(P_{Y_a|X_a})_{a\in\mc{A}}$.}
    \label{fig:Linearchannelprod}
\end{figure*}

To understand why no single-letter exists for the zero-error channel capacity, we study 
the case where the conditional distribution of the channel decomposes into a product $\bigotimes_{a\in\mc{A}} P_{Y_a|X_a}$, as depicted in Fig.~\ref{fig:Linearchannelprod}. This is called the \emph{independent channel coding problem}.

\begin{proposition}[from \cite{shannon1956zero}]\label{prop:charchannelprod}
The zero-error capacity of the independent channel $\bigotimes_{a\in\mc{A}} P_{Y_a|X_a}$
 is given by
\begin{align}
C_0\left(\bigwedge_{a \in \mathcal{A}} G_a \right),
\end{align}
where for all $a \in \mathcal{A}$, $G_a$ is the characteristic graph associated to the conditional distribution $P_{Y_a|X_a}$.
\end{proposition}

In the vanishing error regime, the capacity of independent channels \mlt{is additive} since
\begin{align}
C =& \max_{P_{X_1,\ldots,X_{|\mc{A}|}}} I\big( X_1,\ldots,X_{|\mc{A}|} ; Y_1,\ldots,Y_{|\mc{A}|}  \big) \nonumber\\
=& \sum_{a\in\mc{A}} \max_{P_{X_a}} I( X_a ; Y_a) = \sum_{a\in\mc{A}} C_a.
\end{align}
Therefore, it is optimal to concatenate the optimal  codebooks designed for each channel $P_{Y_a|X_a}$. 

In the zero-error regime, the capacity is super-linear as shown by Shannon in \cite[Theorem 4]{shannon1956zero},
\begin{align}
 C_0(G) + C_0(G') \leq& \, C_0(G \wedge G').\label{eq:linchannelprod_}
\end{align}
Haemers  shows in \cite{haemers1979some} that the inequality  \eqref{eq:linchannelprod_} is strict for the product of the  Schl\"{a}fli graph $S$ and its complementary graph $\overline{S}$, as stated in Theorem~\ref{th:haemers}. An explicit construction of the Schl\"{a}fli graph is provided in \cite[Sec.~6.1]{chudnovsky2005structure}. Haemers's result relies on a bound on the zero-error capacity based on the rank of the adjacency matrix of the graph. Refinements of this bound are developed by Bukh and Cox in \cite{bukh2018fractional}, and by Gao et al. in \cite{gao2022tracial}. 
\begin{theorem}[{{from \cite{haemers1979some}}}]\label{th:haemers}
Let $S$ be the Schl\"{a}fli graph and $\overline{S}$ its complementary graph, then 
\begin{align}
C_0(S) + C_0(\overline{S}) < C_0(S \wedge \overline{S}).
\end{align}
\end{theorem}

Since the \mlt{additivity} does not hold in general, Wigderson and Zuiddam in \cite{WigdersonZuiddam2023} and Schrijver in \cite{schrijver2023shannon}, established a condition under which \mlt{the additivity holds for the zero-error capacity.}
\begin{theorem}[{{from  \cite[Theorem 4.1]{WigdersonZuiddam2023} and \cite[Theorem 2]{schrijver2023shannon}}}] \label{th:schrijver} 
For all graphs $G, G'$,
\begin{align}
 C_0(G) + C_0(G') =& \; C_0(G \wedge G'),\label{eq:linchannelprod_S}\\
\Longleftrightarrow\quad \log\left(2^{C_0(G)} + 2^{C_0(G')}\right)=& \;C_0(G \sqcup G').
\end{align}
\end{theorem}

The zero-error capacity of a disjoint union of graphs $C_0(G \sqcup G')$ is the optimal coding rate when each input letter can be selected from any of the two channels. We establish the connection between the \mlt{additivity} properties of  Theorem~\ref{th:caraclin} and Theorem~\ref{th:schrijver}, by introducing the zero-error capacity $C(G,P_X)$ of a graph relative to a distribution $P_X$, defined by Csisz\'{a}r and K\"{o}rner in \cite{csiszar1981capacity}.

\subsection{Zero-Error Capacity $C(G,P_X)$ of a Graph Relative to a Distribution $P_X$}\label{sec:ZECapacityDistribution}

Theorems~\ref{th:caraclin} and \ref{th:schrijver} both establish the equivalence of \mlt{additivity properties} for the AND product and \mlt{for the} disjoint union of graphs \mlt{but they do so for different underlying quantities:} $\overline{H}(G,P_X)$ and $C_0(G)$. We study these two quantities and we establish the equivalence between \mlt{additivity} properties. To do this, we \mlt{introduce} the zero-error capacity $C(G,P_X)$ of a graph relative to a distribution $P_X\in\Delta(\mc{X})$.

\begin{definition}
A sequence of codes $(\mathcal{C}_n)_n$ is said to be \emph{typical with respect to} $P_X$,
or in short 
\emph{$P_X$-typical}, if 
\begin{align}
\max_{x^n \in \mathcal{C}_n} \|T_{x^n} - P_X \|_\infty \underset{n \rightarrow \infty}{\rightarrow} 0. \label{eq:propCGPbasicA2}
\end{align}
The \emph{zero-error channel capacity relative to the input distribution} $P_X$ is the maximal rate among all \emph{$P_X$-typical} sequence of codes  that satisfy the zero-error property:
\begin{align}
C^{\star}(P_X) \doteq \!\!\!\!  \sup\limits_{ \displaystyle (n, \mathcal{C}_n, \phi_d): (\mathcal{C}_n)_n \ P_X\text{\emph{-typical}}, P_e^{(n)}=0}   \frac{1}{n}\log|\mathcal{C}_n|.\nonumber
\end{align}
\end{definition}



Csisz\'{a}r and K\"{o}rner \cite[equation~(3.2)]{csiszar1981capacity} derived an upper bound on the channel capacity for arbitrarily varying channel with maximal error probability. \mlt{This expression, denoted by $C(G,P_X)$,} corresponds to the zero-error capacity relative to a distribution.
\begin{lemma}[from \cite{csiszar1981capacity}]\label{prop:CGPbasic}
The zero-error capacity of the graph $G = (\mathcal{X}, \mathcal{E})$ relative to the distribution $P_X$ is 
\begin{align}
C^{\star}(P_X)\! = C(G, P_X)\! \doteq \lim_{\varepsilon \rightarrow 0} \limsup_{n \rightarrow \infty} \frac{1}{n} \log \alpha\big(G^{\wedge n}[\mathcal{T}^n_{\varepsilon}(P_X)]\big),
\label{eq:defCGP}
\end{align}
where $G^{\wedge n}[\mathcal{T}^n_{\varepsilon}(P_X)]$ is the subgraph of $G^{\wedge n}$ induced by the set of \emph{typical  sequences} $\mathcal{T}^n_{\varepsilon}(P_X)$ with tolerance $\varepsilon>0$.
\end{lemma}

The superior limit in \eqref{eq:defCGP} can be replaced by a regular limit, thanks to the superadditivity of the sequence $\big(\log \alpha(G^{\wedge n}[\mathcal{T}^n_{\varepsilon}(P_X)])\big)_{n\in\mathbb{N}^\star}$ and Fekete's Lemma \cite[pp. 103]{van2001course}. 
%
%
In \cite[Lemma 1]{marton1993shannon}, Marton established the relationship between the complementary graph entropy $\overline{H}(G,P_X)$ and $C(G,P_X)$.

\begin{theorem}[{{from \cite[Lemma 1]{marton1993shannon}}}]\label{th:marton}
Given a graph $G = (\mathcal{X}, \mathcal{E})$ and a probability distribution $P_X$,
    \begin{align}\label{eq:marton}
        C(G,P_X) + \overline{H}(G,P_X) = H(X).
    \end{align}
\end{theorem}


Equation \eqref{eq:marton} can be seen as an analog for zero-error regime of the formula $I(X;Y) + H(X|Y) = H(X)$. 
\mlt{
\begin{remark}[Zero-error source and channel duality at the code-level]
    We interpret the formula in Theorem~\ref{th:marton} in the following way. The quantities $n \cdot \overline{H}(G,P_X)$ and $n \cdot C(G,P_X)$ represent in bits, respectively, the minimum number of colors and the maximum size of an independent set. A color class, i.e. the set of vertices to which the same color is assigned,  is an independent subset of vertices. In the case where all color classes have the same size, we would need $ n \cdot \log \alpha(G)$ bits to describe the source sequence within its color class. Thus, $n  \cdot C(G,P_X)$ can be seen as the average number of bits needed to specify the index of the source sequence in its color class. These two quantities sum up to $n \cdot H(X)$, which is the information needed to describe the source sequence with zero-error. \ar{However, this interpretation does not allow to conclude about the source and channel duality at the code level. In particular it does not guarantee that all sources are covered by a codeword, i.e. a color and an index within that color class, which is required in the source-coding problem.}
\end{remark}}

We establish below the connection between the \mlt{additivity} properties of $\overline{H}(G,P_X)$ and $C(G,P_X)$, where the equivalences in \eqref{eq:Hbar5341} and \eqref{eq:Hbar5363} follow from Marton's formula in Theorem~\ref{th:marton}, and the equivalence \eqref{eq:Hbar5352} from Theorem~\ref{th:caraclin}. The complete proof is in App.~\ref{section:proofpropmarton}.
\begin{proposition}\label{prop:marton}
Let $\mathcal{A}$ be a finite set, $P_A$ a distribution with full-support, and let 
$(G_a,P_{X_a})_{a \in \mathcal{A}} $ a family of probabilistic graphs. The following equivalences hold:
\begin{align}
& C\left(\bigsqcup_{a \in \mathcal{A}} G_a,\; \sum_{a \in \mathcal{A}} P_A(a) P_{X_a}\right) \nonumber\\
&= H(P_A) + \sum_{a \in \mathcal{A}} P_A(a) C(G_a, P_{X_a}) \\
\Longleftrightarrow \quad & \overline{H}\left(\bigsqcup_{a \in \mathcal{A}} G_{a},\; \sum_{a \in \mathcal{A}} P_A(a) P_{X_a}\right)\nonumber\\
& = \sum_{a \in \mathcal{A}} P_{A}(a) \overline{H}(G_{a}, P_{X_a})\label{eq:Hbar5341}\\
\Longleftrightarrow \quad & \overline{H}\left(\bigwedge_{a \in \mathcal{A}} G_a , \bigotimes_{a \in \mathcal{A}} P_{X_a} \right) = \sum_{a \in \mathcal{A}} \overline{H}(G_a, P_{X_a})\label{eq:Hbar5352}\\
 \Longleftrightarrow \quad    & C\left(\bigwedge_{a \in \mathcal{A}} G_a,\; \bigotimes_{a \in \mathcal{A}} P_{X_a}\right) = \sum_{a \in \mathcal{A}} C(G_a, P_{X_a}). \label{eq:Hbar5363}
    \end{align}
\end{proposition}

\ar{Proposition~\ref{prop:marton} is an important result, as it enables the passage from additivity equivalences for $\overline{H}(G,P_X)$ to those for $C(G,P_X)$, and vice versa.}

\subsection{Capacity Achieving Distributions and Equivalence of the \mlt{Additivity} of $C_0$ and $\overline{H}$ for the AND Product}\label{sec:FinalEquivalenceAND}

\mlt{The key element  of this section} is the set of input distributions $P_X$ that achieve the zero-error capacity in the sense that $C_0(G) = C(G,P_X)$. As in the vanishing error regime, it \mlt{is} optimal to consider codebooks composed of codewords that are typical with respect to the input distribution $P_X$ that maximizes $C(G,P_X)$.


\begin{lemma}[{{from \cite[\mlt{equation} (11.2)]{csiszar2011information}}}]\label{th:simonyi} For all graph $G = (\mathcal{X}, \mathcal{E})$, 
    \begin{align}
        C_0(G) = \max_{P_X } C(G,P_X) .
    \end{align}
\end{lemma}

The proof, stated in \cite[Theorem 2]{gargano1994capacities}, follows  similar arguments to those in the proof of \cite[Lemma~3]{simonyi2003witsenhausen}.
As a consequence of Theorem~\ref{th:marton} and Lemma~\ref{th:simonyi}, the zero-error capacity reformulates
\begin{align}
C_0(G) = \max_{P_X } \Big(H(X) - \overline{H}(G,P_X)\Big).\label{eq:C0HbarH}
\end{align}

%


In order to show the equivalence of the \mlt{additivity of $C_0(G)$ and of $C(G, P_X)$}, we define the set of capacity-achieving distributions.

\begin{definition}
\label{def:Pstar}
    Let $G = (\mathcal{X}, \mathcal{E})$ be a graph. The \emph{set of capacity-achieving distributions} of $G$ is the subset of $\Delta(\mathcal{X})$ defined by 
    \begin{align}
        \mathcal{P}^{\star}(G) \doteq \argmax_{P_X}\; C(G,P_X).
    \end{align}
\end{definition}

 \begin{proposition}\label{prop:Pstar}
    For all graphs $G$, the mapping $P_X \mapsto C(G,P_X)$ is concave and the set of capacity-achieving distributions $\mathcal{P}^{\star}(G)$ is convex, nonempty.
\end{proposition}

The proof of Proposition~\ref{prop:Pstar} is stated in App.~\ref{section:proofpropPstar}, and relies on Lemma~\ref{th:simonyi}. The following Theorem is essential \mlt{to demonstrate} the equivalence of the \mlt{additivity properties} depicted in Fig.~\ref{fig:additivity}. It establishes that if a joint distribution achieves capacity, then the product of its marginals also achieves it.
\begin{theorem}\label{lemma:Pstarprod}
If $P_{X_1, ..., X_{|\mathcal{A}|}} \in \mathcal{P}^{\star}(\bigwedge_{a \in \mathcal{A}} G_a)$, then $\bigotimes_{a \in \mathcal{A}}P_{X_a} \in \mathcal{P}^{\star}(\bigwedge_{a \in \mathcal{A}} G_a)$.
\end{theorem}

The proof of Theorem~\ref{lemma:Pstarprod} is stated in  App.~\ref{section:prooflemmaPstarprod} and relies on a \emph{codebook shifting argument}: given a codebook composed of codewords $(x_1^n,x_2^n)$ that are typical with respect to the joint distribution $P_{X_1,X_2}$, we construct a set of permuted codebooks by applying a cyclic permutation only to the first component $x_1^n$ of each codeword. We concatenate all the permuted codebooks and we replicate them $n$ times so that the codewords length is equal to $n'=n^3$. Then, we remove the codewords $(x_1^{n'},x_2^{n'})$ that are not typical with respect to the product of marginal distributions $P_X \otimes P_{X'}$. We show that this construction has the same rate and preserves the zero-error property. However, it modifies the types of the codewords, which become the product of marginals as wished.



We can now establish the equivalence \mlt{between the additivity of $C_0(G)$ and of $C(G,P_X)$} for the AND product.
\begin{theorem}\label{th:mainC0}
Let $\mathcal{A}$ be a finite set, and $(G_a)_{a \in \mathcal{A}} = (\mathcal{X}_a, \mathcal{E}_a)_{a \in \mathcal{A}}$ be a family of graphs. The following equivalence holds:
    \begin{align}
        & C_0\left(\bigwedge_{a \in \mathcal{A}} G_a\right) = \sum_{a \in \mathcal{A}} C_0(G_a) \label{eq:mainthC_0A}\\
        \Longleftrightarrow \;\;& \exists P_{X_1, ..., X_{|\mathcal{A}|}} \in  \mathcal{P}^{\star}\left(\bigwedge_{a \in \mathcal{A}} G_a \right)\!,\nonumber\\
        &C\left(\bigwedge_{a \in \mathcal{A}} G_a,\; P_{X_1, ..., X_{|\mathcal{A}|}} \right) = \sum_{a \in \mathcal{A}} C(G_a, P_{X_a}).\label{eq:mainthC_0B}
    \end{align}
    Furthermore, any distribution $P_{X_1, ..., X_{|\mathcal{A}|}} \in \mathcal{P}^{\star}\left(\bigwedge_{a \in \mathcal{A}} G_a \right)$ that satisfies \eqref{eq:mainthC_0B} also satisfies $P_{X_a} \in \mathcal{P}^{\star}(G_a)$ for all $a \in \mathcal{A}$.
\end{theorem}

The proof of Theorem~\ref{th:mainC0} is given in App.~\ref{section:proofthmainC0}. \mlt{The main difficulty comes from the decomposition of the capacity-achieving distributions over the AND product of graphs. On the one hand, if $C_0$ is additive, then the product of capacity-achieving distributions $P_{X_a} \in \mathcal{P}^{\star}(G_a)$ is optimal for the graph $\bigwedge_{a \in \mathcal{A}} G_a$ and thus, $C$ is additive. On the other hand, if $C$ is additive for the  graph $\bigwedge_{a \in \mathcal{A}} G_a$ with the capacity-achieving  distribution $P_{X_1, ..., X_{|\mathcal{A}|}}\in \mathcal{P}^{\star}\big(\bigwedge_{a \in \mathcal{A}} G_a \big)$, then $C_0$ is additive. \ar{These} are the statements of Lemma~\ref{lemma:mainC0A} and Lemma~\ref{lemma:mainC0B} in App.~\ref{section:proofthmainC0}.}


\subsection{\mlt{Additivity} of the Sum of Independent Channels}\label{sec:SumChannel}

\begin{figure*}[!ht]
\centering
\resizebox{.95\textwidth}{!}{%
\begin{circuitikz}
\tikzstyle{every node}=[font=\large]
\node [font=\LARGE] at (2,11.25) {};
\draw (9,13.5) to[short, -o] (7.5,13.5) ;
\draw [](9,15.25) to[short, -o] (7.5,15.25) ;
\draw [](9,11.75) to[short, -o] (7.5,11.75) ;
\node [font=\LARGE] at (12.75,15.25) {Channel $P_{Y_1|X_1}$};
\node [font=\LARGE] at (12.75,11.75) {Channel $P_{Y_{|\mathcal{A}|}|X_{|\mathcal{A}|}}$};
\node [font=\LARGE] at (12.75,13.5) {$\vdots$};
\draw  (12.75,15.25) ellipse (3.75cm and 0.75cm);
\draw  (12.75,11.75) ellipse (3.75cm and 0.75cm);
\draw  (12.75,13.5) ellipse (3.75cm and 0.75cm);
\draw [](16.5,11.75) to[short, -o] (18,11.75) ;
\draw [](16.5,15.25) to[short, -o] (18,15.25) ;
\draw [](16.5,13.5) to[short, -o] (18,13.5) ;
\draw [](3.75,13.5) to[short, -o] (6,13.5) ;
\draw [short] (6,13.5) -- (7,14.75);
\draw [](21.75,13.5) to[short, -o] (19.5,13.5) ;
\node [font=\LARGE] at (5,14) {$(X_{a_t})_{t\le n}$};
\draw  (0.75,14.25) rectangle (3.75,12.75);
\node [font=\LARGE] at (2.25,13.5) {Encoder};
\node [font=\LARGE] at (20.5,14) {$(Y_{a_t})_{t\le n}$};
\draw [short] (18.5,14.75) -- (19.5,13.5);
\draw  (21.75,14.25) rectangle (24.75,12.75);
\node [font=\LARGE] at (23.25,13.5) {Decoder};
\node [font=\LARGE] at (27.25,13.5) {$(\widehat{X}_{a_t})_{t \le n}$};
\node [font=\Large] at (6.5,15) {$a_t$};
\draw [->, >=Stealth] (24.75,13.5) -- (26,13.5);
\end{circuitikz}
}%
    \caption{Sum of the $|\mc{A}|$ channels $(P_{Y_a|X_a})_{a\in\mc{A}}$: only the channel $a_t\in \mc{A}$ is used at instant  $t\in \{1,\ldots,n\}$.}
\label{fig:Linearchannelsum}
\end{figure*}
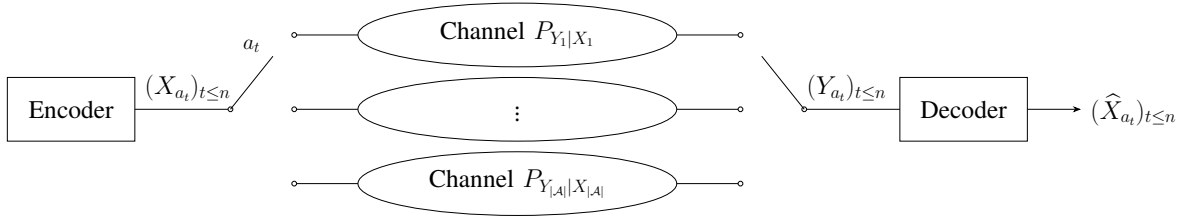

The equivalences in Theorem~\ref{th:schrijver} and Proposition~\ref{prop:marton} rely on the zero-error capacity of a disjoint union of graphs. Such graphs have an operational interpretation in terms of sum of channels \mlt{which is depicted in Fig.~\ref{fig:Linearchannelsum}, see also \cite[pp.~13]{shannon1956zero}.} The codewords are intermingled, each letter in a codeword can be selected from any of the channels $(P_{Y_a|X_a})_{a\in \mc{A}}$. Since the output alphabets of each individual channel are disjoint, the channel output symbol uniquely identifies the channel that is used. This allows to embed additional information with rate $H(P_A)$. 



In the vanishing error regime, the \mlt{additivity} of the capacity for the sum of channels holds since
\begin{align}
C = \log\bigg(\sum_{a \in \mathcal{A}} 2^{C_a}\bigg),
\end{align}
where $C_a \doteq \max_{P_{X_a}} I(X_a;Y_a)$ is the capacity of the channel $P_{Y_a|X_a}$. 

\begin{proposition}[from \cite{shannon1956zero}]\label{prop:charchannelsum}
The zero-error capacity of the sum of channels is given by
\begin{align}
C_0\left(\bigsqcup_{a \in \mathcal{A}} G_a \right). \label{eq:ratechannelsum}
\end{align}
\end{proposition}

In the zero-error regime, Shannon in \cite[Theorem 4]{shannon1956zero} shows that 
\begin{align}
\forall G, G', \quad  \log\left(2^{C_0(G)} + 2^{C_0(G')}\right)\leq& \,C_0(G \sqcup G').
\end{align}

For the sum of channels, a natural coding scheme consists in using the optimal codebooks for each channel in a time sharing manner, with respect to the distribution $P_A$ that maximizes $H(P_A) + \sum_{a \in \mathcal{A}} P_A(a) C_0(G_a)$. {In other words, communicating over the sum channel is equivalent to sending 2 types of information: one related to identification of the chosen channel of rate $H(P_A)$, and the other to the information sent at rate $C_0(G_a)$ over each channel $a\in\mc{A}$.}

\begin{lemma}\label{lemma:conjugate}
The mapping $P_A \mapsto H(P_A) + \sum_{a \in \mathcal{A}} P_A(a) C_0(G_a)$ has a unique maximum 
\begin{align}
    P^{\star}_A \doteq \left(\frac{2^{C_0(G_a)}}{\sum_{a' \in \mathcal{A}} 2^{C_0(G_{a'})}} \right)_{a \in \mathcal{A}},\label{eq:P_Zopti}
\end{align}
which gives
\begin{align}
    H(P^{\star}_A) + \sum_{a \in \mathcal{A}} P^{\star}_A(a) C_0(G_a)  
    & = \log\left(\sum_{a' \in \mathcal{A}} 2^{C_0(G_{a'})} \right). \label{eq:natratechannelsum}
\end{align}
\end{lemma}

The proof of Lemma~\ref{lemma:conjugate} is given in App.~\ref{section:prooflemmaconjugate} and relies on the fact that the function $(w_a)_{a \in \mathcal{A}} \mapsto \log\left(\sum_{a \in \mathcal{A}} 2^{w_a}\right)$ is the Legendre-Fenchel conjugate of the entropy function $P_A \mapsto H(P_A)$, \mlt{see \cite[Chap.~1]{hiriart1993convex}.} 

We consider the time-sharing strategy between the optimal codebooks along with the distribution $P^{\star}_A\in \Delta(\mc{A})$ defined in \eqref{eq:P_Zopti}. If this strategy is optimal, then 
\begin{align}
    C_0\left(\bigsqcup_{a \in \mathcal{A}} G_a \right) = \log\left(\sum_{a \in \mathcal{A}} 2^{C_0(G_a)}\right),\label{eq:linchannelsum}
\end{align}
which means that the \mlt{additivity} holds for the disjoint union of graphs. 


\begin{remark}\label{rq:fullsupp}
Note that $P^{\star}_A\in \Delta(\mc{A})$ \mlt{in \eqref{eq:P_Zopti} is full-support. It can be observed that} $P_A \mapsto H(P_A)$ has an infinite slope at the frontier of $\Delta(\mathcal{A})$, consequently the maximizer of $P_A \mapsto H(P_A) + \sum_{a \in \mathcal{A}} P_A(a) C_0(G_a)$ is always an interior point. In other words, the information carried by the channel index $H(P_A)$ offsets the loss in rate, if the channels with smaller capacities are not chosen too often. Therefore, in the sum of channels setting, always choosing the channel with highest capacity is suboptimal, and never choosing a channel is also suboptimal, even if this channel has \mlt{a capacity equal to $0$.}
\end{remark}

Similar to Theorem~\ref{th:mainC0}, we establish the equivalence between the \mlt{additivity} between $C_0(G)$ and  $C(G,P_X)$ for the disjoint union of a family of graphs $(G_a)_{a\in\mc{A}}$.

\begin{theorem}\label{th:linC_0sqcup}
    The following equivalence holds
    \begin{align}
        & C_0\left(\bigsqcup_{a \in \mathcal{A}} G_a \right) = \log\left(\sum_{a \in \mathcal{A}} 2^{C_0(G_a)}\right) \\
        \Longleftrightarrow \; & \exists P_X \in \mathcal{P}^{\star} \left(\bigsqcup_{a \in \mathcal{A}} G_a \right)\!, \nonumber\\
        &C\left(\bigsqcup_{a \in \mathcal{A}} G_a,\; P_X \right) = H(P_A) + \sum_{a \in \mathcal{A}} P_A(a) C(G_a, P_{X_a}), \label{eq:linC_0sqcup}
    \end{align}
    where \mlt{for all $a \in \mathcal{A}$, $P_A(a) = \sum_{x\in \mathcal{X}_a}P_X(x)$ and $P_{X_a} = P_{X|X \in \mathcal{X}_a}= P_X/P_A(a)$}. Furthermore, any $\sum_{a \in \mathcal{A}} P_A(a) P_{X_a}$ that satisfies \eqref{eq:linC_0sqcup} also satisfies the following for all $a \in \mathcal{A}$:
    \begin{align}
        P_A(a) = \frac{2^{C_0(G_a)}}{\sum_{a' \in \mathcal{A}} 2^{C_0(G_{a'})}}, \text{ and } P_{X_a} \in \mathcal{P}^{\star}(G_a). \label{eq:linC_0sqcup2}
    \end{align}
\end{theorem}

The proof of Theorem~\ref{th:linC_0sqcup} is given in App.~\ref{section:proofthlinC_0sqcup}. \mlt{The key ingredient is the decomposition of the capacity-achieving distributions over the disjoint union of graphs.
The first result is Lemma~\ref{lemma:conjugate} which proves that the distribution given by $P_A(a) = \frac{2^{C_0(G_a)}}{\sum_{a' \in \mathcal{A}} 2^{C_0(G_{a'})}}$ maximizes $P_A \mapsto H(P_A) + \sum_{a \in \mathcal{A}} P_A(a)C_0(G_a)$. On the one hand, if $C_0$ is additive, then the average distribution $\sum_{a\in\mc{A}}P_A(a) P_{X_a}$ with $\forall a\in\mc{A}$, $P_{X_a} \in \mathcal{P}^{\star}(G_a)$ is optimal for the graph $\bigsqcup_{a \in \mathcal{A}} G_a$ and thus, $C$ is additive. On the other hand, if $C$ is additive for the  graph $\bigsqcup_{a \in \mathcal{A}} G_a$ and $\sum_{a\in\mc{A}}P_A(a) P_{X_a}$ is capacity-achieving, then $C_0$ is additive. \ar{These} are the statements of Lemma~\ref{lemma:linC_0sqcup1} and Lemma~\ref{lemma:linC_0sqcup2} in App.~\ref{section:proofthlinC_0sqcup}.}

\ar{\begin{remark}
One could try an alternative route to prove Theorem~\ref{th:linC_0sqcup}, by successively using the equivalences in  Theorem~\ref{th:schrijver}\footnote{i.e. \cite[Theorem 4.1]{WigdersonZuiddam2023} and\cite[Theorem 2]{schrijver2023shannon}}, Theorem~\ref{th:mainC0}, and Proposition~\ref{prop:marton}.
    However, this approach yields the following statement:
\begin{align} 
&\textstyle C_0\left(\textstyle\bigsqcup_{a \in \mathcal{A}} G_a \right) = \log\left(\textstyle\sum_{a \in \mathcal{A}} 2^{C_0(G_a)}\right) \\
\Longleftrightarrow \quad\; & \left\{ \begin{array}{l}
\exists P_A \in \Delta(\mathcal{A}) \text{ full-support},\\ 
\exists P_{X_1, ..., X_{|\mathcal{A}|}} \in \mathcal{P}^{\star} \left(\textstyle\bigwedge_{a \in \mathcal{A}} G_a \right)\!, \\
\textstyle C\left(\textstyle\bigsqcup_{a \in \mathcal{A}} G_a,\; \sum_{a \in \mathcal{A}} P_A(a) P_{X_a} \right) \\
\;\;\; = H(P_A) + \sum_{a \in \mathcal{A}} P_A(a) C(G_a, P_{X_a}), 
\end{array}\right. \label{eq:OtherRoute}
\end{align}
Nevertheless, this does not complete the proof, as we still need to relate the sets of capacity-achieving distributions
$\mathcal{P}^{\star} \left(\bigsqcup_{a \in \mathcal{A}} G_a \right)$ required in Theorem~\ref{th:linC_0sqcup} to $\mathcal{P}^{\star} \left(\bigwedge_{a \in \mathcal{A}} G_a \right)$ appearing in \eqref{eq:OtherRoute}.
\end{remark}}

Theorem~\ref{th:linC_0sqcup} together with Theorem~\ref{th:mainC0}, Theorem~\ref{th:schrijver} and Proposition~\ref{prop:marton}, establish the equivalence of the \mlt{additivity} between $C_0(G)$,  $C(G,P_X)$ and $\overline{H}(G,P_X)$ for the AND product $\wedge$, and for the disjoint union $\sqcup$ of a family of graphs $(G_a)_{a\in\mc{A}}$, as depicted in Fig.~\ref{fig:additivity}.

\section{Main Example and Counter-Examples for the \mlt{Additivity} of Optimal Rates}\label{sec:Examples}

In this section, we exploit the equivalences in the \mlt{additivity properties} 
depicted in Fig.~\ref{fig:additivity}, in order to provide single-letter \mlt{characterizations} of $\overline{H}$ and $C_0$ for several new classes of graphs.


\subsection{Perfect Graphs}\label{sec:PerfectGraphs}

\mlt{In this section,} we show that perfect graphs allow for \mlt{additivity of $\overline{H}$ and $C_0$} with respect to both $\sqcup$ and $\wedge$ with any underlying probability distribution. Perfect graphs are one of the only known examples of graphs with a single-letter formula for \mlt{$\overline{H}$ and $C_0$}. 
Theorem~\ref{th:caraclin} allows us to provide new single-letter \mlt{characterizations} for \mlt{$\overline{H}$ and $C_0$} for \mlt{all} products of perfect graphs, which are not perfect in general. 

\begin{definition}[Graph complement, clique number $\omega$]\label{def:omega}
    For all $G = (\mathcal{X}, \mathcal{E})$, the complementary graph of $G$ is defined by $\overline{G} \doteq (\mathcal{X}, \mathcal{E}^c)$. The clique number of $G$ is defined by $\omega(G) \doteq \alpha(\overline{G})$, where the independence number $\alpha$ is stated in Definition~\ref{def:independent}.
\end{definition}


\begin{definition}[Perfect graph]
    A graph $G = (\mathcal{X}, \mathcal{E})$ is \emph{perfect} if for all subset of vertices $\mathcal{S} \subseteq \mathcal{X}, \; \chi(G[\mathcal{S}]) = \omega(G[\mathcal{S}])$. A probabilistic graph $(G, P_X)$ is perfect if $G$ is perfect. 
\end{definition}


A remarkable property of perfect graphs is their single-letter characterization for zero-error source and channel coding problems. For example, when $G$ is a perfect graph, the optimal rate of the side information problem $\overline{H}(G,P_X)$ is equal to the K\"{o}rner graph entropy introduced in \cite{Korner73}, \mlt{see also \cite[equation (13)]{alon1996source} and \cite[Sec.II.B]{OrlitskyRoche2001},}.

\begin{definition}[K\"{o}rner graph entropy $H_{\kappa}$]\label{def:Hkappa}
For all probabilistic graph $(G,P_X)$, let $\Gamma(G)$ be the collection of independent sets of vertices in $G$. The K\"{o}rner graph entropy of $(G,P_X)$ is defined by
\begin{align}
    H_{\kappa}(G,P_X) = \min_{X \in W \in \Gamma(G)} I(W;X), \label{eq:defHkappa}
\end{align}
where the minimum is taken over all distributions $P_{W|X}$ with the constraint that the random vertex $X$ belongs to the random independent set $W$ with probability one, i.e. $X \in W \in \Gamma(G)$ in \eqref{eq:defHkappa}.
\end{definition}

\begin{theorem}[{{from \cite[Corollary 12]{csiszar1990entropy}}}]\label{th:Hbarperfect}
Let $(G,P_X)$ be a perfect probabilistic graph, then 
\begin{align}
\overline{H}(G,P_X) = H_{\kappa}(G,P_X).
\end{align}
\end{theorem}

Similarly, \mlt{a single-letter characterization exists} for the zero-error capacity of perfect graphs, as stated below. This is a consequence of a more general result due to Shannon (see \cite[Theorem 3]{shannon1956zero}) that states that a graph $G$ whose vertex set can be partitioned into $\alpha(G)$ cliques, i.e. complete induced subgraphs, satisfies $C_0(G) = \log \alpha(G)$. \mlt{A perfect graph $G$ satisfies this property as its complementary $\overline{G}$} is also perfect, and satisfy $\chi(\overline{G}) = \omega(\overline{G}) = \alpha(G)$, where $\omega(\overline{G})$ is the clique  number, see  \cite[pp.~382]{berge73}. 

\begin{theorem}[{{from \cite[Theorem 3]{shannon1956zero}}}]\label{th:C_0perfect}
    If $G$ is a perfect graph, then $C_0(G) = \log \alpha(G)$.
\end{theorem}


\mlt{We now provide single-letter characterizations for $C_0$, $C$, and $\overline{H}$ that were previously unknown.} These characterizations are consequences of the \mlt{additivity} results of Wigderson and Zuiddam   \cite{WigdersonZuiddam2023} and Schrijver \cite{schrijver2023shannon} for $C_0$, and of Theorem~\ref{th:caraclin} for $\overline{H}$. 

More precisely, \mlt{we} consider some perfect graphs. Their disjoint union is perfect, as shown in  Lemma~\ref{lemma:perfectunion}, and $C_0$ \mlt{is additive} since $C_0(G \sqcup G') = \log \alpha(G\sqcup G') = \log(\alpha(G) + \alpha(G')) = \log(2^{C_0(G)}+2^{C_0(G')})$ holds for all perfect graphs $G$, $G'$. According to the \mlt{additivity} result of \cite{WigdersonZuiddam2023} and \cite{schrijver2023shannon}, since $C_0$ \mlt{is additive for the disjoint union}, so does $C_0$ \mlt{for} the AND product $C_0(G\wedge G') = C_0(G) + C_0(G')$. This leads to the following proposition.


\begin{proposition}\label{prop:linC_0perf}
Let $G$ and $G'$ be perfect graphs, then
    \begin{align}
        & C_0(G \sqcup G') = \! \log\left(\! 2^{C_0(G)} + 2^{C_0(G')}\right) \!  = \log(\alpha(G) + \alpha(G')),\\
        & C_0(G \wedge G') = C_0(G) + C_0(G') = \log\alpha(G) + \log\alpha(G').
    \end{align}
\end{proposition}

\begin{figure}[h!]
    \centering
    \begin{tikzpicture}
        \foreach \i in {0,1,2,3,4,5,6}{
        \foreach \j in {0,1,2,3,4}{
        \draw[] ($(\i,\j)$) -- ($(\i+1,\j+1)$);
        \draw[] ($(\i+1,\j)$) -- ($(\i,\j+1)$);
        \draw[] ($(\i,\j)$) -- ($(\i+1,\j)$);
        \draw[] ($(\i,\j)$) -- ($(\i,\j+1)$);
        }
        }
        \foreach \i in {0,1,2,3,4,5,6}{
        \draw[] ($(\i,5)$) -- ($(\i+1,5)$);
        }
        \foreach \j in {0,1,2,3,4}{
        \draw[] ($(7,\j)$) -- ($(7,\j+1)$);
        }
        
        \draw[myBG] ($(2,2)$) -- ($(2,3)$) -- ($(3,4)$) -- ($(4,3)$)  -- ($(5,2)$) -- ($(4,1)$) -- ($(3,1)$) -- ($(2,2)$);
        
        \foreach \i in {0,1,2,3,4,5,6,7}{
        \draw[dashed] ($(\i,5)$) -- ($(\i+.75,5.75)$);
        \draw[dashed] ($(\i,5)$) -- ($(\i,5.75)$);
        \draw[dashed] ($(\i,5)$) -- ($(\i-.75,5.75)$);
        \draw[dashed] ($(\i,0)$) -- ($(\i+.75,-.75)$);
        \draw[dashed] ($(\i,0)$) -- ($(\i,-.75)$);
        \draw[dashed] ($(\i,0)$) -- ($(\i-.75,-.75)$);
        }
        
        \foreach \j in {0,1,2,3,4,5}{
        \draw[dashed] ($(0,\j)$) -- ($(-.75,\j-.75)$);
        \draw[dashed] ($(0,\j)$) -- ($(-.75,\j)$);
        \draw[dashed] ($(0,\j)$) -- ($(-.75,\j+.75)$);
        \draw[dashed] ($(7,\j)$) -- ($(7.75,\j-.75)$);
        \draw[dashed] ($(7,\j)$) -- ($(7.75,\j)$);
        \draw[dashed] ($(7,\j)$) -- ($(7.75,\j+.75)$);
        }
        
        \foreach \i in {0,1,2,3,4,5,6,7}{
        \foreach \j in {0,1,2,3,4,5}{
        \node[shape=circle,draw=black, inner sep = .5pt,fill=white] (\i,\j) at ($(\i,\j)$) {\i,\j};
        }
        }
        
    \end{tikzpicture}
    \caption{A non-perfect AND product of perfect graphs: $C_6 \wedge C_8$ with an induced $C_7$.}
    \label{fig:counterperfect}
\end{figure}
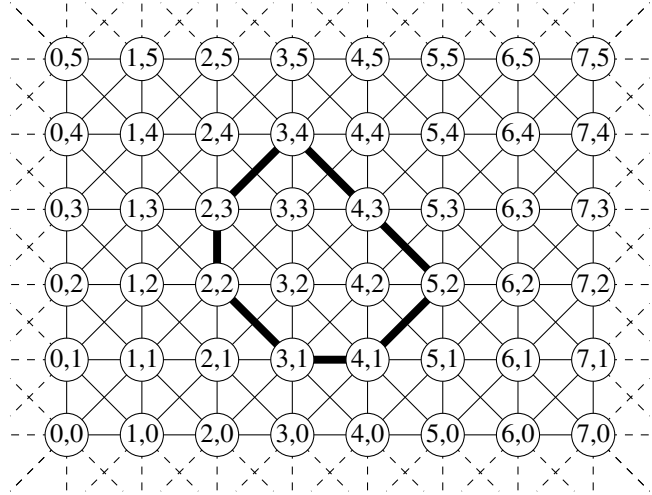

According to the previous proposition, $C_0(G \wedge G')$ can now be computed for any \mlt{pairs} of perfect graphs, as it \mlt{is additive}. This result was previously unknown, as the AND product of perfect graphs is not necessarily perfect. For instance, cycle graphs $C_6$ and $C_8$ are perfect, due to the strong perfect graph Theorem mentioned below, but their AND product is not. \mlt{This is also} due to the strong perfect graph Theorem \mlt{since} it contains an odd cycle of length 7, illustrated in  Fig.~\ref{fig:counterperfect}. 

\begin{theorem}[{{Strong Perfect Graph Theorem, from \cite[Theorem 1.2]{chudnovsky2006strong}}}]\label{th:strongperfect}
A graph $G$ is perfect if and only if neither $G$ nor $\overline{G}$ have an induced odd cycle of length at least 5.
\end{theorem}


Similarly, we show that the \mlt{additivity} property of $\overline{H}$ and $C$ holds for perfect graphs and for all underlying probability distributions, and we provide new single-letter expression  for $\overline{H}$ \mlt{and} $C$ in that case. 
\begin{theorem}\label{th:mainperfect}
When $(G_a,P_{X_a})_{a \in \mathcal{A}}$
is a family of perfect probabilistic graphs, we have the following single-letter characterizations
\begin{align}
&\overline{H}\left(\bigwedge_{a \in \mathcal{A}} G_a, \bigotimes_{a \in \mathcal{A}} P_{X_a}  \right) = \sum_{a \in \mathcal{A}} \overline{H}(G_a,P_{X_a})\nonumber\\
&\quad= \sum_{a \in \mathcal{A}} H_{\kappa}(G_a,P_{X_a}), \\
&\overline{H}\left(\bigsqcup_{a \in \mathcal{A}} G_a, \sum_{a \in \mathcal{A}} P_A(a) P_{X_a}  \right) = \sum_{a \in \mathcal{A}} P_A(a) \overline{H}(G_a,P_{X_a})\nonumber\\
&\quad= \sum_{a \in \mathcal{A}} P_A(a) H_{\kappa}(G_a,P_{X_a}),\\
&C\left(\bigwedge_{a \in \mathcal{A}} G_a, \; \bigotimes_{a \in \mathcal{A}} P_{X_a} \right)  = \sum_{a \in \mathcal{A}} C(G_a, P_{X_a}) \nonumber\\
&\quad= \sum_{a \in \mathcal{A}} \big(H(X_a) - H_{\kappa}(G_a,P_{X_a})\big),\\
&C\left(\bigsqcup_{a \in \mathcal{A}} G_a, \; \sum_{a \in \mathcal{A}} P_A(a) P_{X_a} \right) \nonumber\\
&\quad= H(P_A) + \sum_{a \in \mathcal{A}} P_A(a) C(G_a, P_{X_a}) \nonumber\\
&\quad= H(P_A) + \sum_{a \in \mathcal{A}} P_A(a) \Big( H(X_a) - H_{\kappa}(G_a,P_{X_a}) \Big).
\end{align}
\end{theorem}
The proof of Theorem~\ref{th:mainperfect} is given in App.~\ref{section:proofthmainperfect}. As an example, we consider the AND product of the cycle graphs $C_6$ and $C_8$.
\begin{corollary}\label{coro:C6C8}
\mlt{We} consider the cycle graphs $C_6$ and $C_8$ and \mlt{we} denote by $P_{X_6}$ and $P_{X_8}$ the probability distributions over their vertices. We have 
\begin{align}
\overline{H}(C_6 \wedge C_8, P_{X_6}\otimes P_{X_8}) =& 
H_{\kappa}(C_6,P_{X_6}) + H_{\kappa}(C_8,P_{X_8}),\\
C(C_6 \wedge C_8 , P_{X_6}\otimes P_{X_8} ) =& 
H(P_{X_6}) - H_{\kappa}(C_6,P_{X_6})\nonumber\\
&+ H(P_{X_8}) - H_{\kappa}(C_8,P_{X_8}) .
\end{align}
\end{corollary}


We now explore the combination of a perfect graph with a non-perfect graph. More specifically, we consider the graph $C_5 \sqcup G$ where $G$ is perfect, for which the \mlt{additivity} of $\overline{H}$ was studied by Tuncel et al. in \cite{tuncel2009complementary}. 
The pentagon graph $C_5$ is not perfect, thereby making any disjoint union or AND product involving it non-perfect. However, by using Theorem~\ref{th:caraclin}, we provide a non-perfect example where the \mlt{additivity} property holds, offering a single-letter characterization of $\overline{H}$ for the class of graphs $C_5 \wedge G$ \mlt{when} $G$ is perfect.

\begin{theorem}[{{from \cite[Lemma 3]{tuncel2009complementary}}}]\label{th:G_5}
    Let $s \in [0,1]$, let $(C_5, \Unif(\lbrace 1, ..., 5\rbrace))$ and $(G,P_X)$ be a perfect probabilistic graph, we have
    \begin{align}
        &\overline{H}(C_5 \sqcup G,s\Unif(\lbrace 1, ..., 5\rbrace) + (1-s) P_X) \nonumber\\
        & = s \overline{H}(C_5,\Unif(\lbrace 1, ..., 5\rbrace)) + (1-s)\overline{H}(G,P_X)  \\
        & = \frac{s}{2} \log 5 + (1-s) H_{\kappa}(G,P_X).
    \end{align}
\end{theorem}

\begin{corollary}\label{cor:C_5}
For all perfect probabilistic graph $(G,P_X)$, 
\begin{align}
&\overline{H}(G \wedge C_5,P_X\otimes \Unif(\lbrace 1, ..., 5\rbrace)) \nonumber\\
&= \overline{H}(G,P_X) + \overline{H}(C_5,\Unif(\lbrace 1, ..., 5\rbrace))\nonumber\\
&= H_{\kappa}(G,P_X) + \textstyle\frac{1}{2}\log 5.
    \end{align}
\end{corollary}




\subsection{Vertex Transitive Graphs}

\mlt{We study the class of vertex-transitive graphs, where all the vertices play the same ``role''. For these graphs, we show that the uniform distribution achieves the zero-error capacity.} 

\begin{definition}[Vertex-transitive graph]
    An automorphism of a graph $G = (\mathcal{X}, \mathcal{E})$ is a bijection $\psi : \mathcal{X} \rightarrow \mathcal{X}$ such that for all $x, x' \in \mathcal{X}$, $xx' \in \mathcal{E}$ if and only if $\psi(x)\psi(x') \in \mathcal{E}$. The group of automorphisms of G is denoted by $\Aut(G)$.

    A graph $G = (\mathcal{X}, \mathcal{E})$ is {vertex-transitive} if $\Aut(G)$ acts transitively on its vertices, i.e. for all $x,x' \in \mathcal{X}$, there exists $\psi \in \Aut(G)$ such that $\psi(x) = x'$.
\end{definition}

\begin{proposition}\label{th:vertextrans}
If $G = (\mathcal{X}, \mathcal{E})$ is vertex-transitive, then 
\begin{align}
\Unif(\mathcal{X}) \in \mathcal{P}^{\star}(G).
\end{align}
\end{proposition}

The proof of Proposition~\ref{th:vertextrans} is given in App.~\ref{section:proofthvertextrans}. 
\begin{corollary}\label{cor:vertextrans}
    Let $(G_a)_{a \in \mathcal{A}} = (\mathcal{X}_a, \mathcal{E}_a)_{a \in \mathcal{A}}$ be vertex-transitive graphs, their product is also vertex-transitive and 
    \begin{align}
        \Unif\left(\prod_{a \in \mathcal{A}} \mathcal{X}_a \right) = \bigotimes_{a \in \mathcal{A}} \Unif(\mathcal{X}_a) \in \mathcal{P}^{\star}\left(\bigwedge_{a \in \mathcal{A}} G_a \right).
    \end{align}
\end{corollary}

%
%
%

\subsection{The Schl\"{a}fli Graph}\label{section:maincontexS}

We now study the important case of the Schl\"{a}fli graph $S$ as it offers a counterexample for the \mlt{additivity of $\overline{H}$, $C$ and $C_0$.} In \cite{haemers1979some}, Haemers showed that the \mlt{additivity} property does not hold for the \mlt{zero-error capacity of the} product of the Schl\"{a}fli graph $S$ with its complement $\overline{S}$,
\begin{align}
C_0(S) + C_0(\overline{S}) < C_0(S \wedge \overline{S}).
\end{align}
More specifically, Haemers \mlt{showed} that $C_0(S) = \log_2(3)$, 
$C_0(\overline{S}) \leq \log_2(7)$ and $\log_2(27)\leq C_0(S \wedge \overline{S})$. In this section, we show that a similar conclusion holds for $C(G,P_X)$ and for $\overline{H}(G,P_X)$.

According to \cite[Lemma~3.7]{cameron19806}, the Schl\"{a}fli graph $S$ and its complement $\overline{S}$ are vertex transitive, as well as their product $S \wedge \overline{S}$.
By Proposition~\ref{th:vertextrans}, the uniform distribution is capacity-achieving for $S$, $\overline{S}$, and $S \wedge \overline{S}$. 


\begin{corollary}\label{cor:schlafli}
We denote by $\mathcal{X}_S$ and $\mathcal{X}_{\overline{S}}$ the sets of vertices of the Schl\"{a}fli graph $S$ and its complement $\overline{S}$. Then, 
\begin{align}
&C(S, \Unif(\mathcal{X}_S))= C_0(S),\;  C(\overline{S}, \Unif(\mathcal{X}_S)) = C_0(\overline{S}), \\
&C(S \wedge  \overline{S}, \Unif(\mathcal{X}_S)\otimes \Unif(\mathcal{X}_{\overline{S}})) = C_0(S \wedge  \overline{S}).
\end{align}
\end{corollary}

In Theorem~\ref{th:maincontex}, we extend Haemers's results of \cite{haemers1979some} and we show that the \mlt{additivity of $\overline{H}$ and $C$ does not hold when the distribution $P_X$  is uniform.} By using \eqref{eq:corpartvaluesUnif} in Lemma~\ref{lemma:partvaluesA}, we also equalize $\overline{H}$, and similarly $C_0$, for the AND product $S \wedge  \overline{S}$ and for the disjoint union $S \sqcup  \overline{S}$, up to a certain constant.


\begin{theorem}\label{th:maincontex}
Let $s \in (0,1)$, let $S$  be the Schl\"{a}fli graph and $\overline{S}$ its complementary, with uniform distributions on their vertices. Then, 
\begin{align}
&C(S \wedge  \overline{S}, \Unif(\mathcal{X}_S) \otimes \Unif(\mathcal{X}_{\overline{S}})) \nonumber\\ 
 & > \:C(S, \Unif(\mathcal{X}_S)) + C(\overline{S}, \Unif(\mathcal{X}_{\overline{S}})), \label{eq:ThMainContex1}\\
&C(S \sqcup \overline{S}, s \Unif(\mathcal{X}_S) + (1-s) \Unif(\mathcal{X}_{\overline{S}})) \nonumber\\ 
 &> \: h_b(s) + s C(S, \Unif(\mathcal{X}_S))   + (1-s) C(\overline{S}, \Unif(\mathcal{X}_{\overline{S}})), \label{eq:ThMainContex2}\\
& \overline{H}(S \wedge \overline{S}, \Unif(\mathcal{X}_S) \otimes \Unif(\mathcal{X}_{\overline{S}})) \nonumber\\ 
& < \: \overline{H}(S,\Unif(\mathcal{X}_S)) + \overline{H}(\overline{S},\Unif(\mathcal{X}_{\overline{S}})),\label{eq:ThMainContex3} \\
&\overline{H}(S  \sqcup \overline{S}, s \Unif(\mathcal{X}_S) + (1-s) \Unif(\mathcal{X}_{\overline{S}})) \nonumber\\ 
& < \: s \overline{H}(S,\Unif(\mathcal{X}_S)) + (1-s)\overline{H}(\overline{S},\Unif(\mathcal{X}_{\overline{S}}));\label{eq:ThMainContex4}
\end{align}
where $h_b$ is the binary entropy. 
\end{theorem}


We obtain \eqref{eq:ThMainContex1} from Theorem~\ref{th:mainC0} and Corollary \ref{cor:schlafli}, equations \eqref{eq:ThMainContex2} and \eqref{eq:ThMainContex3} come from Proposition~\ref{prop:marton}, and \mlt{equation} \eqref{eq:ThMainContex4} comes from Theorem~\ref{th:caraclin}.

\begin{remark}
    \mlt{In \cite{alon1998shannon},} Alon has built infinite families of graphs that satisfy $C_0(G \sqcup G') > \log(2^{C_0(G)} + 2^{C_0(G')})$. Similar results as in Theorem~\ref{th:maincontex} can be derived for these graphs, by using their respective capacity-achieving distributions.
\end{remark}

\section{Conclusion}\label{section:conclusion}

We have shown the equivalences of \mlt{additivity properties between $C_0$, $C$, and $\overline{H}$} for the AND product ($\wedge$) and for the disjoint union ($\sqcup$) of graphs, as depicted in Fig.~\ref{fig:additivity}. \mlt{For zero-error problems with same characteristic graphs and capacity-achieving distributions, we show that it is optimal to code separately for independent channels if and only if it is optimal to code separately for independent sources with side-information.} We also state the following open questions:
\begin{itemize}
    \item[-] As pointed out in Theorem~\ref{lemma:Pstarprod}, for all capacity-achieving \mlt{distributions} of a product graph, the product of its marginals is also capacity-achieving. Are these marginals capacity-achieving for the respective graphs in the product, and conversely, if we consider the product of capacity-achieving distributions of graphs, is this distribution capacity-achieving for the product of graphs? In other words,
        \begin{align}
            \mathcal{P}^{\star}\left(\bigwedge_{a \in \mathcal{A}} G_a \right) \cap \bigotimes_{a \in \mathcal{A}} \Delta(\mathcal{X}_a) \overset{?}{=} \bigotimes_{a \in \mathcal{A}} \mathcal{P}^{\star}(G_a).
        \end{align}
    \mlt{Theorem~\ref{th:mainC0} partially answer this question} in the sense that inclusion holds when the \mlt{additivity holds for the AND product.}
    \item[-] \mlt{Theorem~\ref{th:mainC0} and Theorem~\ref{th:linC_0sqcup} show} that the \mlt{additivity} of $C_0$ holds if and only if the \mlt{additivity} of $C$ holds \mlt{when $P_X$ is capacity-achieving.} Can we find graphs $G$ such that the \mlt{additivity} of $C(G,P_X)$ holds when $P_X$ is capacity-achieving, but does not hold for some $P_X$ that is not capacity-achieving? A negative answer would imply that the \mlt{additivity} of $C_0$ is equivalent to the \mlt{additivity of $C(G,P_X)$ and $\overline{H}(G,P_X)$} \emph{for all} $P_X$, similarly to perfect graphs.
    \item[-] \mlt{In the vanishing error regime, Wyner showed in \cite[Sec.~VI.C]{Wyner74} that a linear channel code can be used to construct a source code with side information, where the encoder transmits the syndrome. Similarly in Theorem~\ref{th:marton}, the index of a source sequence can be decomposed into a color class and an index within the color class. Does such a source-channel duality hold at the code level \ar{in the zero-error regime}? Can we show that the decomposition into the color classes is optimal for both zero-error problems?}
    \item[-] Finally, we have seen in Corollary \ref{cor:C_5} that $\overline{H}\big(G\wedge C_5,P_X\otimes P_{X_5} \big)$ \mlt{is additive} when $G$ perfect. Is the non-\mlt{additivity} of $\overline{H}$ tied to specific non-perfect induced subgraphs in each graph in the product? And if so, can we find a minimal \mlt{sub-family for} these graphs?
\end{itemize}

\appendices

\section{Results for the Proof of Theorem~\ref{th:caraclin}} \label{section:proofthcaraclin}

\subsection{Proof of Lemma~\ref{lemma:partvaluesA}}\label{section:prooflemmapartvalues}

We need several lemmas for this result.
Lemma~\ref{lemma:distrib} establishes the distributivity of $\wedge$ with respect to $\sqcup$ for probabilistic graphs, similarly as in \cite{zuiddam2018algebraic} for graphs without underlying distribution. 
Lemma~\ref{lemma:simplificationinduced} states that $\overline{H}$ can be computed with subgraphs induced by sets that have an asymptotic probability one, in particular we will use it with typical sets of vertices. 
Lemma~\ref{lemma:unionisomorph} gives the chromatic entropy of a disjoint union of isomorphic probabilistic graphs.
The proofs of Lemma~\ref{lemma:distrib}, Lemma~\ref{lemma:simplificationinduced} and Lemma~\ref{lemma:unionisomorph} are respectively given in App.~\ref{section:prooflemmadistrib}, App.~\ref{section:prooflemmasimplificationinduced}, and Appendix  \ref{setion:prooflemmaunionisomorph}.

\begin{lemma}\label{lemma:distrib}
    Let $\mathcal{A}, \mathcal{B}$ be finite sets, let $P_A \in \Delta(\mathcal{A})$ and $P_B \in \Delta(\mathcal{B})$. For all $a \in \mathcal{A}$ and $b \in \mathcal{B}$, let $(G_a , P_{X_a})$ and $(G_b, P_{X_b})$ be probabilistic graphs. Then 
    \begin{align}
        \left(\bigsqcup_{a \in \mathcal{A}} G_a\right) \wedge \left(\bigsqcup_{b \in \mathcal{B}} G_b\right) = \bigsqcup_{(a,b) \in \mathcal{A} \times \mathcal{B}} G_a \wedge G_b.\label{eq:distribunion}
    \end{align}
\mlt{For each $(a,b) \in \mathcal{A} \times \mathcal{B}$ and $(x_a,x_b)\in\mc{X}_a \times \mc{X}_b$, the induced probability distribution takes value $P_A(a)P_B(b)P_{X_a}(x_a)P_{X_b}(x_b)$.}    
\end{lemma}

\begin{lemma}\label{lemma:simplificationinduced}
    Let a probabilistic graph $(G, P_X)$, and let $(\mathcal{S}_n)_{n \in \mathbb{N}^\star}$ be a sequence of sets such that for all $n \in \mathbb{N}^\star$, $\mathcal{S}_n \subseteq \mathcal{X}^n$, and $P^{\otimes n}_X(\mathcal{S}_n) \rightarrow 1$ when $n \rightarrow \infty$. Then $\overline{H}(G,P_X) = \lim_{n \rightarrow \infty} \frac{1}{n} H_\chi\big(G^{\wedge n}[\mathcal{S}_n], P'_{X^n}\big)$ where $P'_{X^n}(x^n) = P^{\otimes n}_X(x^n)/P^{\otimes n}_X(\mathcal{S}_n)$ for all $x^n\in \mathcal{S}_n$, and $P'_{X^n}(x^n)=0$ otherwise.
\end{lemma}

\begin{definition}[Isomorphic probabilistic graphs]\label{def:iso}
Let $(G_1, P_{X_1})$ and $(G_2 , P_{X_2})$ be two probabilistic graphs with $G_1 = (\mathcal{X}_1, \mathcal{E}_1)$ and $G_2 = (\mathcal{X}_2, \mathcal{E}_2)$. We say that $(G_1, P_{X_1})$ is isomorphic to $(G_2 , P_{X_2})$ (denoted by $(G_1, P_{X_1}) \simeq (G_2 , P_{X_2})$) if there exists an isomorphism between them, i.e. a bijection $\psi : \mathcal{X}_1 \rightarrow \mathcal{X}_2$ such that:
\begin{itemize}[label = - ]
    \item For all $x_1, x_1' \in \mathcal{X}_1$, $x_1 x'_1 \in \mathcal{E}_1 \Longleftrightarrow \psi(x_1)\psi(x'_1) \in \mathcal{E}_2$,
    \item For all $x_1 \in \mathcal{X}_1$, $P_{X_1}(x_1) = P_{X_2}\big(\psi(x_1)\big)$.
\end{itemize}
\end{definition}

\begin{lemma}\label{lemma:unionisomorph}
    Let $\mathcal{B}$ be a finite set, let $P_B \in \Delta(\mathcal{B})$ and let $(G_b,P_{X_b})_{b \in \mathcal{B}}$ be a family of isomorphic probabilistic graphs, then $H_\chi\big(\bigsqcup_{b' \in \mathcal{B}} G_{b'} ,\sum_{b'\in\mc{B}}P_B(b')P_{X_b'}\big) = H_\chi(G_b,P_{X_b})$ for all $b \in \mathcal{B}$.
\end{lemma}

Now let us prove Lemma~\ref{lemma:partvaluesA}. Let $P_A \in \Delta(\mathcal{A})$, and let $(G,P_X) =\Big( \bigsqcup_{a \in \mathcal{A}} G_a, \sum_{a\in\mc{A}}P_A(a) P_{X_a}\Big)$, by definition we have
\begin{align}
&\overline{H}(G,P_X)\nonumber\\
=& \lim_{n \rightarrow \infty} \frac{1}{n} H_\chi\bigg(\Big(\bigsqcup_{a \in \mathcal{A}} G_a\Big)^{\wedge n}, \Big(\sum_{a\in\mc{A}}P_A(a) P_{X_a}\Big)^{ n}\bigg)\\
=& \lim_{n \rightarrow \infty} \frac{1}{n} H_\chi\bigg(\bigsqcup_{a^n \in \mathcal{A}^n} \bigwedge_{t\leq n} G_{a_t}, \sum_{a^n\in\mc{A}^n}P_A^n(a^n) \bigotimes_{t\leq n}P_{X_{a_t}}\bigg),\label{eq:conclemmapart1yc}
\end{align}
where \eqref{eq:conclemmapart1yc} is a consequence of Lemma~\ref{lemma:distrib}.

We focus our attention on typical sequences, we denote by  $T_{a^n}$ the type of the sequence $a^n$. Let $\varepsilon > 0$, and for all $n \in \mathbb{N}^\star$, we define
\begin{align}
\mathcal{T}^n_{\varepsilon}(P_A) \doteq& \big\lbrace a^n \in \mathcal{A}^n \:\big|\: \| T_{a^n} - P_A \|_\infty \leq \varepsilon \big\rbrace, \label{eq:lemmapartA0} \\
\mathcal{S}_{n,\varepsilon} \doteq& \bigsqcup_{a^n \in \mathcal{T}^n_{\varepsilon}(P_A)} \;\prod_{t \leq n} \mathcal{X}_{a_t}. \nonumber
\end{align}
We introduce the conditional distribution of $A^n$ given $A^n\in \mathcal{T}^n_{\varepsilon}(P_A)$ by
\begin{align}
P'_{A^n}(a^n)\doteq
\begin{cases}
\frac{P^{\otimes n}_A(a^n)}{P^{\otimes n}_A(\mathcal{T}^n_{\varepsilon}(P_A))},&\text{ if }a^n\in \mathcal{T}^n_{\varepsilon}(P_A),\\
0&\text{ otherwise,}
\end{cases}
\end{align}
and the distribution $P'_{X^n}$ induced over the subgraph $G^{\wedge n}[\mathcal{S}_{n,\varepsilon}]$, defined by 
\begin{align}
P'_{X^n} = \sum_{a^n\in\mathcal{T}^n_{\varepsilon}(P_A)}P'_{A^n}(a^n) \bigotimes_{t\leq n}P_{X_{a_t}}.
\end{align}
Since for all $\varepsilon>0$, $P^{\otimes n}_X(\mathcal{S}_{n,\varepsilon}) \rightarrow 1$ when $n \rightarrow \infty$, we have by Lemma~\ref{lemma:simplificationinduced} 
\begin{align}
      \overline{H}(G,P_X) = \lim_{n \rightarrow \infty} \frac{1}{n}H_\chi\Big(G^{\wedge n}[\mathcal{S}_{n,\varepsilon}], P'_{X^n}\Big).\label{eq:conclemmapart1}
\end{align}

Let us study the limit in \eqref{eq:conclemmapart1}. Let $(\overline{a}_n)_{n \in \mathbb{N}^\star} \in \mathcal{A}^{\mathbb{N}^\star}$ be a sequence such that $\overline{a}^n \in \mathcal{T}^n_{\varepsilon}(P_A)$ as $T_{\overline{a}^n} \rightarrow P_A$. Therefore, for all $a^n \in \mathcal{T}^n_{\varepsilon}(P_A)$, $a \in \mathcal{A}$, and $n$ large enough, we have
\begin{align}
    \big|T_{\overline{a}^n}(a)-T_{a^n}(a)\big|\leq 2\varepsilon.\label{eq:lemmapartB0}
\end{align}

We have on one hand
\begin{align}
& H_\chi\big(G^{\wedge n}[\mathcal{S}_{n,\varepsilon}],P'_{X^n}\big)\nonumber\\
=& H_\chi\bigg(\Big(\bigsqcup_{a^n \in \mathcal{A}^n} \bigwedge_{t\leq n} G_{a_t}\Big)\Big[\mathcal{S}_{n,\varepsilon}\Big],\nonumber\\
&\qquad \sum_{a^n\in\mathcal{T}^n_{\varepsilon}(P_A)}P'_{A^n}(a^n) \bigotimes_{t\leq n}P_{X_{a_t}}\bigg)\label{eq:lemmapartB1}\\
=& H_\chi\bigg(\bigsqcup_{a^n \in \mathcal{T}^n_{\varepsilon}(P_A)} \bigwedge_{a\in\mc{A}} G_{a}^{\wedge n T_{a^n}(a)}, \nonumber\\
&\qquad\sum_{a^n\in\mathcal{T}^n_{\varepsilon}(P_A)}P'_{A^n}(a^n) \bigotimes_{a\in \mc{A}}P_{X_{a}}^{\otimes n T_{a^n}(a)}\bigg)\label{eq:lemmapartB2} \\
\leq& H_\chi\bigg(\bigsqcup_{a^n \in \mathcal{T}^n_{\varepsilon}(P_A)} \bigwedge_{a\in\mc{A}} G_{a}^{\wedge n T_{\overline{a}^n}(a) + \lceil 2 n\varepsilon \rceil}, \nonumber\\
&\qquad \sum_{a^n\in\mathcal{T}^n_{\varepsilon}(P_A)}P'_{A^n}(a^n) \bigotimes_{a\in \mc{A}}P_{X_{a}}^{\otimes n T_{\overline{a}^n}(a) + \lceil 2 n\varepsilon \rceil}\bigg)\label{eq:lemmapartB3} \\
=& H_\chi\bigg( \bigwedge_{a\in\mc{A}} G_{a}^{\wedge n T_{\overline{a}^n}(a) + \lceil 2 n\varepsilon \rceil},  \bigotimes_{a\in \mc{A}}P_{X_{a}}^{\otimes n T_{\overline{a}^n}(a) + \lceil 2 n\varepsilon \rceil}\bigg)\label{eq:lemmapartB4} \\
\leq& H_\chi\bigg( \bigwedge_{a\in\mc{A}} G_{a}^{\wedge n T_{\overline{a}^n}(a)},  \bigotimes_{a\in \mc{A}}P_{X_{a}}^{\otimes n T_{\overline{a}^n}(a)}\bigg) \nonumber\\
& + H_\chi\bigg( \bigwedge_{a\in\mc{A}} G_{a}^{\wedge \lceil 2 n\varepsilon \rceil},  \bigotimes_{a\in \mc{A}}P_{X_{a}}^{\otimes \lceil 2 n\varepsilon \rceil}\bigg)\label{eq:lemmapartB5}\\
\leq& H_\chi\bigg( \bigwedge_{a\in\mc{A}} G_{a}^{\wedge n T_{\overline{a}^n}(a)},  \bigotimes_{a\in \mc{A}}P_{X_{a}}^{\otimes n T_{\overline{a}^n}(a)}\bigg) + \lceil 2 n\varepsilon \rceil|\mathcal{A}|\log|\mathcal{X}|,\label{eq:lemmapartB6}
\end{align}
where \eqref{eq:lemmapartB1} comes from Lemma~\ref{lemma:distrib}; \eqref{eq:lemmapartB2} comes from the definition of $\mathcal{S}_{n,\varepsilon}$ in \eqref{eq:lemmapartA0} and a rearrangement of the terms inside the product; \eqref{eq:lemmapartB3} comes from \eqref{eq:lemmapartB0} and the fact that $H_{\chi}(G_1,P_{X_1}) \leq H_{\chi}(G_1\wedge G_2, P_{X_1}\otimes P_{X_2})$ for all pair of probabilistic graphs $(G_1,P_{X_1})$ and $(G_2,P_{X_2})$; \eqref{eq:lemmapartB4} follows from Lemma~\ref{lemma:unionisomorph}, the graphs $\big(\bigwedge_{a' \in \mathcal{A}} G_{a'}^{\wedge n T_{\overline{a}^n}(a) + \lceil 2 n\varepsilon \rceil}\big)_{a^n \in \mathcal{T}^n_{\varepsilon}(P_A)}$ are isomorphic as they do not depend on $a^n$; \eqref{eq:lemmapartB5} follows from the subadditivity of $H_\chi$; and \eqref{eq:lemmapartB6} is the upper bound on $H_\chi$ given by the highest entropy of a coloring.

On the other hand, we obtain by using similar arguments
\begin{align}
&H_\chi\big(G^{\wedge n}[\mathcal{S}_{n,\varepsilon}],P'_{X^n}\big)\nonumber\\
\geq& H_\chi\bigg( \bigwedge_{a\in\mc{A}} G_{a}^{\wedge n T_{\overline{a}^n}(a) - \lfloor 2 n\varepsilon \rfloor},  \bigotimes_{a\in \mc{A}}P_{X_{a}}^{\otimes n T_{\overline{a}^n}(a) - \lfloor 2 n\varepsilon \rfloor}\bigg) \label{eq:lemmapartC3} \\
\geq& H_\chi\bigg( \bigwedge_{a\in\mc{A}} G_{a}^{\wedge n T_{\overline{a}^n}(a)},  \bigotimes_{a\in \mc{A}}P_{X_{a}}^{\otimes n T_{\overline{a}^n}(a)}\bigg) \nonumber\\
& - H_\chi\bigg( \bigwedge_{a\in\mc{A}} G_{a}^{\wedge  \lfloor 2 n\varepsilon \rfloor},  \bigotimes_{a\in \mc{A}}P_{X_{a}}^{\otimes  \lfloor 2 n\varepsilon \rfloor}\bigg) \label{eq:lemmapartC4a} \\
\geq& H_\chi\bigg( \bigwedge_{a\in\mc{A}} G_{a}^{\wedge n T_{\overline{a}^n}(a)},  \bigotimes_{a\in \mc{A}}P_{X_{a}}^{\otimes n T_{\overline{a}^n}(a)}\bigg) - \lfloor 2 n\varepsilon \rfloor
|\mathcal{A}|\log|\mathcal{X}|.\label{eq:lemmapartC5}
\end{align}
Note that \eqref{eq:lemmapartC4a} also comes from the subadditivity of $H_\chi$, as $H_\chi(G_2,P_{X_2}) \geq H_\chi(G_1 \wedge G_2, P_{X_1} \otimes P_{X_2}) - H_\chi(G_1,P_{X_1})$ for all $(G_1,P_{X_1})$ and $(G_2,P_{X_2})$.

By combining \eqref{eq:lemmapartB6} and \eqref{eq:lemmapartC5} we obtain
\begin{align}
    & \bigg|\lim_{n \rightarrow \infty} \frac{1}{n} H_\chi(G^{\wedge n}[\mathcal{S}_{n,\varepsilon}], P'_{X^n}) -\nonumber\\
    & \;\;\lim_{n \rightarrow \infty} \frac{1}{n} H_\chi\left(\textstyle\bigwedge_{a \in \mathcal{A}} G_{a}^{\wedge n T_{\overline{a}^n}(a)}, \bigotimes_{a\in \mc{A}}P_{X_{a}}^{\otimes n T_{\overline{a}^n}(a)}\right)\bigg| \nonumber \\
    & \leq 2\varepsilon|\mathcal{A}|\log|\mathcal{X}|.\label{eq:conclemmapart2}
\end{align}

As this holds for all $\varepsilon > 0$, combining \eqref{eq:conclemmapart1} and \eqref{eq:conclemmapart2} yields the result of \eqref{eq:lemmapartvaluesA}.

Now let us prove \eqref{eq:corpartvalues}. We assume that $P_A$ is a type. Let $(\overline{a}_n)_{n \in \mathbb{N}^\star}$ be a $k$-periodic sequence such that $T_{\overline{a}^k} = P_A$, then $T_{\overline{a}^{nk}} = T_{\overline{a}^k}$ for all $n \in \mathbb{N}^\star$, and $T_{\overline{a}^n} \underset{n \rightarrow \infty}{\rightarrow} P_A$. We can use Lemma~\ref{lemma:partvaluesA} and consider every $k$-th term in the limit:
\begin{align}
    &\overline{H}\Big(\textstyle\bigsqcup_{a \in \mathcal{A}} G_a, P_X\Big) \nonumber\\
    & = \lim\limits_{n \rightarrow \infty} \frac{1}{kn}H_\chi\Big(\textstyle\bigwedge_{a \in \mathcal{A}} G_a^{\wedge kn T_{\overline{a}^{kn}} (a)}, \bigotimes_{a\in \mc{A}}P_{X_{a}}^{\otimes kn T_{\overline{a}^{kn}}(a)}  \Big) \\
    & = \lim\limits_{n \rightarrow \infty} \frac{1}{kn}H_\chi\Big(\Big(\textstyle\bigwedge_{a \in \mathcal{A}} G_a^{\wedge k T_{\overline{a}^{k}} (a)}, \bigotimes_{a\in \mc{A}}P_{X_{a}}^{\otimes k T_{\overline{a}^k}(a)} \Big)^{\wedge n}\Big) \\
    & = \frac{1}{k} \overline{H}\Big(\textstyle\bigwedge_{a \in \mathcal{A}} G_a^{\wedge k P_A (a)}, \bigotimes_{a\in \mc{A}}P_{X_{a}}^{\otimes k T_{\overline{a}^k}(a)} \Big). &
\end{align}

%
%
\subsection{Proof of Lemma~\ref{lemma:distrib}} \label{section:prooflemmadistrib}

    The probabilistic graphs in both sides of \eqref{eq:distribunion} have 
    \begin{align}
        \left(\textstyle\bigsqcup_{a \in \mathcal{A}} \mathcal{X}_a\right) \times \left(\textstyle\bigsqcup_{b \in \mathcal{B}} \mathcal{X}_b\right) = \textstyle\bigsqcup_{(a,b) \in \mathcal{A}\times \mathcal{B}} \mathcal{X}_a \times \mathcal{X}_b
    \end{align}
    as set of vertices, with underlying distribution 
    \begin{align}
        & \left(\textstyle\sum_{a \in \mathcal{A}} P_A(a) P_{X_a}\right)\left(\textstyle\sum_{b \in \mathcal{B}} P_B(b) P_{X_b}\right) \nonumber\\
        & = \textstyle\sum_{(a,b) \in \mathcal{A}\times \mathcal{B}} P_A(a) P_B(b) P_{X_a} P_{X_b}.
    \end{align}
    
    Now let us show that these two graphs have the same edges. Let $(x_{\mathcal{A}},x_{\mathcal{B}}),(x'_{\mathcal{A}},x'_{\mathcal{B}}) \in \left(\bigsqcup_{a \in \mathcal{A}} \mathcal{X}_a\right) \times \left(\bigsqcup_{b \in \mathcal{B}} \mathcal{X}_b\right)$, let $a_*, a'_* \in \mathcal{A}$ and $b_*,b'_* \in \mathcal{B}$ be the unique indexes such that
    \begin{align}
        (x_{\mathcal{A}},x_{\mathcal{B}}) \in \mathcal{X}_{a_*} \times \mathcal{X}_{b_*} \quad \text{and} \quad (x'_{\mathcal{A}},x'_{\mathcal{B}}) \in \mathcal{X}_{a'_*} \times \mathcal{X}_{b'_*}.
    \end{align}
    We have:
    \begin{align}
        &  \!\!\! (x_{\mathcal{A}},x_{\mathcal{B}}),(x'_{\mathcal{A}},x'_{\mathcal{B}}) \text{ adjacent in } \left(\textstyle\bigsqcup_{a \in \mathcal{A}} G_a\right) \wedge \left(\textstyle\bigsqcup_{b \in \mathcal{B}} G_b\right)\\
        \Longleftrightarrow & \; x_{\mathcal{A}}, x'_{\mathcal{A}} \text{ adjacent in } \textstyle\bigsqcup_{a \in \mathcal{A}} G_a\nonumber\\
        &\text{ and } x_{\mathcal{B}}, x'_{\mathcal{B}} \text{ adjacent in } \textstyle\bigsqcup_{b \in \mathcal{B}} G_b\\
        \Longleftrightarrow & \; a_* = a'_* \text{ and } x_{\mathcal{A}} x'_{\mathcal{A}} \in \mathcal{E}_{a_*} \text{ and } b_* = b'_* \text{ and } x_{\mathcal{B}} x'_{\mathcal{B}} \in \mathcal{E}_{b_*} \\
        \Longleftrightarrow & \; (a_*,b_*) = (a'_*, b'_*) \text{ and } \nonumber\\
        &(x_{\mathcal{A}},x_{\mathcal{B}}),(x'_{\mathcal{A}},x'_{\mathcal{B}}) \text{ are adjacent in } G_{a_*} \wedge G_{b_*} \\
        \Longleftrightarrow & \; (x_{\mathcal{A}},x_{\mathcal{B}}),(x'_{\mathcal{A}},x'_{\mathcal{B}}) \text{ are adjacent in } \textstyle\bigsqcup_{(a,b) \in \mathcal{A} \times \mathcal{B}} G_a \wedge G_b.
    \end{align}

\subsection{Proof of Lemma~\ref{lemma:simplificationinduced}} \label{section:prooflemmasimplificationinduced}

In order to prove Lemma~\ref{lemma:simplificationinduced}, we need Lemma~\ref{lemma:induced} which gives upper and lower bounds on the chromatic entropy of an induced subgraph $G[\mathcal{S}]$, using the chromatic entropy of the whole graph $G$ and the probability $P_X(\mathcal{S})$. The core idea is that if $P_X(\mathcal{S})$ is close to $1$ and $H_\chi(G)$ is large, then $H_\chi(G[\mathcal{S}])$ is close to $H_\chi(G)$. The proof of Lemma~\ref{lemma:induced} is given in App.~\ref{section:prooflemmainduced}

\begin{lemma}\label{lemma:induced}
        Let a probabilistic graph $(G , P_X)$ and $\mathcal{S} \subseteq \mathcal{X}$, then 
        \begin{align}
           & H_\chi(G, P_X) - 1 - (1-P_X(\mathcal{S}))\log|\mathcal{X}| \nonumber\\
        &\leq H_\chi(G[\mathcal{S}], P_X/P_X(\mathcal{S})) \leq\frac{H_\chi(G, P_X)}{P_X(\mathcal{S})}.
        \end{align}
\end{lemma}

\begin{remark}
    $H_\chi(G[\mathcal{S}], P_X/P_X(\mathcal{S}))$ can be greater than $H_\chi(G, P_X)$, even if $G[\mathcal{S}]$ has less vertices and inherits the structure of $G$. This stems from the normalized distribution $P_X/P_X(\mathcal{S})$ on the vertices of $G[\mathcal{S}]$ which gives more weight to the vertices in $\mathcal{S}$. For example, consider
\begin{align}
G = N_5\sqcup 
 K_5,\nonumber
\end{align}
 where $K_{n}$ (resp. $N_n$) is the complete (resp. empty) graph with $n$ vertices, i.e. there is an edge (resp. no edge) between any pair of distinct vertices, and with $\mathcal{S}$ being the vertices in the connected component $K_5$ in $G$. Then $H_\chi(G,\Unif(\lbrace 1, ..., 5\rbrace)) = \varepsilon \log 5$ and $H_\chi(G[\mathcal{S}],\Unif(\lbrace 1, ..., 5\rbrace)/\varepsilon) = \log 5$.
\end{remark}

Now let us prove Lemma~\ref{lemma:simplificationinduced}. By Lemma~\ref{lemma:induced}, we have for all $n \in \mathbb{N}^\star$:
\begin{align}
    & H_\chi(G^{\wedge n},P_X^n) - 1 - (1-P^{\otimes n}_X(\mathcal{S}_n))\log|\mathcal{X}| \nonumber\\
    \leq \: & H_\chi(G^{\wedge n}[\mathcal{S}^n],P_X^n/P_X^n(\mathcal{S}_n)) \leq \frac{H_\chi(G^{\wedge n},P_X^n)}{P^{\otimes n}_X(\mathcal{S}_n)}.
\end{align}
Since $P^{\otimes n}_X(\mathcal{S}_n) \rightarrow 1$, and $H_\chi(G^{\wedge n},P_X^n) = n \overline{H}(G,P_X) + o(n)$ when $n \rightarrow \infty$, the desired results follows immediately by normalization and limit.

\subsection{Proof of Lemma~\ref{lemma:unionisomorph}}\label{setion:prooflemmaunionisomorph}

    Let $(\tilde{G}_i,P_{i})_{i \leq N}$ be isomorphic probabilistic graphs and $G$ such that $G = \bigsqcup_i \tilde{G}_i$. Let $c_1^{\star} : \mathcal{X}_1 \rightarrow \mathcal{C}$ be the coloring of $\tilde{G}_1$ with minimal entropy, and let $c^{\star}$ be the coloring of $G$ defined by
    \begin{align}
        c^{\star} : \: & \mathcal{X} \rightarrow \mathcal{C} \\
        & x \mapsto c_1^{\star} \circ \psi_{i_x \rightarrow 1}(x),
    \end{align}
    where $i_x$ is the unique integer such that $x \in \mathcal{X}_{i_x}$, and $\psi_{i_x \rightarrow 1} : \mathcal{X}_{i_x} \rightarrow \mathcal{X}_1$ is an isomorphism between $\tilde{G}_{i_x}$ and $\tilde{G}_1$. In other words $c^{\star}$ applies the same coloring pattern $c^{\star}_1$ on each connected component of $G$. We have
    \begin{align}
    H_\chi(G,P_X) & \leq H(c^{\star}(X)) \label{eq:proofA01}\\
    & = h\Big(\textstyle\sum_{j \leq N} P_{i_X}(j) P_{c^{\star}(X_j)}\Big) \\
    & = h\Big(\textstyle\sum_{j \leq N} P_{i_X}(j) P_{c_1^{\star}(X_1)}\Big) \label{eq:proofA03}\\
    & = H(c^{\star}_1(X_1)) \\
    & = H_\chi(\tilde{G}_1),\label{eq:proofA05}
    \end{align}
    where $h$ denotes the entropy of a distribution; \eqref{eq:proofA03} comes from the definition of $c^{\star}$; and \eqref{eq:proofA05} comes from the definition of $c_1^{\star}$.
    
    Now let us prove the upper bound on $H_\chi(\tilde{G}_1)$. Let $c$ be a coloring of $G$, and let $i^{\star} \doteq \argmin_i H(c(X_i))$ (i.e. $i^{\star}$ is the index of the connected component for which the entropy of the coloring induced by $c$ is minimal). We have
    \begin{align}
    H(c(X)) & = h\Big(\textstyle\sum_{j \leq N} P_{i_X}(j) P_{c(X_j)}\Big) \label{eq:proofA10} \\
    & \geq \textstyle\sum_{j \leq N} P_{i_X}(j) h(P_{c(X_j)}) \label{eq:proofA11} \\
    & \geq \textstyle\sum_{j \leq N} P_{i_X}(j) H(c(X_{i^{\star}})) \label{eq:proofA12} \\
    & \geq H_\chi(\tilde{G}_{i^{\star}}),\label{eq:proofA13} \\
    & = H_\chi(\tilde{G}_1,P_{X_1}),\label{eq:proofA14}
    \end{align}
    where \eqref{eq:proofA11} follows from the concavity of $h$; \eqref{eq:proofA12} follows from the definition of $i^{\star}$; \eqref{eq:proofA13} comes from the fact that $c$ induces a coloring of $\tilde{G}_{i^{\star}}$; \eqref{eq:proofA14} comes from the fact that $\tilde{G}_1$ and $\tilde{G}_{i^{\star}}$ are isomorphic. Now, we can combine the bounds \eqref{eq:proofA05} and \eqref{eq:proofA14}: for all coloring $c$ of $G$ we have
    \begin{align}
        H_\chi(G,P_{X}) \leq H_\chi(\tilde{G}_1,P_{X_1}) \leq H(c(X)),
    \end{align}
    which yields the desired equality when taking the infimum over $c$.

\subsection{Proof of Lemma~\ref{lemma:induced}}\label{section:prooflemmainduced}

Let $c^{\star} : \mathcal{X} \rightarrow \mathcal{C}$ and $c^{\star}_\mathcal{S} : \mathcal{S} \rightarrow \mathcal{C}$ be the optimal colorings of $G$ and $G[\mathcal{S}]$, respectively. Consider the coloring $c : \mathcal{X} \rightarrow \mathcal{C} \sqcup \mathcal{X}$ of $G$ defined by $c(x) = c^{\star}_\mathcal{S}$ if $x \in \mathcal{S}$, $c(x) = x$ otherwise.\newline

\textbf{(Lower bound)} On one hand, we have
\begin{align}
    &H_\chi(G,P_{X}) \nonumber\\
    \leq &\: H(c(X), \mathds{1}_{X \in \mathcal{S}}) \label{eq:prooflemmainducedA1}\\
    = &\: H(\mathds{1}_{X \in \mathcal{S}}) + P_X(\mathcal{S}) H(c(X)|X \in \mathcal{S}) \nonumber\\
     &+ (1-P_X(\mathcal{S})) H(c(X)|X \notin \mathcal{S}) \label{eq:prooflemmainducedA2}\\
    \leq &\: 1 + H(c^{\star}_\mathcal{S}(X)|X \in \mathcal{S}) + (1-P_X(\mathcal{S})) \log |\mathcal{X}| \label{eq:prooflemmainducedA3}\\
    = &\: H_\chi(G[\mathcal{S}],P_{X}/P_X(\mathcal{S})) + 1 + (1-P_X(\mathcal{S})) \log |\mathcal{X}|;\label{eq:prooflemmainducedA4}
\end{align}
where \eqref{eq:prooflemmainducedA1} comes from the fact that $c$ is a coloring of $G$; \eqref{eq:prooflemmainducedA2} is a decomposition using conditional entropies; \eqref{eq:prooflemmainducedA3} comes from the construction of $c$: $c|_\mathcal{S} = c^{\star}_{\mathcal{S}}$; \eqref{eq:prooflemmainducedA4} follows from the optimality of $c^{\star}_\mathcal{S}$ as a coloring of $G[\mathcal{S}]$.\newline

\textbf{(Upper bound)} On the other hand, 
\begin{align}
&H_\chi(G[\mathcal{S}],P_{X}/P_X(\mathcal{S}))\nonumber\\
    \leq & \: H(c^{\star}(X) | X \in \mathcal{S}) \label{eq:prooflemmainducedB1} \\
    = & \, \frac{1}{P_X(\mathcal{S})}\Big(\!H(c^{\star}(X)| \mathds{1}_{X \in \mathcal{S}})\nonumber\\
    & - (1-P_X(\mathcal{S}))H(c^{\star}(X) | X \notin \mathcal{S})\!\Big) \label{eq:prooflemmainducedB2} \\
    \leq & \: \frac{H(c^{\star}(X))}{P_X(\mathcal{S})} = \frac{H_\chi(G,P_X)}{P_X(\mathcal{S})} \label{eq:prooflemmainducedB3} 
\end{align}
where \eqref{eq:prooflemmainducedB1} comes from the fact that $c^{\star}$ induces a coloring of $G[\mathcal{S}]$; \eqref{eq:prooflemmainducedB2} is a decomposition using conditional entropies; \eqref{eq:prooflemmainducedB3} results from the elimination of negative terms and the optimality of $c^{\star}$.

\subsection{Proof of Lemma~\ref{th:etaconvex}}\label{section:proofetaconvex}

In order to prove Lemma~\ref{th:etaconvex}, we need Lemma~\ref{lemma:partvaluesA} which can be found in 
App.~\ref{section:prooflemmapartvalues}; and Lemma~\ref{lemma:typesplit}, which is a generalization for infinite sequences of the following observation: if $T_{\overline{a}^n} = P_A \in \Delta_n(\mathcal{A})$ satisfies $P_A = \frac{i}{n}P'_A + \frac{n-i}{n}P''_A$ with $P'_A \in \Delta_i(\mathcal{A})$ and $P''_A \in \Delta_{n-i}(\mathcal{A})$, then $\overline{a}^n$ can be separated into two subsequences $a'^i$ and $a''^{n-i}$ such that $T_{a'^i} = P'_A$ and $T_{a''^{n-i}} = P''_A$. 

\begin{lemma}[Type-splitting lemma]\label{lemma:typesplit}
    Let $(\overline{a}_n)_{n \in \mathbb{N}^\star} \in \mathcal{A}^{\mathbb{N}^\star}$ be a sequence such that $T_{\overline{a}^n} \rightarrow P_A \in \Delta(\mathcal{A})$ when $n \rightarrow \infty$, let $\beta \in (0,1)$ and $P'_A, P''_A \in \Delta(\mathcal{A})$ such that
    \begin{align}
        P_A = \beta P'_A + (1-\beta)P''_A.
    \end{align}
    Then there exists a sequence $(b_n)_{n \in \mathbb{N}^\star} \in \lbrace 0, 1 \rbrace^{\mathbb{N}^\star}$ such that the two extracted sequences $a' \doteq (\overline{a}_n)_{\substack{n\in \mathbb{N}^\star, \\ b_n = 0}}$ and $a'' \doteq (\overline{a}_n)_{\substack{n\in \mathbb{N}^\star, \\ b_n = 1}}$ satisfy
    \begin{align}
        & T_{b^n} \underset{n \rightarrow \infty}{\rightarrow} (\beta, 1-\beta), & & \\
        & T_{a'^n} \underset{n \rightarrow \infty}{\rightarrow} P'_A, & & T_{a''^n} \underset{n \rightarrow \infty}{\rightarrow} P''_A.
    \end{align}
\end{lemma}

The proof of Lemma~\ref{lemma:typesplit} is given in App.~\ref{section:prooflemmatypesplit}. Now let us prove Lemma~\ref{th:etaconvex}. We recall the definition of the function
\begin{align}
    \eta : P_A \mapsto \overline{H}\Bigg(\bigsqcup_{a \in \mathcal{A}} G_a,  \sum_{a \in \mathcal{A}} P_A(a) P_{X_a} \Bigg).
\end{align}

\textbf{($\eta$ Lipschitz)} Let us first prove that the function $\eta$ is Lipschitz. For all $P_A$, $P'_A \in \Delta(\mathcal{A})$ we need to bound the quantity $|\eta(P_A)-\eta(P'_A)|$; by Lemma~\ref{lemma:partvaluesA} this is equivalent to bounding
\begin{align}
     \lim_{n \rightarrow \infty} \frac{1}{n}\bigg|&H_\chi\left(\textstyle\bigwedge_{a \in \mathcal{A}} G_a^{\wedge n T_{\overline{a}^n} (a)},\bigotimes_{a \in \mathcal{A}} P_{X_a}^{n T_{\overline{a}^n} (a)} \right) \nonumber\\
     &- H_\chi\left(\textstyle\bigwedge_{a \in \mathcal{A}} G_a^{\wedge n T_{\overline{a}'^n} (a)},\bigotimes_{a \in \mathcal{A}} P_{X_a}^{n T_{\overline{a}'^n} (a)}   \right)\bigg|\label{eq:prooflemmastaconvexA0}
    \end{align}
where $(T_{\overline{a}^n}, T_{\overline{a}'^n}) \rightarrow (P_A, P'_A)$ when $n \rightarrow \infty$.

Fix $n \in \mathbb{N}^\star$, we assume that the quantity inside $|\cdot|$ in \eqref{eq:prooflemmastaconvexA0} is positive; the other case can be treated with the same arguments by symmetry of the roles. We have
\begin{align}
    & H_\chi\bigg(\textstyle\bigwedge_{a \in \mathcal{A}} G_a^{\wedge n T_{\overline{a}^n} (a)},\bigotimes_{a \in \mathcal{A}} P_{X_a}^{n T_{\overline{a}^n} (a)} \bigg)\nonumber\\
& - H_\chi\bigg(\textstyle\bigwedge_{a \in \mathcal{A}} G_a^{\wedge n T_{\overline{a}'^n} (a)},\bigotimes_{a \in \mathcal{A}} P_{X_a}^{n T_{\overline{a}'^n} (a)}  \bigg) \label{eq:Th3A}\\
    \leq & \; H_\chi\bigg(\textstyle\bigwedge_{a \in \mathcal{A}} G_a^{\wedge n T_{\overline{a}^n} (a)},\bigotimes_{a \in \mathcal{A}} P_{X_a}^{n T_{\overline{a}^n} (a)} \bigg) \nonumber\\
    &- H_\chi\bigg(\textstyle\bigwedge_{a \in \mathcal{A}} G_a^{\wedge n \min(T_{\overline{a}^n}(a),T_{\overline{a}'^n}(a))} , \nonumber\\
&\qquad \qquad \bigotimes_{a \in \mathcal{A}} P_{X_a}^{n \min(T_{\overline{a}^n}(a),T_{\overline{a}'^n}(a))}   \bigg) \label{eq:Th3B}\\
    = & \; H_\chi\bigg(\textstyle\bigwedge_{a \in \mathcal{A}} G_a^{\wedge n \min(T_{\overline{a}^n}(a),T_{\overline{a}'^n}(a))}\nonumber\\
&\qquad \qquad \textstyle\bigwedge_{a \in \mathcal{A}} G_a^{\wedge n |T_{\overline{a}^n}(a)-T_{\overline{a}'^n}(a)|_+},  \bigotimes_{a \in \mathcal{A}} P_{X_a}^{n T_{\overline{a}^n} (a)}  \bigg) \nonumber\\
    & - H_\chi\bigg(\textstyle\bigwedge_{a \in \mathcal{A}} G_a^{\wedge n \min(T_{\overline{a}^n}(a),T_{\overline{a}'^n}(a))} ,\nonumber\\
&\qquad \qquad \bigotimes_{a \in \mathcal{A}} P_{X_a}^{n \min(T_{\overline{a}^n}(a),T_{\overline{a}'^n}(a))}  \bigg) \label{eq:Th3C}\\
    \leq & \; H_\chi\bigg(\textstyle\bigwedge_{a \in \mathcal{A}} G_a^{\wedge n |T_{\overline{a}^n}(a)-T_{\overline{a}'^n}(a)|_+},\nonumber\\
&\qquad \qquad  \bigotimes_{a \in \mathcal{A}} P_{X_a}^{n  |T_{\overline{a}^n}(a)-T_{\overline{a}'^n}(a)|_+}   \bigg) \label{eq:Th3D}\\
    \leq & \; \log\bigg(\max_a |\mathcal{X}_a|\bigg) \textstyle\sum_{a \in \mathcal{A}} n |T_{\overline{a}^n}(a)-T_{\overline{a}'^n}(a)|_+ \label{eq:Th3E}\\
    \leq & \; n \log\bigg(\max_a |\mathcal{X}_a|\bigg) \|T_{\overline{a}^n}-T_{\overline{a}'^n}\|_1, \label{eq:Th3F}
\end{align}
where $|\cdot|_+ = \max(\cdot,0)$ and $\|T_{\overline{a}^n}-T_{\overline{a}'^n}\|_1 = \sum_{a \in \mathcal{A}} |T_{\overline{a}^n}(a)-T_{\overline{a}'^n}(a)|$; \eqref{eq:Th3B} follows from the removal of terms in the second product, as $H_\chi(G\wedge G',P_X\otimes P'_X) \geq H_\chi(G,P_X)$ for all probabilistic graphs $(G,P_X)$ and $(G',P'_X)$; \eqref{eq:Th3C} is an arrangement of the terms in the first product, as $\min(s,t) + \max(s-t,0) = s$ for all real numbers $s,t$; \eqref{eq:Th3D} comes from the subadditivity of $H_\chi$; \eqref{eq:Th3E} follows from $H_\chi(G_a,P_{X_a}) \leq \log \max_{a'} |\mathcal{X}_{a'}|$ for all $a \in \mathcal{A}$; \eqref{eq:Th3F} results from $|T_{\overline{a}^n}(a)-T_{\overline{a}'^n}(a)|_+ \leq |T_{\overline{a}^n}(a)-T_{\overline{a}'^n}(a)|$ for all $a\in \mathcal{A}$.

By normalization and limit, it follows that
\begin{align}
    |\eta(P_A)-\eta(P'_A)| & \leq \lim_{n \rightarrow \infty} \log\left(\max_a |\mathcal{X}_a|\right)\cdot\|T_{\overline{a}^n}-T_{\overline{a}'^n}\|_1 \\
    & = \log\left(\max_a |\mathcal{X}_a|\right)\cdot\|P_A - P'_A\|_1.\label{eq:Th3G}
\end{align}
Hence $\eta$ is $(\log \max_a |\mathcal{X}_a|)$-Lipschitz.\newline

\textbf{($\eta$ convex)} Let us now prove that $\eta$ is convex. Let $P'_A, P''_A \in \Delta(\mathcal{A})$, and $\beta \in (0,1)$, we have by Lemma~\ref{lemma:partvaluesA}
\begin{align}
&\eta\big(\beta P'_A + (1-\beta) P''_A\big) \nonumber\\
=& \lim_{n \rightarrow \infty} \frac{1}{n} H_\chi\left(\textstyle\bigwedge_{a \in \mathcal{A}} G_a^{\wedge n T_{\overline{a}^n} (a)}, \bigotimes_{a \in \mathcal{A}} P_{X_a}^{n T_{\overline{a}^n} (a)}  \right),\label{eq:Th3H}
\end{align}

where $T_{\overline{a}^n} \rightarrow \beta P'_A + (1-\beta) P''_A$ when $n \rightarrow \infty$. By Lemma~\ref{lemma:typesplit}, there exists $(b_n)_{n \in \mathbb{N}^\star} \in \lbrace 0,1\rbrace^{\mathbb{N}^\star}$ such that the decomposition of $(\overline{a}_n)_{n \in \mathbb{N}^\star}$ into two subsequences $a' \doteq (\overline{a}_n)_{\substack{n\in \mathbb{N}^\star, \\ b_n = 0}}$ and $a'' \doteq (\overline{a}_n)_{\substack{n\in \mathbb{N}^\star, \\ b_n = 1}}$ satisfies
    \begin{align}
        & T_{b^n} \underset{n \rightarrow \infty}{\rightarrow} (\beta,1-\beta), & &  \label{eq:Th3J}\\
        & T_{a'^n} \underset{n \rightarrow \infty}{\rightarrow} P'_A, & & T_{a''^n} \underset{n \rightarrow \infty}{\rightarrow} P''_A.\label{eq:Th3K}
    \end{align}

For all $n \in \mathbb{N}^\star$, let $\Xi(n) \doteq n T_{b^n}(0)$, we have
\begin{align}
    & \eta\big(\beta P'_A + (1-\beta) P''_A\big) \\
    = & \; \lim_{n \rightarrow \infty} \frac{1}{n} H_\chi\bigg(\textstyle\bigwedge_{a \in \mathcal{A}} G_a^{\wedge \Xi(n) T_{a'^{\Xi(n)}}(a) + (n-\Xi(n)) T_{a''^{n-\Xi(n)}}(a) } , \nonumber\\
&\qquad \quad \bigotimes_{a \in \mathcal{A}} P_{X_a}^{n  \Xi(n) T_{a'^{\Xi(n)}}(a) + (n-\Xi(n)) T_{a''^{n-\Xi(n)}}(a)  }  \bigg)\label{eq:Th3L} \\
    \leq & \; \lim_{n \rightarrow \infty} \frac{\Xi(n)}{n} \frac{1}{\Xi(n)} H_\chi\bigg(\textstyle\bigwedge_{a \in \mathcal{A}} G_a^{\wedge \Xi(n) T_{a'^{\Xi(n)}}(a)},\nonumber\\
&\qquad \quad \bigotimes_{a \in \mathcal{A}} P_{X_a}^{\Xi(n) T_{a'^{\Xi(n)}}(a)} \bigg) \\
    & + \frac{n-\Xi(n)}{n} \frac{1}{n-\Xi(n)}  H_\chi\bigg(\textstyle\bigwedge_{a \in \mathcal{A}} G_a^{\wedge (n-\Xi(n)) T_{a''^{n-\Xi(n)}}(a)} , \nonumber\\
&\qquad \quad \bigotimes_{a \in \mathcal{A}} P_{X_a}^{(n-\Xi(n)) T_{a''^{n-\Xi(n)}}(a)}  \bigg) \label{eq:Th3M}\\
    = & \; \beta \eta(P'_A) + (1-\beta)\eta(P''_A); \label{eq:Th3O}
\end{align}
where \eqref{eq:Th3L} comes from \eqref{eq:Th3H}; \eqref{eq:Th3M} follows from the subadditivity of $H_\chi$; \eqref{eq:Th3O} comes from \eqref{eq:Th3J}, \eqref{eq:Th3K} and Lemma~\ref{lemma:partvaluesA}. Since \eqref{eq:Th3O} holds for all $P'_A, P''_A \in \Delta(\mathcal{A})$ and $\beta \in (0,1)$, we have that $\eta$ is convex.


\subsection{Proof of Lemma~\ref{lemma:typesplit}}\label{section:prooflemmatypesplit}

Let $(\overline{a}_n)_{n \in \mathbb{N}^\star} \in \mathcal{A}^{\mathbb{N}^\star}$ be a sequence such that $T_{\overline{a}^n} \rightarrow P_A = \beta P'_A + (1-\beta)P''_A$ when $n \rightarrow \infty$.

Consider a sequence $(B_n)_{n \in \mathbb{N}^\star}$ of independent Bernoulli random variables such that for all $n \in \mathbb{N}^\star$,
\begin{align}
    \mathbb{P}(B_n = 0) = \frac{\beta P'_A(\overline{a}_n)}{P_A(\overline{a}_n)}.
\end{align}
By the strong law of large numbers,
\begin{align}
    \mathbb{P}\left(T_{B^n,\overline{a}^n} \underset{n \rightarrow \infty}{\rightarrow} (\beta P'_A, (1-\beta)P''_A) \right) = 1.
\end{align}
Therefore, there exists at least one realization $(b_n)_{n \in \mathbb{N}^\star}$ of $(B_n)_{n \in \mathbb{N}^\star}$ such that $T_{b^n,\overline{a}^n}$ converges to $\big(\beta P'_A, (1-\beta)P''_A \big)$. The convergences of marginal and conditional types yield
    \begin{align}
        & T_{b^n} \underset{n \rightarrow \infty}{\rightarrow} (\beta, 1-\beta),  \\
        & T_{a'^n} \underset{n \rightarrow \infty}{\rightarrow} P'_A, \qquad T_{a''^n} \underset{n \rightarrow \infty}{\rightarrow} P''_A,
    \end{align}
where $a' \doteq (\overline{a}_n)_{\substack{n\in \mathbb{N}^\star, \\ b_n = 0}}$ and $a'' \doteq (\overline{a}_n)_{\substack{n\in \mathbb{N}^\star, \\ b_n = 1}}$ are the extracted sequences.

\subsection{Proof of Lemma~\ref{lemma:convexanalysis2}}\label{section:prooflemmaconvexanalysis}

It can be easily observed that
\begin{align}
    & \exists P_A \in \interior(\Delta(\mathcal{A})), \, \gamma(P_A) = \textstyle\sum_{a \in \mathcal{A}} P_A(a) \gamma(\mathds{1}_a) \label{eq:prooflemmaconvexanA1}\\
    \Longleftarrow \;\; & \forall P_A \in \Delta(\mathcal{A}), \, \gamma(P_A) = \textstyle\sum_{a \in \mathcal{A}} P_A(a) \gamma(\mathds{1}_a). \label{eq:prooflemmaconvexanA2}
\end{align}
Now let us prove \eqref{eq:prooflemmaconvexanA1} $\Rightarrow$ \eqref{eq:prooflemmaconvexanA2}. Let $P^{\star}_A \in \interior\Delta(\mathcal{A})$ such that $\gamma(P^{\star}_A) = \sum_{a \in \mathcal{A}} P^{\star}_A(a) \gamma(\mathds{1}_a)$. Let $m : \Delta(\mathcal{A}) \rightarrow \mathbb{R}$ linear such that $m(P^{\star}_A) = \gamma(P^{\star}_A)$ and $\forall P_A \in \Delta(\mathcal{A}),\, m(P_A) \leq \gamma(P_A)$. We have
\begin{align}
    0 = \gamma(P^{\star}_A) - m(P^{\star}_A) = \textstyle\sum_{a \in \mathcal{A}} P^{\star}_A(a) \big(\gamma(\mathds{1}_a) - m(\mathds{1}_a)\big);
\end{align}
and therefore $\gamma(\mathds{1}_a) = m(\mathds{1}_a)$ for all $a \in \mathcal{A}$, as $\gamma - m \geq 0$ and $P^{\star}_A(a) > 0$ for all $a \in \mathcal{A}$. For all $P_A \in \Delta(\mathcal{A})$, we have
\begin{align}
    f(P_A) & \leq \textstyle\sum_{a \in \mathcal{A}} P_A(a) \gamma(\mathds{1}_a) \\
    & = \textstyle\sum_{a \in \mathcal{A}} P_A(a) m(\mathds{1}_a) = m(P_A),
\end{align}
hence $\gamma = m$ and $\gamma$ is linear.

\section{Proof of Proposition~\ref{prop:marton}}\label{section:proofpropmarton}

In order to prove Proposition~\ref{prop:marton}, we need Lemma~\ref{lemma:martonunion}, which is a consequence of Marton's formula in Theorem~\ref{th:marton} applied to a disjoint union. The proof of Lemma~\ref{lemma:martonunion} can be found in App.~\ref{section:proofmartonunion}.

\begin{lemma}\label{lemma:martonunion}
    Let $P_A \in \Delta(\mathcal{A})$, then
    \begin{align}
&\overline{H}\left(\bigsqcup_{a \in \mathcal{A}} G_a , \sum_{a \in \mathcal{A}} P_A(a) P_{X_a} \right) \nonumber\\
&+ C\left(\bigsqcup_{a \in \mathcal{A}} G_a,\; \sum_{a \in \mathcal{A}} P_A(a) P_{X_a} \right) \nonumber\\
&= H(P_A) + \sum_{a \in \mathcal{A}} P_A(a) H(P_{X_a}).
    \end{align}
    \;
\end{lemma}

Let us prove Proposition~\ref{prop:marton}. We have on one hand:
\begin{align}
    & H(P_A) + \textstyle\sum_{a \in \mathcal{A}} P_A(a) C(G_a, P_{X_a}) \\
    =& = H(P_A) - \textstyle\sum_{a \in \mathcal{A}} P_A(a)\overline{H}(G_a, P_{X_a})\nonumber\\
& +  \textstyle\sum_{a \in \mathcal{A}}P_A(a) H(P_{X_a}) \label{eq:proofpropmartonA1}\\
     \leq& H(P_A) - \overline{H}\left(\textstyle\bigsqcup_{a \in \mathcal{A}} G_a ,\; \sum_{a \in \mathcal{A}} P_A(a) P_{X_a}\right) \nonumber\\
&+ \textstyle\sum_{a \in \mathcal{A}} P_A(a) H(P_{X_a}) \label{eq:proofpropmartonA2}\\
     =& C\left(\textstyle\bigsqcup_{a \in \mathcal{A}} G_a,\; \sum_{a \in \mathcal{A}} P_A(a) P_{X_a} \right); \label{eq:proofpropmartonA3}
\end{align}
where \eqref{eq:proofpropmartonA1} comes from Theorem~\ref{th:marton}; \eqref{eq:proofpropmartonA2} follows from \eqref{eq:linUnionTuncel}, see \cite[Theorem 2]{tuncel2009complementary}; and \eqref{eq:proofpropmartonA3} follows from Lemma~\ref{lemma:martonunion}. Therefore, 
\begin{align}
    & C\left(\textstyle\bigsqcup_{a \in \mathcal{A}} G_a,\; \textstyle\sum_{a \in \mathcal{A}} P_A(a) P_{X_a} \right)\nonumber\\
& = H(P_A) + \textstyle\sum_{a \in \mathcal{A}} P_A(a) C(G_a, P_{X_a}) \\
    \Longleftrightarrow \; & \overline{H}\left(\textstyle\bigsqcup_{a \in \mathcal{A}} G_a ,\; \sum_{a \in \mathcal{A}} P_A(a) P_{X_a}\right) \nonumber\\
&= \textstyle\sum_{a \in \mathcal{A}} P_A(a)\overline{H}(G_a,P_{X_a}).
\end{align}

On the other hand:
\begin{align}
    &\textstyle\sum_{a \in \mathcal{A}} C(G_a, P_{X_a}) \nonumber\\
    & = -\textstyle\sum_{a \in \mathcal{A}}  \overline{H}(G_a,P_{X_a}) + H(P_{X_a}) \label{eq:proofpropmartonB1}\\
    & \leq -\overline{H}\left(\textstyle\bigwedge_{a \in \mathcal{A}} G_a,\; \textstyle\bigotimes_{a \in \mathcal{A}} P_{X_a}  \right) + \textstyle\sum_{a \in \mathcal{A}} H(P_{X_a}) \label{eq:proofpropmartonB2}\\
    & = -\overline{H}\left(\textstyle\bigwedge_{a \in \mathcal{A}} G_a,\; \textstyle\bigotimes_{a \in \mathcal{A}} P_{X_a}  \right) + H\left(\textstyle\bigotimes_{a \in \mathcal{A}} P_{X_a}\right) \label{eq:proofpropmartonB3}\\
    & = C\left(\textstyle\bigwedge_{a \in \mathcal{A}} G_a,\; \textstyle\bigotimes_{a \in \mathcal{A}} P_{X_a} \right); \label{eq:proofpropmartonB4}
\end{align}
where \eqref{eq:proofpropmartonB1} comes from Theorem~\ref{th:marton}; \eqref{eq:proofpropmartonB2} follows from \eqref{eq:linprodTuncel}, see \cite[Theorem 2]{tuncel2009complementary}; and \eqref{eq:proofpropmartonB4} also follows from Theorem~\ref{th:marton}. Therefore, 
\begin{align}
    & \textstyle\sum_{a \in \mathcal{A}} C(G_a, P_{X_a}) = C\left(\textstyle\bigwedge_{a \in \mathcal{A}} G_a,\; \bigotimes_{a \in \mathcal{A}} P_{X_a} \right) \\
    \Longleftrightarrow \; & \overline{H}\left(\textstyle\bigwedge_{a \in \mathcal{A}} G_a ,\; \textstyle\bigotimes_{a \in \mathcal{A}} P_{X_a}  \right) = \textstyle\sum_{a \in \mathcal{A}} \overline{H}(G_a, P_{X_a}).
\end{align}

\subsection{Proof of Lemma~\ref{lemma:martonunion}}\label{section:proofmartonunion}

The graph $\bigsqcup_{a \in \mathcal{A}} G_a$ has $\sum_{a \in \mathcal{A}} P_A(a) P_{X_a}$ as underlying distribution. Let $A,X$ be two random variables such that $A$ is drawn with $P_A$, and $X$ is drawn with $P_{X|A}(\cdot|a) \doteq P_{X_a}$, so that 
\begin{align}
   P_X = \textstyle\sum_{a \in \mathcal{A}} P_A(a) P_{X_a}.\label{eq:proofmartonunionA0}
\end{align}
We have
\begin{align}
    & \overline{H}\left(\textstyle\bigsqcup_{a \in \mathcal{A}} G_a , \textstyle\sum_{a \in \mathcal{A}} P_A(a) P_{X_a}\right) \nonumber\\
&+ C\left(\textstyle\bigsqcup_{a \in \mathcal{A}} G_a, \textstyle\sum_{a \in \mathcal{A}} P_A(a) P_{X_a} \right)  = H(X) \label{eq:proofmartonunionA2}\\
    =&  H(A,X) = H(A) + H(X|A)\label{eq:proofmartonunionA3}\\
    =& H(P_A) + \textstyle\sum_{a \in \mathcal{A}} P_A(a) H(P_{X_a});\label{eq:proofmartonunionA5}
\end{align}
where \eqref{eq:proofmartonunionA2} comes from Theorem~\ref{th:marton} and \eqref{eq:proofmartonunionA0}; and \eqref{eq:proofmartonunionA3} comes from the fact that $A$ can be written as a function of $X$: by definition, the vertex set of $\bigsqcup_{a \in \mathcal{A}} G_a$ \mlt{is given by} $\mathcal{X} = \bigsqcup_{a \in \mathcal{A}} \mathcal{X}_a$ and $\supp P_{X_a} \subseteq \mathcal{X}_a$, therefore $A$ is the unique index such that $X\in \mathcal{X}_A$.

\section{Proof of Proposition~\ref{prop:Pstar}}\label{section:proofpropPstar}

Let us show that for all graph $G = (\mathcal{X}, \mathcal{E})$, the function $P_X \mapsto C(G,P_X)$ is concave. Let $P_X, P'_X \in \Delta(\mathcal{X})$ and $\beta \in [0,1]$. Let $(b_n)_{n \in \mathbb{N}}$ be a sequence of integers such that $\frac{b_n}{n} \underset{n \rightarrow \infty}{\rightarrow} \beta$. 

By Lemma~\ref{prop:CGPbasic}, there exists two sequences $(\mathcal{C}_n)_{n \in \mathbb{N}}$ and $(\mathcal{C}'_n)_{n \in \mathbb{N}}$ that satisfy the following:
\begin{align}
    \forall n \in \mathbb{N}^\star, \; \mathcal{C}_n \subseteq \mathcal{X}^{n} \text{ and } \mathcal{C}'_n \subseteq \mathcal{X}^{n} \text{ are independent in } G^{\wedge n};
\end{align}
and
    \begin{align}
        & \frac{\log|\mathcal{C}_n|}{n} \underset{n \rightarrow \infty}{\rightarrow} C(G,P_X), & &  \frac{\log|\mathcal{C}'_n|}{n} \underset{n \rightarrow \infty}{\rightarrow} C(G,P'_X), &\\
        & \max_{x^n \in \mathcal{C}_n} \|T_{x^n} - P_X \|_\infty \underset{n \rightarrow \infty}{\rightarrow} 0, & & \max_{x^n \in \mathcal{C}'_n} \|T_{x^n} - P_X \|_\infty \underset{n \rightarrow \infty}{\rightarrow} 0. &
    \end{align}

Let us build a sequence of codebooks $(\mathcal{C}''_n)_{n \in \mathbb{N}^\star}$ adapted to the distribution $\beta P_X + (1-\beta) P'_X$ by using a time-sharing between $(\mathcal{C}_n)_{n \in \mathbb{N}^\star}$ and $(\mathcal{C}'_n)_{n \in \mathbb{N}^\star}$. For all $n \in \mathbb{N}^\star$, let 
\begin{align}
    \mathcal{C}''_n \doteq \mathcal{C}_n^{b_n} \times \mathcal{C}'^{n-b_n}_n.
\end{align}
For all $n \in \mathbb{N}^\star$, $\mathcal{C}''_n \subseteq \mathcal{X}^{n^2}$ is independent in $G^{\wedge n^2}$ as a product of independent sets.

The rate associated to $\mathcal{C}''_n$ \mlt{is given by}
\begin{align}
    \frac{\log|\mathcal{C}''_n|}{n^2} & = \frac{b_n \log|\mathcal{C}_n| + (n-b_n) \log|\mathcal{C}'_n|}{n^2} \\
    & = \frac{b_n}{n}\frac{\log|\mathcal{C}_n|}{n} + \frac{n-b_n}{n} \frac{\log|\mathcal{C}'_n|}{n} \\
    & \underset{n \rightarrow \infty}{\rightarrow} \beta C(G,P_X) + (1-\beta) C(G,P'_X);
\end{align}
and the types of the codewords in $\mathcal{C}''_n$ satisfy
\begin{align}
    & \max_{x^{n^2} \in \mathcal{C}''_n} \bigg\| T_{x^{n^2}} - \beta P_X - (1-\beta) P'_X \bigg\|_\infty \\
    = \: & \max_{x^{n b_n} \in \mathcal{C}_n} \max_{x'^{n(n - b_n)} \in \mathcal{C}'_n} \bigg\| \frac{n b_n}{n^2}T_{x^{n b_n}} + \frac{n(n- b_n)}{n^2}T_{x'^{n(n - b_n)}}\nonumber\\
& \qquad \quad - \beta P_X - (1-\beta) P'_X \bigg\|_\infty \\
    \leq \: & \max_{x^{n b_n} \in \mathcal{C}_n} \bigg\| \frac{b_n}{n}T_{x^{n b_n}} - \beta P_X\bigg\|_\infty \nonumber\\
&+ \max_{x'^{n(n - b_n)} \in \mathcal{C}'_n} \bigg\| \frac{n- b_n}{n}T_{x'^{n(n - b_n)}} - (1-\beta) P'_X \bigg\|_\infty \\
    = \: & \beta \max_{x^{n b_n} \in \mathcal{C}_n} \bigg\| T_{x^{n b_n}} - P_X + o(1) T_{x^{n b_n}}\bigg\|_\infty  + (1-\beta)  \nonumber\\
    &\times \max_{x'^{n(n - b_n)} \in \mathcal{C}'_n} \bigg\| T_{x'^{n(n - b_n)}} - P'_X  + o(1)T_{x'^{n(n - b_n)}}\bigg\|_\infty \\
    \leq \: & \beta \max_{x^{n b_n} \in \mathcal{C}_n} \bigg\| T_{x^{n b_n}} - P_X \bigg\|_\infty + o(1)\bigg\| T_{x^{n b_n}}\bigg\|_\infty \nonumber\\
    & + (1-\beta) \max_{x'^{n(n - b_n)} \in \mathcal{C}'_n} \bigg\| T_{x'^{n(n - b_n)}} - P'_X \bigg\|_\infty \nonumber\\
&+ o(1) \bigg\| T_{x'^{n(n - b_n)}}\bigg\|_\infty   \underset{n \rightarrow \infty}{\rightarrow} 0.
\end{align}

By Lemma~\ref{prop:CGPbasic}, $\lim_{n \rightarrow \infty} \frac{\log|\mathcal{C}''_n|}{n^2} \leq C(G,\beta P_X + (1-\beta) P'_X)$, thus
\begin{align}
    \beta C(G,P_X) + (1-\beta) C(G,P'_X) \leq C(G,\beta P_X + (1-\beta) P'_X).
\end{align}

The function $P_X \mapsto C(G,P_X)$ is concave on the convex compact set $\Delta(\mathcal{X})$, therefore its set of maximizers $\mathcal{P}^{\star}(G) = \argmax_{P_X \in \Delta(\mathcal{X})} C(G,P_X)$ is convex. Furthermore, by Lemma~\ref{th:simonyi}, the set $\mathcal{P}^{\star}(G)$ is nonempty and satisfies
\begin{align}
    \forall P_X \in \mathcal{P}^{\star}(G), \; C(G,P_X) = C_0(G).
\end{align}

\section{Proof of Theorem~\ref{lemma:Pstarprod}}\label{section:prooflemmaPstarprod}


Let us start by showing that Theorem~\ref{lemma:Pstarprod} is true when $\mathcal{A}$ has two elements. Let $G = (\mathcal{X}, \mathcal{E})$, and $G' = (\mathcal{X}', \mathcal{E'})$ be two graphs, and let $P_{X,X'} \in \mathcal{P}^{\star}(G \wedge G')$. We will prove that $P_{X} \otimes P_{X'}$ is also capacity-achieving by building an adequate sequence of codebooks.

For all $n \in \mathbb{N}^\star$, let $\mathcal{C}_n \subseteq (\mathcal{X} \times \mathcal{X}')^{n}$ such that $\mathcal{C}_n$ is an independent set in $(G \wedge G')^{\wedge n}$, and
\begin{align}
    & \frac{1}{n}\log|\mathcal{C}_n| \underset{n \rightarrow \infty}{\rightarrow} C_0(G \wedge G'),\label{eq:proofthPstarprodV0}\\
    & \max_{(x^n,x'^n) \in \mathcal{C}_n} \|T_{x^n, x'^n} - P_{X,X'}\|_\infty \underset{n \rightarrow \infty}{\rightarrow} 0.\label{eq:proofthPstarprodV1}
\end{align}
The existence of such a sequence is given by Lemma~\ref{prop:CGPbasic}, and Proposition~\ref{prop:Pstar}. Let 
\begin{align}
    Q^{(n)}_{X,X'} \doteq \frac{1}{|\mathcal{C}_n|} \sum_{(x^n, x'^n) \in \mathcal{C}_n} T_{x^n, x'^n}.\label{eq:proofthPstarprodV2}
\end{align}
An immediate observation is that
\begin{align}
    Q^{(n)}_{X,X'} \underset{n \rightarrow \infty}{\rightarrow} P_{X,X'} \label{eq:proofthPstarprodW0}
\end{align}
as a consequence of \eqref{eq:proofthPstarprodV1}.

Let us build a sequence of codebooks with asymptotic rate $C_0(G \wedge G')$, such that the type of their codewords converge uniformly to $P_{X} \otimes P_{X'}$:
\begin{align}
    \mathcal{C}^{\star}_{n^3} \doteq \mathcal{T}^{n^3}_{\varepsilon_n}(P_{X} \otimes P_{X'}) \cap \left(\textstyle\prod_{t \leq n} \mathcal{C}^{(t)}_{n}\right)^n; \label{eq:proofthPstarprodW1}
\end{align}
where 
\begin{align}
    \varepsilon_n \doteq \| Q^{(n)}_{X} \otimes Q^{(n)}_{X'} - P_{X} \otimes P_{X'} \|_\infty + \textstyle\frac{1}{\sqrt[4]{n}}\label{eq:proofthPstarprodW1a};
\end{align}
and where for all $t \leq n$, the shifted codebook $\mathcal{C}_n^{(t)}$ is defined by
\begin{align}
    &\mathcal{C}^{(t)}_n \doteq \nonumber\\
&\Big\lbrace \Big((x_t, x_{t+1}, ..., x_n, x_1, ..., x_{t-1}), x'^n \Big) \:\Big|\: (x^n, x'^n) \in \mathcal{C}_n \Big\rbrace. \label{eq:proofthPstarprodW2}
\end{align}

By construction, $\mathcal{C}^{\star}_{n^3} \subseteq \mathcal{T}^{n^3}_{\varepsilon_n}(P_{X} \otimes P_{X'})$ thanks to \eqref{eq:proofthPstarprodW1}, and $\varepsilon_n \underset{n \rightarrow \infty}{\rightarrow} 0$ thanks to \eqref{eq:proofthPstarprodW1a} and \eqref{eq:proofthPstarprodW0}; therefore we have
\begin{align}
    \max_{x^{n^3} \in \mathcal{C}^{\star}_n} \| T_{x^{n^3}} - P_X \otimes P_{X'} \|_\infty \underset{n \rightarrow \infty}{\rightarrow} 0.\label{eq:proofthPstarprodA0}
\end{align}
Furthermore, $\mathcal{C}^{\star}_{n^3}$ is an independent set in $(G \wedge G')^{\wedge n^3}$, as it is contained in the product independent set $\left(\textstyle\prod_{t \leq n} \mathcal{C}^{(t)}_{n}\right)^n$; note that this holds because the shifted codebook $\mathcal{C}^{(t)}_n$ is an independent set in $(G \wedge G')^{\wedge n}$ for all $t \leq n$.

Now let us prove that $\frac{\log|\mathcal{C}^{\star}_{n^3}|}{n^3} \underset{n \rightarrow \infty}{\rightarrow} C_0(G \wedge G')$. Let us draw a codeword uniformly from $\left(\textstyle\prod_{t \leq n} \mathcal{C}^{(t)}_{n}\right)^n$:
\begin{align}
    C^{n^3} \doteq (C^{n^2}_1, ..., C^{n^2}_n),\label{eq:proofthPstarprodB0}
\end{align}
where for all $t \leq n$, $C^{n^2}_t$ is a random $n \times n$-sequence drawn uniformly from $\textstyle\prod_{t \leq n} \mathcal{C}^{(t)}_{n}$. We want to prove that $C^{n^3} \in \mathcal{T}^{n^3}_{\varepsilon_n}(P_{X} \otimes P_{X'})$ with high probability. 

On one hand we have to determine the average type of the random variables $(C^{n^2}_t)_{t \leq n}$ which are iid copies of $C^{n^2} = (C^n_1, ..., C^n_n)$; where each $C^n_{t}$ is drawn uniformly from $\mathcal{C}^{(t)}_n$, and the $(C^n_{t})_{t \leq n}$ are mutually independent.
\begin{align}
    \mathbb{E}\left[T_{C^{n^2}_t}\right] & = \frac{1}{n}\sum_{t \leq n} \mathbb{E}\left[T_{C^{n}_{t}}\right] \label{eq:proofthPstarprodC0}\\
    & = \frac{1}{n}\sum_{t \leq n} \frac{1}{|\mathcal{C}^{(t)}_n|}\sum_{(x^n,x'^n) \in \mathcal{C}^{(t)}_n} T_{x^n,x'^n} \label{eq:proofthPstarprodC1}\\
    & = \frac{1}{n}\sum_{t \leq n} \frac{1}{|\mathcal{C}_n|}\sum_{(x^n,x'^n) \in \mathcal{C}_n} T_{\sigma_t(x^n),x'^n} \label{eq:proofthPstarprodC1a}\\
    & = \frac{1}{|\mathcal{C}_n|}\sum_{(x^n,x'^n) \in \mathcal{C}_n} \frac{1}{n}\sum_{t \leq n} T_{\sigma_t(x^n),x'^n} \label{eq:proofthPstarprodC1b}\\
    & = \frac{1}{|\mathcal{C}_n|}\sum_{(x^n,x'^n) \in \mathcal{C}_n} T_{x^n} \otimes T_{x'^n} \label{eq:proofthPstarprodC2}\\
    & 
    = Q^{(n)}_{X} \otimes Q^{(n)}_{X'}, \label{eq:proofthPstarprodC3}
\end{align}
where $\sigma_t(x^n) = (x_t, x_{t+1}, ..., x_n, x_1, ..., x_{n-1})$; \eqref{eq:proofthPstarprodC1a} comes from the construction of $\mathcal{C}^{(t)}_n$ in \eqref{eq:proofthPstarprodW2}; and \eqref{eq:proofthPstarprodC2} comes from the following observation: 
\begin{align}
    & \sum_{t \leq n} T_{\sigma_{t}(x^n),x'^n} = \sum_{t \leq n} \sum_{s \leq n} T_{x_{s+t},x'_s} = \sum_{s \leq n} \sum_{t \leq n} T_{x_{s+t},x'_s} \\
    & = \sum_{s \leq n} T_{x^n,(x'_s, ..., x'_s)} = \sum_{s \leq n} T_{x^n} \otimes T_{x'_s} = T_{x^n} \otimes T_{x'^n},
\end{align}
where the index $s + t$ is taken modulo $n$.

On the other hand we have
\begin{align}
    & \frac{|\mathcal{C}^{\star}_{n^3}|}{\left| \left(\textstyle\prod_{t \leq n} \mathcal{C}^{(t)}_{n}\right)^n\right|} \\
    & = \frac{\left| \mathcal{T}^{n^3}_{\varepsilon_n}(P_{X} \otimes P_{X'}) \cap \left(\textstyle\prod_{t \leq n} \mathcal{C}^{(t)}_{n}\right)^n \right|}{\left| \left(\textstyle\prod_{t \leq n} \mathcal{C}^{(t)}_{n}\right)^n\right|} \label{eq:proofthPstarprodD0} \\
    & = \mathbb{P}\left(C^{n^3} \in \mathcal{T}^{n^3}_{\varepsilon_n}(P_{X} \otimes P_{X'})\right) \label{eq:proofthPstarprodD1}\\
    & = \mathbb{P}\left(\left\|\textstyle\frac{1}{n} \sum_{t \leq n} T_{C_t^{n^2}} - P_{X} \otimes P_{X'}\right\|_\infty \leq \varepsilon_n\right) \label{eq:proofthPstarprodD2}\\
    & \geq \mathbb{P}\bigg(\left\|\textstyle\frac{1}{n} \sum_{t \leq n} T_{C_t^{n^2}} - Q^{(n)}_X \otimes Q^{(n)}_{X'} \right\|_\infty \nonumber\\
&+ \left\| Q^{(n)}_X \otimes Q^{(n)}_{X'} - P_{X} \otimes P_{X'} \right\|_\infty \leq \varepsilon_n\bigg) \label{eq:proofthPstarprodD3}\\
    & = \mathbb{P}\left(\left\|\textstyle \sum_{t \leq n} T_{C_t^{n^2}} - nQ^{(n)}_X \otimes Q^{(n)}_{X'} \right\|_\infty \leq n^{3/4}\right) \label{eq:proofthPstarprodD4}\\
    & \geq 1 - \textstyle\sum_{(x, x') \in \mathcal{X} \times \mathcal{X}'} \mathbb{P}\bigg(\bigg|\textstyle\sum_{t \leq n} T_{C_t^{n^2}}(x,x') \nonumber\\
&  \qquad \quad   - n Q^{(n)}_{X} \otimes Q^{(n)}_{X'}(x,x') \bigg| > n^{3/4}\bigg) \label{eq:proofthPstarprodD5}\\
    & \geq 1 - \textstyle\sum_{(x, x') \in \mathcal{X} \times \mathcal{X}'}  \frac{1}{n^{3/2}}\mathbb{X}\left[\textstyle\sum_{t \leq n} T_{C_t^{n^2}}(x,x')\right] \label{eq:proofthPstarprodD6}\\
    & \geq 1 - \textstyle\frac{|\mathcal{X}||\mathcal{X}'|}{n^{1/2}} \underset{n \rightarrow \infty}{\rightarrow} 1; \label{eq:proofthPstarprodD7}
\end{align}
where \eqref{eq:proofthPstarprodD1} and \eqref{eq:proofthPstarprodD2} come from the construction of $C^{n^3}$; \eqref{eq:proofthPstarprodD4} comes from the construction of $\varepsilon_n$; \eqref{eq:proofthPstarprodD5} follows from the union bound; \eqref{eq:proofthPstarprodD6} comes from Chebyshev's inequality and \eqref{eq:proofthPstarprodC3}; and \eqref{eq:proofthPstarprodD7} comes from the fact that $\mathbb{V}\left[\sum_{t \leq n} T_{C_t^{n^2}}(x,x')\right] = \sum_{t \leq n} \mathbb{V}\left[T_{C_t^{n^2}}(x,x')\right] \leq n$, as the random variables $T_{C_t^{n^2}}(x,x')$ are iid and takes values in $[0,1]$. Hence 
\begin{align}
    &\lim_{n \rightarrow \infty} \frac{\log |\mathcal{C}^{\star}_{n^3}|}{n^3} = \lim_{n \rightarrow \infty} \frac{\log \left|\left(\textstyle\prod_{t \leq n} \mathcal{C}^{(t)}_{n}\right)^n\right|}{n^3} \nonumber\\
&= \lim_{n \rightarrow \infty} \frac{\log |\mathcal{C}_n|}{n} = C_0(G \wedge G');\label{eq:proofthPstarprodL0}
\end{align}
where the second equality holds as the shifted codebooks $(\mathcal{C}^{(t)}_n)_{t \leq n}$ all have cardinality $|\mathcal{C}_n|$.

Thus, by combining \eqref{eq:proofthPstarprodL0}, Lemma~\ref{prop:CGPbasic}, and Proposition~\ref{prop:Pstar} we obtain
\begin{align}
    C_0(G \wedge G') =& \lim_{n \rightarrow \infty} \frac{\log|\mathcal{C}^{\star}_{n^3}|}{n^3} \leq C(G \wedge G', P_X \otimes P_{X'})\nonumber\\
\leq& C_0(G \wedge G'),
\end{align}
hence $P_X \otimes P_{X'} \in \mathcal{P}^{\star}(G\wedge G')$.

Therefore, Theorem~\ref{lemma:Pstarprod} is proved when $\mathcal{A}$ has two elements:
\begin{align}
    P_{X, X'} \in \mathcal{P}^{\star}(G \wedge G') \Longrightarrow P_{X} \otimes P_{X'} \in \mathcal{P}^{\star}(G \wedge G'). \label{eq:resultat}
\end{align}

Now let us consider the case where $\mathcal{A}$ has a cardinality greater than 2. Let $P_{X_1, ..., X_{\mathcal{A}}} \in \mathcal{P}^{\star}(\bigwedge_{a \in \mathcal{A}} G_a)$. By considering the product graphs 
\begin{align}
    \textstyle\bigwedge_{a \in \mathcal{A}} G_a = \Big(\bigwedge_{1 \leq i < i^{\star}} G_i\Big) \wedge \Big(\bigwedge_{i^{\star} \leq i \leq |\mathcal{A}|} G_i\Big);
\end{align}
 for all $i^{\star} \leq |\mathcal{A}|$, and applying \eqref{eq:resultat} successively, we obtain 
\begin{align}
    &P_{X_1, ..., X_{\mathcal{A}}} \in \mathcal{P}^{\star}\left(\textstyle\bigwedge_{a \in \mathcal{A}} G_a\right) \nonumber\\
    & \Longrightarrow P_{X_1} \otimes P_{X_2, ..., X_{|\mathcal{A}|}} \in \mathcal{P}^{\star}\left(\textstyle\bigwedge_{a \in \mathcal{A}} G_a\right) \\
& \Longrightarrow (P_{X_1} \otimes P_{X_2}) \otimes P_{X_3, ..., X_{|\mathcal{A}|}} \in \mathcal{P}^{\star}\left(\textstyle\bigwedge_{a \in \mathcal{A}} G_a\right) \\
& \Longrightarrow \textstyle\bigotimes_{a \in \mathcal{A}} P_{X_a} \in \mathcal{P}^{\star}\left(\textstyle\bigwedge_{a \in \mathcal{A}} G_a\right).
\end{align}

\section{Results on Capacity-Achieving Distributions}\label{section:proofcap}

\subsection{Proof of Proposition~\ref{th:vertextrans}}\label{section:proofthvertextrans}

Let $G$ be a vertex-transitive graph, and let $P_X \in \mathcal{P}^{\star}(G)$. Let $\psi \in \Aut(G)$, we first prove that $P_{\psi(X)} \in \mathcal{P}^{\star}(G)$, then we will conclude by using the convexity of $\mathcal{P}^{\star}(G)$.

Let $(\mathcal{C}_n)_{n \in \mathbb{N}^\star}$ be a sequence such that
    \begin{align}
        & \forall n \in \mathbb{N}^\star, \; \mathcal{C}_n \subseteq \mathcal{X}^n \text{ is an independent set in } G^{\wedge n},\\ 
        & \max_{x^n \in \mathcal{C}_n} \|T_{x^n} - P_X \|_\infty \underset{n \rightarrow \infty}{\rightarrow} 0,\\
        & \frac{\log|\mathcal{C}_n|}{n} \underset{n \rightarrow \infty}{\rightarrow} C(G,P_X) = C_0(G).\label{eq:proofthvertextransZ3}
    \end{align}
The existence of such a sequence is given by Lemma~\ref{prop:CGPbasic}. Note that the last equality in \eqref{eq:proofthvertextransZ3} comes from the assumption $P_X \in \mathcal{P}^{\star}(G)$.

Now, for all $n \in \mathbb{N}^\star$ the codebook
\begin{align}
    \psi(\mathcal{C}_n) \doteq \lbrace (\psi(x_1), ..., \psi(x_n)) \:|\: x^n \in \mathcal{C}_n \rbrace
\end{align}
is also independent in $G^{\wedge n}$, as $\psi$ is a graph automorphism and therefore preserves adjacencies. We have by construction
\begin{align}
    \max_{x^n \in \psi(\mathcal{C}_n)} \|T_{x^n} - P_{\psi(X)} \|_\infty \underset{n \rightarrow \infty}{\rightarrow} 0.
\end{align}
Furthermore, since $\psi$ is a bijection we have $|\psi(\mathcal{C}_n)| = |\mathcal{C}_n|$ for all $n \in \mathbb{N}^\star$, thus
\begin{align}
    \frac{\log|\psi(\mathcal{C}_n)|}{n} =  \frac{\log|\mathcal{C}_n|}{n} \underset{n \rightarrow \infty}{\rightarrow} C_0(G).
\end{align}
Hence 
\begin{align}
    P_{\psi(X)} \in \mathcal{P}^{\star}(G).\label{eq:proofthvertextransB0}
\end{align}

Now, for all $x\neq x' \in \mathcal{X}$, denote by $\mathcal{S}_{x' \rightarrow x} \subseteq \Aut(G)$ the set of automorphisms that map $x'$ to $x$; note that this set is nonempty thanks to the vertex-transitivity of $G$. We have for all $x \in \mathcal{X}$
\begin{align}
    \Aut(G) = \textstyle\bigsqcup_{x' \in \mathcal{X}} \mathcal{S}_{x' \rightarrow x}. \label{eq:proofthvertextransC0}
\end{align}
Furthermore, for all $x \in \mathcal{X}$, all the sets $(\mathcal{S}_{x' \rightarrow x})_{x' \rightarrow x}$ have the same cardinality: for all $x', x'' \in \mathcal{X}$, 
\begin{align}
    \mathcal{S}_{x'' \rightarrow x} \circ \psi_1 \subseteq \mathcal{S}_{x' \rightarrow x},
\end{align}
where $\psi_1 \in \mathcal{S}_{x' \rightarrow x''}$. It follows that for all $x, x' \in \mathcal{X}$,
\begin{align}
    |\mathcal{S}_{x' \rightarrow x}| = \frac{|\Aut(G)|}{|\mathcal{X}|}. \label{eq:proofthvertextransD0}
\end{align}
Therefore, for all $x \in \mathcal{X}$ we have 
\begin{align}
    & \frac{1}{|\Aut(G)|} \sum_{\psi \in \Aut(G)} P_{\psi(X)} \qquad \in \mathcal{P}^{\star}(G) \label{eq:proofthvertextransA0}\\
    = \; & \bigg(\frac{1}{|\Aut(G)|} \sum_{\psi \in \Aut(G)} P_{X}(\psi^{-1}(x))\bigg)_{x \in \mathcal{X}} \\
    = \; & \bigg(\frac{1}{|\Aut(G)|} \sum_{x' \in \mathcal{X}} |\mathcal{S}_{x' \rightarrow x}| P_{X}(x')\bigg)_{x \in \mathcal{X}} \label{eq:proofthvertextransA1}\\
    = \; & \bigg(\frac{1}{|\Aut(G)|} \sum_{x' \in \mathcal{X}} \frac{|\Aut(G)|}{|\mathcal{X}|} P_X(x')\bigg)_{x \in \mathcal{X}}\label{eq:proofthvertextransA2}\\
    = \; & \Unif(\mathcal{X});\label{eq:proofthvertextransA3}
\end{align}
where \eqref{eq:proofthvertextransA0} comes from the convexity of $\mathcal{P}^{\star}(G)$ given by Proposition~\ref{prop:Pstar} and \eqref{eq:proofthvertextransB0}; \eqref{eq:proofthvertextransA1} comes from \eqref{eq:proofthvertextransC0}; and \eqref{eq:proofthvertextransA2} comes from \eqref{eq:proofthvertextransD0}.

\subsection{Proof of Lemma~\ref{lemma:conjugate}}\label{section:prooflemmaconjugate}
Let $(w_a)_{a \in \mathcal{A}} \in \mathbb{R}^{|\mathcal{A}|}$, 
and maximize 
    \begin{align}
        \zeta : P_A \mapsto H(P_A) + \sum_{a \in \mathcal{A}} P_A(a) w_a.
    \end{align}
It can be easily observed that $\zeta$ is strictly concave, hence the existence and uniqueness of the maximum. We have 
    \begin{align}
        \nabla \zeta (P_A) = \left(-\log P_A(a) - \frac{1}{\ln 2} + w_a\right)_{a \in \mathcal{A}},
    \end{align}
hence 
    \begin{align}
       & \nabla \zeta (P_A) \perp \Delta(\mathcal{A}) \nonumber\\
        & \Longleftrightarrow \exists C \in \mathbb{R}, \, \nabla \zeta (P_A) = (C, ..., C) \\
        & \Longleftrightarrow \exists C' \in \mathbb{R}, \, (-\log P_A(a) + w_a)_{a \in \mathcal{A}} = (C', ..., C') \\
        & \Longleftrightarrow \exists C' \in \mathbb{R}, \, P_A = 2^{-C'} \left( 2^{w_a} \right)_{a \in \mathcal{A}}
    \end{align}
    The value of $C'$ can be deduced from the fact that $P_A$ is a probability distribution: $2^{C'}$ is the normalization constant $\sum_{a' \in \mathcal{A}} 2^{w_{a'}}$. Hence the maximum of $\zeta$ \mlt{is given by}
        \begin{align}
            P^{\star}_A = \left(\frac{2^{w_a}}{\sum_{a' \in \mathcal{A}} 2^{w_{a'}}}\right)_{a \in \mathcal{A}};
        \end{align}
        and we have
        \begin{align}
            \zeta(P^{\star}_A) & = \sum_{a \in \mathcal{A}} \ P^{\star}_A(a) \left(\log\left(\frac{\sum_{a' \in \mathcal{A}} 2^{w_{a'}}}{2^{w_a}} \right) + w_a \right) \\
            & = \log\left(\sum_{a' \in \mathcal{Z}} 2^{w_{a'}} \right).
        \end{align}

\section{Proof of Theorem~\ref{th:mainC0}}\label{section:proofthmainC0}


We prove Theorem~\ref{th:mainC0} in two steps, which are Lemma~\ref{lemma:mainC0A} and Lemma~\ref{lemma:mainC0B}. The proofs are respectively given in App.~\ref{section:prooflemmamainC0A} and \ref{section:prooflemmamainC0B}.

\begin{lemma}\label{lemma:mainC0A}
\begin{align}
     & C_0\left(\bigwedge_{a \in \mathcal{A}} G_a \right) = \sum_{a \in \mathcal{A}} C_0(G_a) \\
     \Longrightarrow \; & \forall (P^{\star}_{X_{a}})_{a \in \mathcal{A}} \in \prod_{a \in \mathcal{A}} \mathcal{P}^{\star}(G_a),\nonumber\\
&     \begin{cases}
     \bigotimes_{a \in \mathcal{A}} P^{\star}_{X_a} \in \mathcal{P}^{\star}\left(\bigwedge_{a \in \mathcal{A}} G_a\right) \text{ and } \\ C\left(\bigwedge_{a \in \mathcal{A}} G_a, \; \bigotimes_{a \in \mathcal{A}} P^{\star}_{X_a}\right) = \sum_{a \in \mathcal{A}} C(G_a,P^{\star}_{X_a}).
     \end{cases}
\end{align}
\end{lemma}

\begin{lemma}\label{lemma:mainC0B}
For all $P_{X_1, ..., X_{|\mathcal{A}|}} \in \mathcal{P}^{\star}\left(\bigwedge_{a \in \mathcal{A}} G_a \right)$, the following holds
\begin{align}
    & C\left(\bigwedge_{a \in \mathcal{A}} G_a, \; P_{X_1, ..., X_{|\mathcal{A}|}}\right) = \sum_{a \in \mathcal{A}} C(G_a,P_{X_a}) \\
    \Longrightarrow \; & C_0 \left(\bigwedge_{a \in \mathcal{A}} G_a\right) = \sum_{a \in \mathcal{A}} C_0(G_a)\nonumber\\
& \text{ and } \forall a \in \mathcal{A}, \; P_{X_a} \in \mathcal{P}^{\star}(G_a).
\end{align}
\;
\end{lemma}

Let us prove Theorem~\ref{th:mainC0}. We consider a family of distributions $P_{X_a} \in \mathcal{P}^{\star}(G_a)$ with $a \in \mathcal{A}$.
By Lemma~\ref{lemma:mainC0A}, we have
\begin{align}
    & C_0\left(\textstyle\bigwedge_{a \in \mathcal{A}} G_a \right) = \textstyle\sum_{a \in \mathcal{A}} C_0(G_a) \\
    \Longrightarrow \; & \textstyle\bigotimes_{a \in \mathcal{A}} P^{\star}_{X_a} \in \mathcal{P}^{\star}\left(\bigwedge_{a \in \mathcal{A}} G_a\right) \text{ and }  \nonumber\\
&C\left(\bigwedge_{a \in \mathcal{A}} G_a, \; \bigotimes_{a \in \mathcal{A}} P^{\star}_{X_a}\right) = \sum_{a \in \mathcal{A}} C(G_a,P^{\star}_{X_a}) \\
    \Longrightarrow \; & \exists P_{X_1, ..., X_{|\mathcal{A}|}} \in  \mathcal{P}^{\star}\left(\textstyle\bigwedge_{a \in \mathcal{A}} G_a \right)\!,\nonumber\\
& C\left(\textstyle\bigwedge_{a \in \mathcal{A}} G_a,\; P_{X_1, ..., X_{|\mathcal{A}|}}  \right) = \textstyle\sum_{a \in \mathcal{A}} C(G_a, P_{X_a}).
\end{align}
Conversely, by Lemma~\ref{lemma:mainC0B} we have
\begin{align}
    & \exists P_{X_1, ..., X_{|\mathcal{A}|}} \in  \mathcal{P}^{\star}\left(\textstyle\bigwedge_{a \in \mathcal{A}} G_a \right)\!,\nonumber\\
&C\left(\textstyle\bigwedge_{a \in \mathcal{A}} G_a,\; P_{X_1, ..., X_{|\mathcal{A}|}}  \right) = \textstyle\sum_{a \in \mathcal{A}} C(G_a, P_{X_a}) \label{eq:proofmainC0W0}\\
    \Longrightarrow \; & C_0 \left(\textstyle\bigwedge_{a \in \mathcal{A}} G_a\right) = \textstyle\sum_{a \in \mathcal{A}} C_0(G_a). 
\end{align}
Moreover, all distribution $P_{X_1, ..., X_{|\mathcal{A}|}}$ that satisfies \eqref{eq:proofmainC0W0}, also satisfies $P_{X_a} \in \mathcal{P}^{\star}(G_a)$, for all $a \in \mathcal{A}$.

\subsection{Proof of Lemma~\ref{lemma:mainC0A}}\label{section:prooflemmamainC0A}

For all family of graphs $(G_a)_{a \in \mathcal{A}}$, and family of distributions $P_{X_a} \in \mathcal{P}^{\star}(G_a)$ with $a \in \mathcal{A}$, we have
\begin{align}
    C_0\left(\textstyle\bigwedge_{a \in \mathcal{A}} G_a \right) 
    & = \max_{P_{X_1, ..., X_{|\mathcal{A}|}}} \; C\left(\textstyle\bigwedge_{a \in \mathcal{A}} G_a, \; P_{X_1, ..., X_{|\mathcal{A}|}} \right) \label{eq:prooflemmamainC0AA1}\\
    & \geq C\left(\textstyle\bigwedge_{a \in \mathcal{A}} G_a, \; \bigotimes_{a \in \mathcal{A}} P^{\star}_{X_a} \right) \label{eq:prooflemmamainC0AA2}\\
    & \geq \textstyle\sum_{a \in \mathcal{A}} C(G_a, P^{\star}_{X_a}) \label{eq:prooflemmamainC0AA3}\\
    & = \textstyle\sum_{a \in \mathcal{A}} C_0(G_a);\label{eq:prooflemmamainC0AA4}
\end{align}
where \eqref{eq:prooflemmamainC0AA1} comes from Lemma~\ref{th:simonyi}; 
\eqref{eq:prooflemmamainC0AA3} comes from Proposition~\ref{prop:marton}; and \eqref{eq:prooflemmamainC0AA4} follows from the hypothesis $P_{X_a} \in \mathcal{P}^{\star}(G_a)$ with $a \in \mathcal{A}$.

Now assume that $\sum_{a \in \mathcal{A}} C_0(G_a) = C_0\left(\bigwedge_{a \in \mathcal{A}} G_a \right)$, then equality holds between the left-hand side of  \eqref{eq:prooflemmamainC0AA1} and the term in \eqref{eq:prooflemmamainC0AA4}. Therefore, we have
\begin{align}
    & C_0\left(\textstyle\bigwedge_{a \in \mathcal{A}} G_a \right) = C\left(\textstyle\bigwedge_{a \in \mathcal{A}} G_a, \; \bigotimes_{a \in \mathcal{A}} P^{\star}_{X_a} \right), \nonumber\\
\Longrightarrow& \textstyle\bigotimes_{a \in \mathcal{A}} P^{\star}_{X_a} \in \mathcal{P}^{\star}\left(\textstyle\bigwedge_{a \in \mathcal{A}} G_a\right), \nonumber \\
    & C\left(\textstyle\bigwedge_{a \in \mathcal{A}} G_a, \; \bigotimes_{a \in \mathcal{A}} P^{\star}_{X_a}\right) = \textstyle\sum_{a \in \mathcal{A}} C(G_a,P^{\star}_{X_a}).
\end{align}

\subsection{Proof of Lemma~\ref{lemma:mainC0B}}\label{section:prooflemmamainC0B}

Let $P_{X_1, ..., X_{|\mathcal{A}|}} \in \mathcal{P}^{\star}(\bigwedge_{a \in \mathcal{A}} G_a)$, 
and let $P^{\star}_{X_a} \in \mathcal{P}^{\star}(G_a)$, for all $a \in \mathcal{A}$. 
The following holds 
\begin{align}
    C_0\left(\textstyle\bigwedge_{a \in \mathcal{A}} G_a \right) 
    &=  C\left(\textstyle\bigwedge_{a \in \mathcal{A}} G_a, \; P_{X_1, ..., X_{|\mathcal{A}|}} \right)\label{eq:prooflemmamainC0BA0}\\
    & \geq C\left(\textstyle\bigwedge_{a \in \mathcal{A}} G_a, \; \textstyle\bigotimes_{a \in \mathcal{A}} P^{\star}_{X_a} \right) \label{eq:prooflemmamainC0BA1}\\
    & \geq \textstyle\sum_{a \in \mathcal{A}} C(G_a, P^{\star}_{X_a}) \label{eq:prooflemmamainC0BA2}\\
    & \geq \textstyle\sum_{a \in \mathcal{A}} C(G_a, P_{X_a}),\label{eq:prooflemmamainC0BA3}
\end{align}
where 
\eqref{eq:prooflemmamainC0BA1} comes from 
the hypothesis $P_{X_1, ..., X_{|\mathcal{A}|}} \in \mathcal{P}^{\star}(\bigwedge_{a \in \mathcal{A}} G_a)$; 
\eqref{eq:prooflemmamainC0BA2} comes from Proposition~\ref{prop:marton}; and \eqref{eq:prooflemmamainC0BA3} comes from the hypothesis $P^{\star}_{X_a} \in \mathcal{P}^{\star}(G_a)$, for all $a \in \mathcal{A}$.

Now assume that 
\begin{align}
    C\left(\textstyle\bigwedge_{a \in \mathcal{A}} G_a, \; P_{X_1, ..., X_{|\mathcal{A}|}} \right) = \textstyle\sum_{a \in \mathcal{A}} C(G_a, P_{X_a}).\label{eq:prooflemmamainC0BB1}
\end{align}
Then equality holds in between the right-hand side of \eqref{eq:prooflemmamainC0BA0} and the term in \eqref{eq:prooflemmamainC0BA3}. In particular, we have for all $a \in \mathcal{A}$
\begin{align}
    C(G_a, P_{X_a}) = C(G_a, P^{\star}_{X_a}),\label{eq:prooflemmamainC0BB0}
\end{align}
which implies that $P_{X_a}$ also maximizes $C(G_a, \cdot)$ for all $a \in \mathcal{A}$:
\begin{align}
    \forall a \in \mathcal{A}, \; P_{X_a} \in \mathcal{P}^{\star}(G_a).\label{eq:prooflemmamainC0BC0}
\end{align}

Furthermore,
\begin{align}
    C_0 \left(\textstyle\bigwedge_{a \in \mathcal{A}} G_a\right) & = C\left(\textstyle\bigwedge_{a \in \mathcal{A}} G_a, \; P_{X_1, ..., X_{|\mathcal{A}|}} \right) \label{eq:prooflemmamainC0BD0}\\
    & = \textstyle\sum_{a \in \mathcal{A}} C(G_a, P_{X_a}) \label{eq:prooflemmamainC0BD1}\\
    & = \textstyle\sum_{a \in \mathcal{A}} C_0(G_a); \label{eq:prooflemmamainC0BD2}
\end{align}
where 
\eqref{eq:prooflemmamainC0BD1} is a consequence of the equality in equations \eqref{eq:prooflemmamainC0BA0}-\eqref{eq:prooflemmamainC0BA3}, and
\eqref{eq:prooflemmamainC0BD2} comes from \eqref{eq:prooflemmamainC0BC0}.

\section{Proof of Theorem~\ref{th:linC_0sqcup}}\label{section:proofthlinC_0sqcup}

The techniques used in this proof are the same as in the proof of Theorem~\ref{th:mainC0}. 
We prove Theorem~\ref{th:linC_0sqcup} in two steps, which are Lemma~\ref{lemma:linC_0sqcup1} and Lemma~\ref{lemma:linC_0sqcup2}; their proofs are respectively given in App.~\ref{section:prooflemmalinC_0sqcup1} and \ref{section:prooflemmalinC_0sqcup2}.
    
\begin{lemma}\label{lemma:linC_0sqcup1}
    Let 
    \begin{align}
        P^{\star}_A \doteq \left(\frac{2^{C_0(G_a)}}{\sum_{a' \in \mathcal{A}} 2^{C_0(G_{a'})}} \right)_{a \in \mathcal{A}}.
    \end{align}
We have
    \begin{align}
        & C_0\left(\bigsqcup_{a \in \mathcal{A}} G_a\right) = \log\left(\sum_{a \in \mathcal{A}} 2^{C_0(G_a)}\right) \nonumber\\
        \Longrightarrow  & \forall (P^{\star}_{X_a})_{a \in \mathcal{A}} \in \prod_{a \in \mathcal{A}} \mathcal{P}^{\star}(G_a), \nonumber\\
& \sum_{a \in \mathcal{A}} P^{\star}_A(a)P^{\star}_{X_a} \in \mathcal{P}^{\star}\left(\bigsqcup_{a \in \mathcal{A}} G_a \right) \text{ and } \nonumber\\
        & C\left(\bigsqcup_{a \in \mathcal{A}} G_a, \; \sum_{a \in \mathcal{A}} P^{\star}_A(a) P^{\star}_{X_a}\right)        \nonumber\\
&= H(P^{\star}_A) + \sum_{a \in \mathcal{A}} P^{\star}_A(a) C(G_a, P^{\star}_{X_a}),\label{eq:proofthlinC_0sqcupD0}
    \end{align}
    \end{lemma}
    
    \begin{lemma}\label{lemma:linC_0sqcup2}
    Let 
    \begin{align}
        & P^{\star}_A \doteq \left(\frac{2^{C_0(G_a)}}{\sum_{a' \in \mathcal{A}} 2^{C_0(G_{a'})}} \right)_{a \in \mathcal{A}}. 
    \end{align}
    For all $\sum_{a \in \mathcal{A}} P_A(a) P_{X_a} \in \mathcal{P}^{\star}\left(\bigsqcup_{a \in \mathcal{A}} G_a\right)$ the following holds 
    \begin{align}
        & C\left(\bigsqcup_{a \in \mathcal{A}} G_a,\; \sum_{a \in \mathcal{A}} P_A(a) P_{X_a}\right) \nonumber\\
&= H(P_A) + \sum_{a \in \mathcal{A}} P_A(a) C(G_a, P_{X_a})\nonumber \\
        \Longrightarrow  & C_0\left(\bigsqcup_{a \in \mathcal{A}} G_a\right) = \log\left(\sum_{a \in \mathcal{A}} 2^{C_0(G_a)}\right)\!,\nonumber\\
& (P_{X_a})_{a \in \mathcal{A}} \in \prod_{a \in \mathcal{A}} \mathcal{P}^{\star}(G_a), \text{ and } P_A = P^{\star}_A.
    \end{align}
    \end{lemma}

    Now let us prove Theorem~\ref{th:linC_0sqcup}. Let $(P^{\star}_{X_a})_{a \in \mathcal{A}} \in \prod_{a \in \mathcal{A}} \mathcal{P}^{\star}(G_a)$, 
    we have by Lemma~\ref{lemma:linC_0sqcup1}
    \begin{align}
        & C_0\left(\textstyle\bigsqcup_{a \in \mathcal{A}} G_a \right) = \log\left(\textstyle\sum_{a \in \mathcal{A}} 2^{C_0(G_a)}\right) \\
        \Longrightarrow \; & \textstyle\sum_{a \in \mathcal{A}} P^{\star}_A(a)P^{\star}_{X_a} \in \mathcal{P}^{\star}\left(\textstyle\bigsqcup_{a \in \mathcal{A}} G_a \right) \text{ and } \nonumber\\
        & C\left(\textstyle\bigsqcup_{a \in \mathcal{A}} G_a, \; \textstyle\sum_{a \in \mathcal{A}} P^{\star}_A(a) P^{\star}_{X_a}\right) \nonumber\\
&        = H(P^{\star}_A) + \textstyle\sum_{a \in \mathcal{A}} P^{\star}_A(a) C(G_a, P^{\star}_{X_a}), \\
        \Longrightarrow \; & \exists P_X \in \mathcal{P}^{\star}\left(\textstyle\bigsqcup_{a \in \mathcal{A}} G_a \right)\!, \, \nonumber\\
        & C\left(\textstyle\bigsqcup_{a \in \mathcal{A}} G_a, \; \textstyle\sum_{a \in \mathcal{A}} P^{\star}_A(a) P^{\star}_{X_a}\right)\nonumber\\
& = H(P^{\star}_A) + \textstyle\sum_{a \in \mathcal{A}} P^{\star}_A(a) C(G_a, P^{\star}_{X_a}),\nonumber
    \end{align}
    where $P^{\star}_{X_a} = P_{X|X \in \mathcal{X}_a}$ and $P^{\star}_A(a) = P_X(\mathcal{X}_a)$ for all $a \in \mathcal{A}$. 
    
    Conversely, by Lemma~\ref{lemma:linC_0sqcup2} we have
    \begin{align}
        & \exists P_X \in \mathcal{P}^{\star}\left(\textstyle\bigsqcup_{a \in \mathcal{A}} G_a \right)\!, \label{eq:proofC0linsqcupW0}\\
        & C\left(\textstyle\bigsqcup_{a \in \mathcal{A}} G_a,\; \textstyle\sum_{a \in \mathcal{A}} P_A(a) P_{X_a}\right) \nonumber\\
&= H(P_A) + \textstyle\sum_{a \in \mathcal{A}} P_A(a) C(G_a, P_{X_a}) \nonumber\\
        \Longrightarrow \; & C_0\left(\textstyle\bigsqcup_{a \in \mathcal{A}} G_a\right) = \log\left(\textstyle\sum_{a \in \mathcal{A}} 2^{C_0(G_a)}\right), 
    \end{align}
    and any $P_X = \sum_{a \in \mathcal{A}} P_A(a) P_{X_a}$ that satisfies \eqref{eq:proofC0linsqcupW0} also satisfies 
    \begin{align}
        &(P_{X_a})_{a \in \mathcal{A}} \in \prod_{a \in \mathcal{A}} \mathcal{P}^{\star}(G_a), \nonumber\\
&\text{ and } P_A = \left(\frac{2^{C_0(G_a)}}{\sum_{a' \in \mathcal{A}} 2^{C_0(G_{a'})}} \right)_{a \in \mathcal{A}}.
    \end{align}

\subsection{Proof of Lemma~\ref{lemma:linC_0sqcup1}}\label{section:prooflemmalinC_0sqcup1}

Now let us prove Lemma~\ref{lemma:linC_0sqcup1}. Let
    \begin{align}
         & \textstyle\sum_{a \in \mathcal{A}} P_A(a) P_{X_a} \in \mathcal{P}^{\star}\left(\textstyle\bigsqcup_{a \in \mathcal{A}} G_a\right), \label{eq:proofthlinC_0sqcupA1}\\
         & (P^{\star}_{X_a})_{a \in \mathcal{A}} \in \textstyle\prod_{a \in \mathcal{A}} \mathcal{P}^{\star}(G_a), \label{eq:proofthlinC_0sqcupA2}\\
         & P^{\star}_A \doteq \left(\frac{2^{C_0(G_a)}}{\sum_{a' \in \mathcal{A}} 2^{C_0(G_{a'})}} \right)_{a \in \mathcal{A}}.\label{eq:proofthlinC_0sqcupA3}
    \end{align}
    We have 
    \begin{align}
        C_0\left(\textstyle\bigsqcup_{a \in \mathcal{A}} G_a\right) & = C\left(\textstyle\bigsqcup_{a \in \mathcal{A}} G_a,\; \sum_{a \in \mathcal{A}} P_A(a) P_{X_a}\right) \label{eq:proofthlinC_0sqcupB2}\\
        & \geq C\left(\textstyle\bigsqcup_{a \in \mathcal{A}} G_a,\; \sum_{a \in \mathcal{A}} P^{\star}_A(a) P^{\star}_{X_a}\right) \label{eq:proofthlinC_0sqcupB3}\\
        & \geq H(P^{\star}_A) + \textstyle\sum_{a \in \mathcal{A}} P^{\star}_A(a) C(G_a, P^{\star}_{X_a}) \label{eq:proofthlinC_0sqcupB4}\\
        & = H(P^{\star}_A) + \textstyle\sum_{a \in \mathcal{A}} P^{\star}_A(a) C_0(G_a) \label{eq:proofthlinC_0sqcupB5}\\
        & = \log\left(\textstyle\sum_{a \in \mathcal{A}} 2^{C_0(G_a)}\right);\label{eq:proofthlinC_0sqcupB6}
    \end{align}
    where \eqref{eq:proofthlinC_0sqcupB2} and \eqref{eq:proofthlinC_0sqcupB3} come from \eqref{eq:proofthlinC_0sqcupA1} and Proposition~\ref{prop:Pstar}; \eqref{eq:proofthlinC_0sqcupB4} comes from Proposition~\ref{prop:marton}; \eqref{eq:proofthlinC_0sqcupB5} comes from \eqref{eq:proofthlinC_0sqcupA2} and Proposition~\ref{prop:Pstar}; and \eqref{eq:proofthlinC_0sqcupB6} comes from \eqref{eq:proofthlinC_0sqcupA3} and Lemma~\ref{lemma:conjugate}.
    
    Assume that $C_0\left(\textstyle\bigsqcup_{a \in \mathcal{A}} G_a\right) = \log\left(\textstyle\sum_{a \in \mathcal{A}} 2^{C_0(G_a)}\right)$, then equality holds in \eqref{eq:proofthlinC_0sqcupB2} to \eqref{eq:proofthlinC_0sqcupB6}, therefore the following holds:
    \begin{align}
        & C_0\left(\textstyle\bigsqcup_{a \in \mathcal{A}} G_a\right) = \log\left(\textstyle\sum_{a \in \mathcal{A}} 2^{C_0(G_a)}\right) \nonumber\\
        \Longrightarrow \; & \forall (P^{\star}_{X_a})_{a \in \mathcal{A}} \in \textstyle\prod_{a \in \mathcal{A}} \mathcal{P}^{\star}(G_a),\nonumber\\
& \textstyle\sum_{a \in \mathcal{A}} P^{\star}_A(a)P^{\star}_{X_a} \in \mathcal{P}^{\star}\left(\textstyle\bigsqcup_{a \in \mathcal{A}} G_a\right) \text{ and } \nonumber\\
        & C\left(\textstyle\bigsqcup_{a \in \mathcal{A}} G_a,\; \sum_{a \in \mathcal{A}} P^{\star}_A(a) P^{\star}_{X_a}\right)\nonumber\\
& = H(P^{\star}_A) + \textstyle\sum_{a \in \mathcal{A}} P^{\star}_A(a) C(G_a, P^{\star}_{X_a}).
    \end{align}
    
\subsection{Proof of Lemma~\ref{lemma:linC_0sqcup2}}\label{section:prooflemmalinC_0sqcup2}

Let 
\begin{align}
         & \textstyle\sum_{a \in \mathcal{A}} P_A(a) P_{X_a} \in \mathcal{P}^{\star}\left(\textstyle\bigsqcup_{a \in \mathcal{A}} G_a\right), \label{eq:proofthlinC_0sqcupA1_}\\
         & (P^{\star}_{X_a})_{a \in \mathcal{A}} \in \textstyle\prod_{a \in \mathcal{A}} \mathcal{P}^{\star}(G_a), \label{eq:proofthlinC_0sqcupA2_}\\
         & P^{\star}_A \doteq \left(\frac{2^{C_0(G_a)}}{\sum_{a' \in \mathcal{A}} 2^{C_0(G_{a'})}} \right)_{a \in \mathcal{A}}.\label{eq:proofthlinC_0sqcupA3_}
    \end{align}
    We have
    \begin{align}
        &C\left(\textstyle\bigsqcup_{a \in \mathcal{A}} G_a,\; \sum_{a \in \mathcal{A}} P_A(a) P_{X_a}\right) \nonumber\\
        & = C_0\left(\textstyle\bigsqcup_{a \in \mathcal{A}} G_a\right) \label{eq:proofthlinC_0sqcupE2}\\
        & \geq \log\left(\textstyle\sum_{a \in \mathcal{A}} 2^{C_0(G_a)}\right)\label{eq:proofthlinC_0sqcupE3} \\
        & = H(P^{\star}_A) + \textstyle\sum_{a \in \mathcal{A}} P^{\star}_A(a) C_0(G_a) \label{eq:proofthlinC_0sqcupE4}\\
        & \geq H(P_A) + \textstyle\sum_{a \in \mathcal{A}} P_A(a) C_0(G_a) \label{eq:proofthlinC_0sqcupE5}\\
        & = H(P_A) + \textstyle\sum_{a \in \mathcal{A}} P_A(a) C(G_a, P^{\star}_{X_a}) \label{eq:proofthlinC_0sqcupE6} \\
        & \geq H(P_A) + \textstyle\sum_{a \in \mathcal{A}} P_A(a) C(G_a, P_{X_a}); \label{eq:proofthlinC_0sqcupE7}
    \end{align}
    where \eqref{eq:proofthlinC_0sqcupE2} comes from \eqref{eq:proofthlinC_0sqcupA1_} and Proposition~\ref{prop:Pstar}; \eqref{eq:proofthlinC_0sqcupE3} comes from \eqref{eq:linchannelprod_}, see \cite[Theorem 4]{shannon1956zero};
    \eqref{eq:proofthlinC_0sqcupE4} and \eqref{eq:proofthlinC_0sqcupE5} come from \eqref{eq:proofthlinC_0sqcupA3_} and Lemma~\ref{lemma:conjugate}, which can be found in App.~\ref{section:prooflemmalinC_0sqcup1}; \eqref{eq:proofthlinC_0sqcupE6} and \eqref{eq:proofthlinC_0sqcupE7} come from \eqref{eq:proofthlinC_0sqcupA2_} and Proposition~\ref{prop:Pstar}.
    
    Assume that $C\left(\textstyle\bigsqcup_{a \in \mathcal{A}} G_a,\; \sum_{a \in \mathcal{A}} P_A(a) P_{X_a}\right) = H(P_A) + \textstyle\sum_{a \in \mathcal{A}} P_A(a) C(G_a, P_{X_a})$, then equality holds in \eqref{eq:proofthlinC_0sqcupE2} to \eqref{eq:proofthlinC_0sqcupE7}. In particular $P_A = P^{\star}_A$ as a consequence of the equality between \eqref{eq:proofthlinC_0sqcupE4} and \eqref{eq:proofthlinC_0sqcupE5}; and $(P_{X_a})_{a \in \mathcal{A}} \in \prod_{a \in \mathcal{A}} \mathcal{P}^{\star}(G_a)$ as a consequence of the equality between \eqref{eq:proofthlinC_0sqcupE6} and \eqref{eq:proofthlinC_0sqcupE7}. Thus, for all $\sum_{a \in \mathcal{A}} P_A(a) P_{X_a} \in \mathcal{P}^{\star}\left(\textstyle\bigsqcup_{a \in \mathcal{A}} G_a\right)$ the following holds: 
    \begin{align}
        & \textstyle C\left(\textstyle\bigsqcup_{a \in \mathcal{A}} G_a,\; \sum_{a \in \mathcal{A}} P_A(a) P_{X_a}\right) \nonumber\\
&= H(P_A) + \textstyle\sum_{a \in \mathcal{A}} P_A(a) C(G_a, P_{X_a})\nonumber \\
        \Longrightarrow \; & C_0\left(\textstyle\bigsqcup_{a \in \mathcal{A}} G_a\right) = \log\left(\textstyle\sum_{a \in \mathcal{A}} 2^{C_0(G_a)}\right),\nonumber\\
& (P_{X_a})_{a \in \mathcal{A}} \in \prod_{a \in \mathcal{A}} \mathcal{P}^{\star}(G_a), \text{ and } P_A = P^{\star}_A.
    \end{align}

\section{Proof of Theorem~\ref{th:mainperfect}}\label{section:proofthmainperfect}

Lemma~\ref{lemma:Hkappasplit} comes from \cite[Corollary 1]{simonyi2001perfect}, and states that the function $P_A \mapsto H_{\kappa}\big(\bigsqcup_{a \in \mathcal{A}} G_a , \sum_{a \in \mathcal{A}} P_A(a) P_{X_a} \big)$, defined analogously to $P_A \mapsto \overline{H}\big(\bigsqcup_{a \in \mathcal{A}} G_a , \sum_{a \in \mathcal{A}} P_A(a) P_{X_a}  \big)$, is always linear. The proof of Lemma~\ref{lemma:perfectunion} is given in App.~\ref{section:prooflemmaperfectunion}.

\begin{lemma}[{{from \cite[Corollary 3.4]{simonyi1995graph}}}]\label{lemma:Hkappasplit}
For all probabilistic graphs $(G_a,P_{X_a})_{a \in \mathcal{A}}$ and $P_A \in \Delta(\mathcal{A})$, we have 
\begin{align}
H_{\kappa}\bigg(\bigsqcup_{a\in\mathcal{A}} G_a , \sum_{a \in \mathcal{A}} P_A(a) P_{X_a} \bigg) = \sum_{a \in \mathcal{A}} P_A(a) H_{\kappa}(G_a,P_{X_a}).
\end{align}
\end{lemma}

\begin{lemma}\label{lemma:perfectunion}
The  graph $\bigsqcup_{a\in\mathcal{A}} G_a$ is perfect if and only if  $G_a$ is perfect for all $a \in \mathcal{A}$.
\end{lemma}

Now let us prove Theorem~\ref{th:mainperfect}.

For all $a \in \mathcal{A}$, let $(G_a,P_{X_a})$ a probabilistic graph where $G_a = (\mathcal{X}_a, \mathcal{E}_a)$ is a perfect  graph. By Lemma~\ref{lemma:perfectunion}, $\bigsqcup_{a \in \mathcal{A}} G_a$ is also perfect; and we have 
\begin{align}
\overline{H}\bigg(\bigsqcup_{a \in \mathcal{A}} G_a, \sum_{a \in \mathcal{A}} P_A(a) P_{X_a} \! \bigg) \!  = \! H_{\kappa}\bigg(\bigsqcup_{a \in \mathcal{A}} G_a, \sum_{a \in \mathcal{A}} P_A(a) P_{X_a} \bigg)
\end{align}
 by Theorem~\ref{th:Hbarperfect}. We also have 
\begin{align}
&H_{\kappa}\bigg(\bigsqcup_{a \in \mathcal{A}} G_a, \sum_{a \in \mathcal{A}} P_A(a) P_{X_a} \bigg)\nonumber\\
& = \sum_{a \in \mathcal{A}} P_A(a) H_{\kappa}(G_a,P_{X_a}) = \sum_{a \in \mathcal{A}} P_A(a) \overline{H}(G_a,P_{X_a})
\end{align}
by Lemma~\ref{lemma:Hkappasplit} and Theorem~\ref{th:Hbarperfect} used on the perfect graphs $(G_a)_{a \in \mathcal{A}}$. Thus 
\begin{align}
    \overline{H}\bigg(\bigsqcup_{a \in \mathcal{A}} G_a, \sum_{a \in \mathcal{A}} P_A(a) P_{X_a} \bigg) = \textstyle\sum_{a \in \mathcal{A}} P_A(a) \overline{H}(G_a,P_{X_a}).
\end{align}

By Theorem~\ref{th:caraclin}, it follows that $\overline{H}\left(\bigwedge_{a \in \mathcal{A}} G_a,\bigotimes_{a\in\mc{A}}P_{X_a} \right) = \sum_{a \in \mathcal{A}} \overline{H}(G_a,P_{X_a}) = \sum_{a \in \mathcal{A}} H_{\kappa}(G_a,P_{X_a})$, where the last equality comes from Theorem~\ref{th:Hbarperfect}.

\subsection{Proof of Lemma~\ref{lemma:perfectunion}}\label{section:prooflemmaperfectunion}
$(\Longrightarrow)$ Let $G = \bigsqcup_{a \in \mathcal{A}} G_a$ be a perfect  graph. Let $a' \in \mathcal{A}$ and $\mathcal{S}_{a'} \subset \mathcal{X}_{a'}$. We have $\chi\big(\big(\bigsqcup_{a \in \mathcal{A}} G_a \big)[\mathcal{S}_{a'}]\big) = \omega\big(\big(\bigsqcup_{a \in \mathcal{A}} G_a \big)[\mathcal{S}_{a'}]\big)$ since $G$ is perfect, and therefore $\chi(G_{a'}[\mathcal{S}_{a'}]) = \omega(G_{a'}[\mathcal{S}_{a'}])$, as $\big(\bigsqcup_{a \in \mathcal{A}} G_a \big)[\mathcal{S}_{a'}] = G_{a'}[\mathcal{S}_{a'}]$. Thus all the graphs $(G_a)_{a \in \mathcal{A}}$ are perfect.

$(\Longleftarrow)$ Conversely, assume that for all $a \in \mathcal{A}$, $G_a = (\mathcal{X}_a, \mathcal{E}_a)$ is perfect. Then for all $\mathcal{S} \subset \bigsqcup_{a \in \mathcal{A}} \mathcal{X}_a$, $\mathcal{S}$ can be written as $\bigsqcup_{a \in \mathcal{A}} \mathcal{S}_a$ where $\mathcal{S}_a \subset \mathcal{X}_a$ for all $a \in \mathcal{A}$, and we have for all $P_A \in \Delta(\mathcal{A})$:
\begin{align}
    &\chi\left(\left(\textstyle\bigsqcup_{a \in \mathcal{A}} G_a \right)[\mathcal{S}]\right)  = \chi\left(\textstyle\bigsqcup_{a \in \mathcal{A}} G_a[\mathcal{S}_a] \right) 
    \nonumber\\
&= \max_{a \in \mathcal{A}} \chi\left(G_a[\mathcal{S}_a]\right) 
    = \max_{a \in \mathcal{A}} \omega\left(G_a[\mathcal{S}_a]\right), \label{eq:proofmainperfectZ3}
\end{align}
and similarly, $\omega\big(\big(\bigsqcup_{a \in \mathcal{A}} G_a\big)[\mathcal{S}]\big) = \max_{a \in \mathcal{A}} \omega\left(G_a[\mathcal{S}_a]\right)$.
Hence $\bigsqcup_{a \in \mathcal{A}} G_a$ is also perfect.




\begin{IEEEbiographynophoto}{Nicolas~Charpenay}
received the M.Sc. degree in Optimization from the ENS Paris-Scalay and the University of Paris-Saclay in 2019. In 2019-2020 he was visiting Bernard Ries at University of Fribourg, Switzerland. He received the PhD degree from the University of Rennes in 2023 under the supervision of Aline~Roumy and Ma\"{e}l~Le~Treust at IRISA UMR 6074 and Inria Center of Rennes University.
\end{IEEEbiographynophoto}

\begin{IEEEbiographynophoto}{Ma\"{e}l~Le~Treust}
is researcher at CNRS since Oct. 2013, working at IRISA UMR 6074 in Rennes, France, since Sept. 2022. He serves as an associate editor for the IEEE Transactions on Information Theory since 2021. Since 2023, he co-organizes  the weekly Game Theory Seminar at Institut Henri Poincar\'{e} in Paris, and the mentorship program at IRISA in Rennes. Since 2022, he serves in the technical program committees of the annual IEEE International Symposium on Information Theory (ISIT). In 2025 he co-organizes the workshop ``30 Years of Game Theory at Institut Henri Poincar\'{e}''. In 2024 and in 2019, he was elected as a member of the scientific council of CNRS Informatics, of which he was secretary from 2019 to 2023. From Oct. 2013 to Aug. 2022, he was researcher at ETIS laboratory UMR 8051, CY Cergy Paris Universit\'{e}, ENSEA, CNRS, where he received the HDR in Dec. 2022. He received the M.Sc. degree in Optimization, Game Theory and Economics (OJME) from the Universit\'{e} de Paris VI in 2008 and the Ph.D. degree in Information Theory from the Universit\'{e} de Paris Sud XI, L2S UMR 8506 Gif-sur-Yvette, in 2011. In 2012-2013, he was a post-doctoral researcher at LIGM UMR 8049 in Marne-la-Vall\'{e}e, France, and at Universit\'{e} INRS and McGill University in Montr\'{e}al, Canada. His research interests include Information Theory, Game Theory, Optimization, Physical Layer Security and Wireless Communications.
\end{IEEEbiographynophoto}

\begin{IEEEbiographynophoto}{Aline~Roumy}
is a Research Director at the Inria Center at Rennes University, France, where she leads a research group on data compression. She received her Ph.D. degree in 2000 from the University of Cergy-Pontoise, France. From 2000 to 2001, she was a research associate at Princeton University, Princeton, NJ, and she joined Inria in 2001. She has held visiting positions at Eurecom and the University of California, Berkeley. Her research interests include statistical signal and image processing, coding theory, and information theory. From 2018 to 2023, she served as an Associate Editor for the IEEE Transactions on Image Processing and received the Outstanding Editorial Board Member Award in 2023. Since 2023, she has been a Senior Associate Editor for the IEEE Transactions on Image Processing.
\end{IEEEbiographynophoto}

\end{document}